\documentclass[aps,prc,twocolumn,floatfix,showpacs,a4paper,
nofootinbib,amsmath,amssymb]{revtex4-2}
\usepackage{graphicx}
\usepackage{dcolumn}
\usepackage{bm}
\usepackage{color}
\usepackage[colorlinks=true, allcolors=blue]{hyperref}
\usepackage{csquotes}
\usepackage{diagbox}
\newcommand{\be}{\begin{equation}}
\newcommand{\ee}{\end{equation}}
\newcommand{\ba}{\begin{eqnarray}}
\newcommand{\ea}{\end{eqnarray}}
\newcommand{\bd}{\begin{displaymath}}
\newcommand{\ed}{\end{displaymath}}
\newcommand{\bea}{\begin{eqnarray}}
\newcommand{\eea}{\end{eqnarray}}

\renewcommand{\vec}[1]{\mbox{\boldmath$#1$}}

\begin{document}

\title{Generalized Effective String Rope Model for the initial stages of Ultra-Relativistic Heavy Ion Collisions}

\author{A. Reina Ramírez$^{1,2}$, V.K. Magas$^{1,2}$, L.P. Csernai$^{3,4,5}$, D. Strottman$^6$}
\smallskip

\affiliation{
$^1$Departament de Fisica Quantica i Astrofisica,  
Universitat\! de\! Barcelona,  Martí i Franquès 1, 08028 Barcelona, Spain\\
$^2$Institut de Ciències del Cosmos,  
Universitat\! de\! Barcelona,  Martí i Franquès 1, 08028 Barcelona, Spain\\
$^3$Institute of Physics and Technology, University of Bergen,
Allegaten 55, 5007 Bergen, Norway\\
$^4$Frankfurt Institute for Advanced Studies (FIAS), Ruth-Moufang-Str. 1, 60438, Frankfurt am Main, Germany\\
$^5$Wigner Research Centre for Physics (RCP), XII. Konkoly Thege Miklós út 29-33, Postbox 49, 1121 Budapest, Hungary\\
$^6$Los Alamos National Laboratory, Los Alamos, 87545 New Mexico, USA
}

\begin{abstract}
We present a Generalized Effective String Rope Model (GESRM), which aims to describe the initial stage of relativistic heavy ion collisions and to give us an initial state for further hydrodynamical calculations. We start with the Effective String Rope Model (ESRM) \cite{PRC6401492001,NPA7121672002} and generalize it in order to take into account fluctuations in the initial state following the Glauber Monte Carlo approach. Results from symmetric nucleus-nucleus collisions at different impact parameters are presented at RHIC and LHC energies; aditionally we study asymmetric A+Au head on collisions.  We also compare the results obtained in the GESRM on event-by-event basis with those received averaging initial states over many events.
\end{abstract}
\date{\today}

\maketitle

\section{Introduction}\label{SEC.I Int}

Relativistic heavy ion collisions allow one to create ultra dense and hot systems, which can undergo a confinement-deconfinement phase transition, leading to the creation of a new state of matter called quark-gluon plasma (QGP) \cite{QGPdiscovery,AKC2014,WF2010}. In the last two decades with the construction of high energy colliders, such as the Relativistic Heavy Ion Collider (RHIC) at BNL and the Large Hadron Collider (LHC) at CERN, as well as with the development of new detectors and new data storing and analyzing methods, the study of heavy ion collisions event-by-event has become possible.

With growing collision energy, the amount of produced particles per event has increased considerably. For example, for central collisions at LHC energies, about $10^4$ particles are produced in a single event. This allows sufficiently precise statistical event-by-event analysis of the fluctuations of observables, such as charged particle multiplicity, particle species ratios, transverse momentum, etc. The analysis of these observables would be able to reveal important information about the properties of the system such as, for example, the order of the phase transition, the presence of a critical point, etc. Thus, the study of fluctuations provides us information, which would be unavailable in the study of averages over a large statistical sample of events.

\begin{figure}[ht]
	\begin{center}
		\resizebox{0.98\columnwidth}{!}
		{\includegraphics{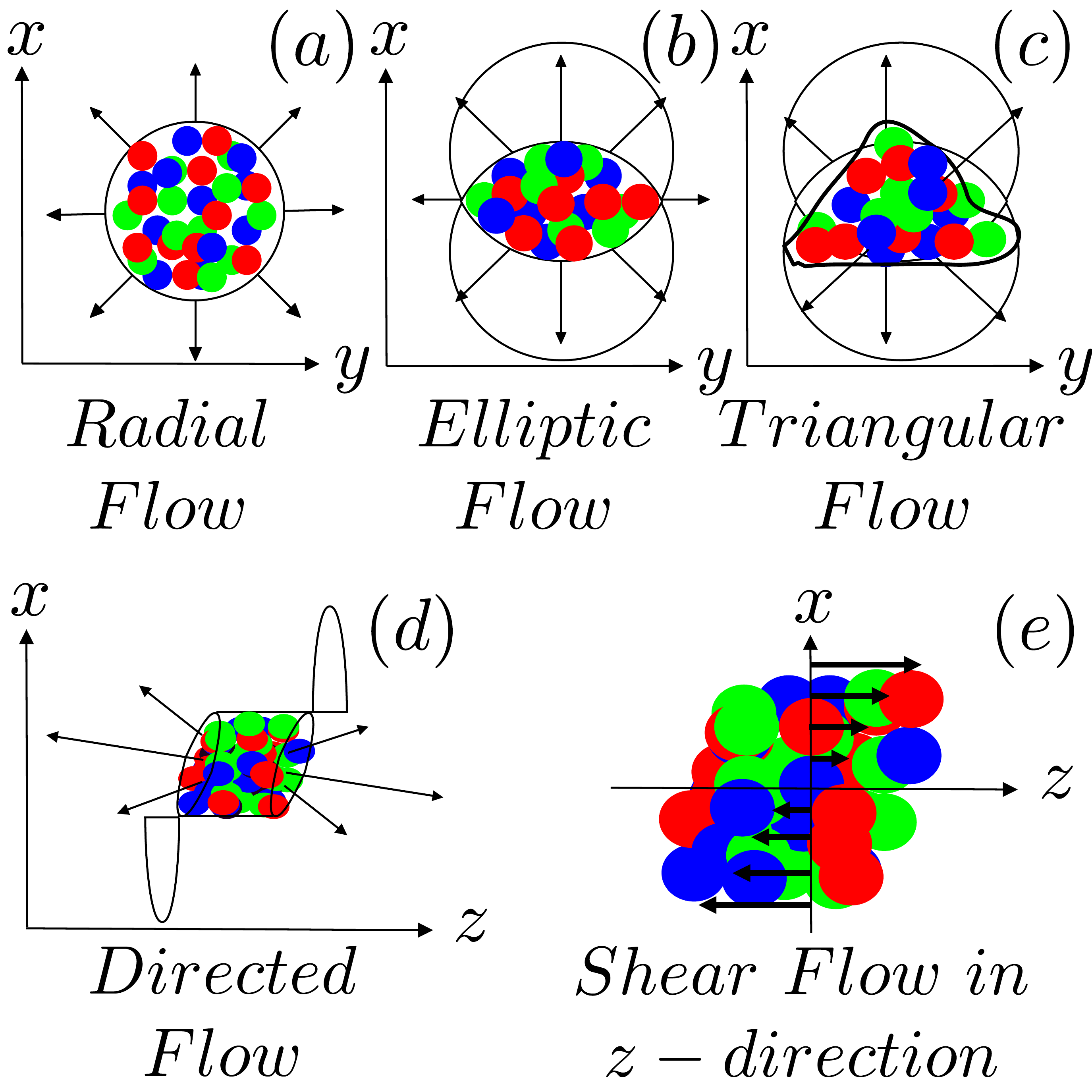}}
		\caption{ (color online)
			Schematic representation of the first four terms in the Fourier expansion of the  azimuthal distribution of emitted particles from high energy heavy ion collisions, Eq. (\ref{MAD}) . The top plots schematically show (in the transverse plane, [xy]-plane)  the radial flow (n=0) [left plot], the elliptic flow (n=2) [center plot], the triangular flow (n=3) [right plot]. The bottom plots schematically show the directed flow (n=1) in the reaction plane ([xz]-plane) [left plot] and the initial flow velocity distribution in the reaction plane [right plot].
		}
		\label{Flows}
	\end{center}
\end{figure}

An important observable in relativistic heavy ion collisions is the azimuthal distribution of the emitted particles. There is a principal difference whether the analysis is done on event-by-event basis or averaging over many events. For example, in head-on collisions, i.e. at zero impact parameter, the average overlapping region between the two colliding nuclei possesses azimuthal symmetry while at non-zero impact parameter it has an \textquotedblleft almond shape\textquotedblright\ giving rise to azimuthal asymmetry (see left and center top plots of Fig. \ref{Flows}). On an event-by-event basis, however, the overlapping region fluctuates around the average geometry, leading to azimuthal anisotropy even at zero impact parameter. 

The azimuthal anisotropy of emitted particles can be quantified by the Fourier expansion of the particles azimuthal distribution as \cite{Niemi:2015qia}:
\be
\frac{dN}{d(\phi-\psi_n)} = \frac{N}{2\pi}\left[1 + 2\sum_n v_n cos\left[n\left(\phi - \psi_n\right)\right] \right], 
\label{MAD}
\ee
where $N$ is the  multiplicity of the produced particles of a given type, $\phi$ is the azimuthal angle of these particles  and $\psi_n$ defines the \textit{n}th-order event plane. For a smooth matter distribution, the second-order event plane $\psi_2$ coincides with the reaction plane (the plane defined by the beam axis and the impact parameter vector). The coefficients $v_n$ are the so-called flow coefficients: $v_0$ is the radial flow, $v_1$ is the direct flow, $v_2$ is the elliptic flow, $v_3$ is the triangular flow, etc. In Fig. \ref{Flows} are sketched  the first four terms of the Fourier expansion (\ref{MAD}). Left-top plot corresponds to the radial flow (n=0), center-top plot to the elliptic flow (n=2), right-top plot to the triangular flow (n=3); all upper plots correspond to the transversal plane. The bottom plots schematically show directed flow (n=1) in the reaction plane.

The flow coefficients $v_1$ and $v_3$ usually separate into two different components: an odd and even component, labeled by $v_{1,3}^{odd}$ and $v_{1,3}^{even}$, respectively. They are dependent on the rapidity $y$ of the emitted particles: $v_{1,3}(y) = v_{1,3}^{odd}(y) + v_{1,3}^{even}(y)$. The odd component of the directed flow $v_1^{odd}(y) = -v_1^{odd}(-y)$ describes the sideward motion of the collective flow and is generated by the pressure gradient in hydrodynamical expansion of the matter created in the heavy ion collisions. The even component $v_{1,3}^{even}(y)$ has its origin in the fluctuations of the initial configuration of nucleons of the colliding nuclei \cite{PRL1200623012018,EPJ204030092019,PRC840249142011,Csernai:2012mh}.  

In relativistic heavy ion collisions at non-zero impact parameter, a gradient in the longitudinal component of the flow velocity is present along the impact parameter direction (see right-bottom plot of Fig. \ref{Flows}), the so-called shear flow in z-direction. The largest values of the flow velocity are reached at the extremes of the system, i.e. close to the projectile and target spectators. 
Such an initial flow velocity configuration, known as a firestreak scenario \cite{Myers:1978bz,Gosset:1978pqf}, has to do with an extremely high angular momentum, which is present in non-central heavy ion collisions (for more recent picture see Ref. \cite{Becattini:2016gvu}). 
Due to angular momentum conservation, in the further evolution of the system this may lead to overall rotation of the reaction volume \cite{PRC840249142011} or/and to a large vorticity of the collective flow, which can manifest itself via polarization of emitted particles \cite{LPC2013,Becattini:2013vja,Csernai:2014ywa,PangEA2016,Karp_Bec1,Ivanov:2020udj,Fu:2020oxj}.  This is supported by experiments like STAR collaboration, which has reported observations of global polarization of $\Lambda$($\bar{\Lambda}$) hyperons at non-zero impact parameter in Au+Au collisions \cite{Nature548622017}.

In sec. \ref{SECII} we discuss the initial state of ultra-relativistic heavy ion collisions and review the Effective String Rope Model (ESRM) developed to describe such state. In sec. \ref{SECIII} we present the Generalized Effective String Rope Model (GESRM) and  explain how the fluctuations have been implemented into the ESRM. In Sec. \ref{SECIV} we present the results obtained from the GESRM simulations for symmetric Au+Au and Pb+Pb collisions and asymmetric A+Au head on collisions at RHIC and LHC energies. We compare the results obtained on event-by-event basis with those obtained averaging over some number of events, $N_{\mathrm{events}}$ (with $N_{\mathrm{events}}$ ranging from 10 to 500000). Finally, in Sec. \ref{CD} we summarize and discuss the main results.

\section{INITIAL STATE OF ULTRA-RELATIVISTIC HEAVY ION COLLISIONS}\label{SECII}

The evolution of the system in relativistic heavy ion collisions is commonly divided into three different stages: an initial stage or pre-equilibrium state, an intermediate stage and a final stage or freeze-out.

The initial state describes the first moments of the collision, i.e. from the time when colliding nuclei pass through each other till the time when local equilibration (or at least some pressure isotropization) is established. The equilibration  is achieved through the collisions among the constituents of the fireball produced in the initial hard parton collisions. Extreme temperatures and densities generated inside the fireball in ultra-relativistic heavy ion collisions lead to formation of QGP. 

An intermediate stage describes the evolution of the fireball from the initial thermalized QGP until the freeze-out stage, and nowadays it is usually simulated within a relativistic hydrodynamical model. The outward thermal pressure of the thermalized QGP acts against the inward pressure exerted by the QCD vacuum. Due to the resulting pressure gradients the fireball will undergo a 3-dimensional hydrodynamic expansion. As it expands, the temperature decreases and when it drops below a certain value, the so-called critical temperature $T_c$, a deconfinement-confinement phase transition will take place and quarks and gluons will hadronize. 
The system will further expand and cool down. At some moment the average distance between hadrons will become larger than the strong interaction range and, thus, the number of such interactions will drastically decrease, and finally hadrons will freely move to the detectors. This process is called freeze-out. At this stage a hydrodynamic description of the system is no longer valid and some transport model, such as the Ultra-relativistic Quantum Molecular Dynamics model (UrQMD) \cite{JPG2518591999}, should be used to describe the evolution of hadron spectra during the freeze-out process. One of the extreme, but frequently used, assumptions is that freeze out happens on an infinitely narrow hypersurface in the space-time. On its inner part we have a collective matter, described by hydrodynamics, on its outer part we have an ideal gas of different hadron species with momentum distributions generated according to the Cooper-Frye formula \cite{PRD101861974}. 

These different stages of the ultra-relativistic heavy ion collision clearly manifest their presence in the models, which intend to simulate such reactions.  For example, most of the models, which account for the initial angular momentum and flow vorticity, and therefore can reproduce the observed
polarization of $\Lambda$ hyperons,  \cite{LPC2013,Csernai:2014ywa,PangEA2016,JPCS6120120522015,Karp_Bec1,Ivanov:2020udj,Fu:2020oxj}, do actually have a so-called Multi-Module structure, introduced in \cite{MMM,PRC6401492001}, where each stage of the collision is described with different, most suitable theoretical approaches. In particular:\\
- Ref. \cite{LPC2013,Csernai:2014ywa}: ESRM + (3+1)D hydro;\\
- Ref. \cite{PangEA2016}: event-by-event fluctuating initial conditions from A MultiPhase Transport Model + (3+1)D viscous hydro;\\
- Ref. \cite{JPCS6120120522015}:  UrQMD + (3+1)D viscous hydro + UrQMD;\\
- Ref. \cite{Karp_Bec1}:  Monte Carlo Glauber model + (3+1)D viscous hydro;\\
- Ref. \cite{Csernai:2012mh,Fu:2020oxj} event-by-event fluctuating initial conditions from A MultiPhase Transport Model + (3+1)D viscous hydro+UrQMD.\\
Only in Ref. \cite{Ivanov:2020udj} simulations are performed within a single model based on three-fluid dynamics, which takes into account non-equilibrium at the early stage of nuclear collisions by means of two counter-streaming baryon-rich fluids. 

As we can see, relativistic fluid dynamics is commonly used 
to describe the intermediate stage of the reaction, in which the QGP is assumed to be in a local thermal equilibrium. The final stage of the reaction is simulated either based on freeze out hypersurface or using the  hadron cascade model UrQMD.  
The initial non-equilibrium stage of the collision is, however, the most problematic to be simulated. The main differences among the models are in selecting initial conditions at which the hydrodynamical description becomes valid. 

On the other hand we can see that the majority of the most recent models use event-by-event fluctuating initial state. This allows them to analyze simulated events in the same way as the real experiments are performed, as it was discussed in the Introduction, and thus maximize the obtained information. Observing such a trend we decided to update the ESRM for the initial state \cite{PRC6401492001,NPA7121672002} by combining it with the Glauber Monte Carlo approach.

\subsection{The Effective String Rope Model}\label{ESRM}
The early stage dynamics of the nucleus-nucleus reaction at low energies (few GeV/nucl) is rather different from that at higher energies. Simplifying the situation - at low energies stopping dominates (Landau model), while at high energies we start seeing the signs of transparency. On the other hand, the over-idealized Bjorken hydrodynamics, based on complete transparency, which leads to boost invariance and zero baryon chemical potential at mid-rapidity, is not directly applicable even at ALICE at LHC energies. 

In order to find the golden mean between Landau and Bjorken initial state scenarios at the beginning of the RHIC era a new Effective String Rope Model \cite{MMM,PRC6401492001,NPA7121672002} was proposed  to produce an initial state for further 3+1D relativistic fluid dynamical evolution. 
Its initial point is the transparency of the colliding nuclei, but then the baryon recoil is taken into account via longitudinal chromo-electric string fields\footnote{Historically the first attempt of this type was done in \cite{Gyulassy:1986fk} for pA collisions.}, called “string ropes”, because these appear to be an order of magnitude stronger than the classical hadronic strings with approximately 1 GeV/fm string tension. Thus, ESRM can be applied  for RHIC energies or higher ones, for example in  \cite{PRC840249142011} it was applied for Pb+Pb reactions at ALICE at LHC.

The big advantage of this initial state, in comparison to others available at that time, was that it reflected correctly not only the energy-momentum, but also angular momentum conservation laws. Consequently, such an initial state for  non-central ultra-relativistic heavy ion collisions showed a rather large flow vorticity \cite{LPC2013,Becattini:2013vja,Csernai:2014ywa}, and even an
effective rotation of the whole fireball has been observed once the ESRM was applied to simulate Pb+Pb collisions at ALICE at LHC \cite{PRC840249142011}.

\begin{figure}[ht!]
 \begin{center}
     \resizebox{0.8\columnwidth}{!}
    {\includegraphics[scale=1]{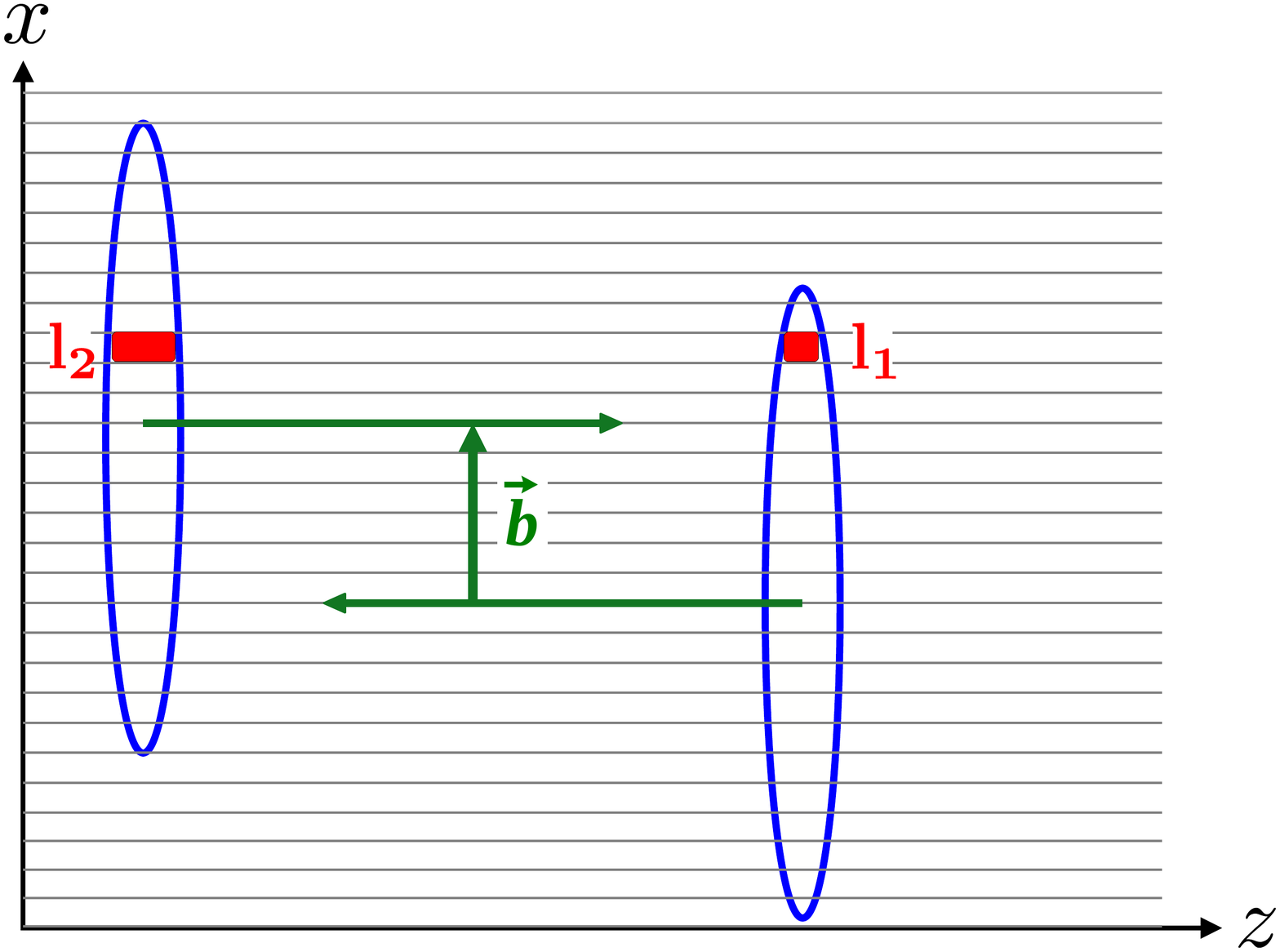}}\\
    \resizebox{0.9\columnwidth}{!}
    {\includegraphics[scale=1]{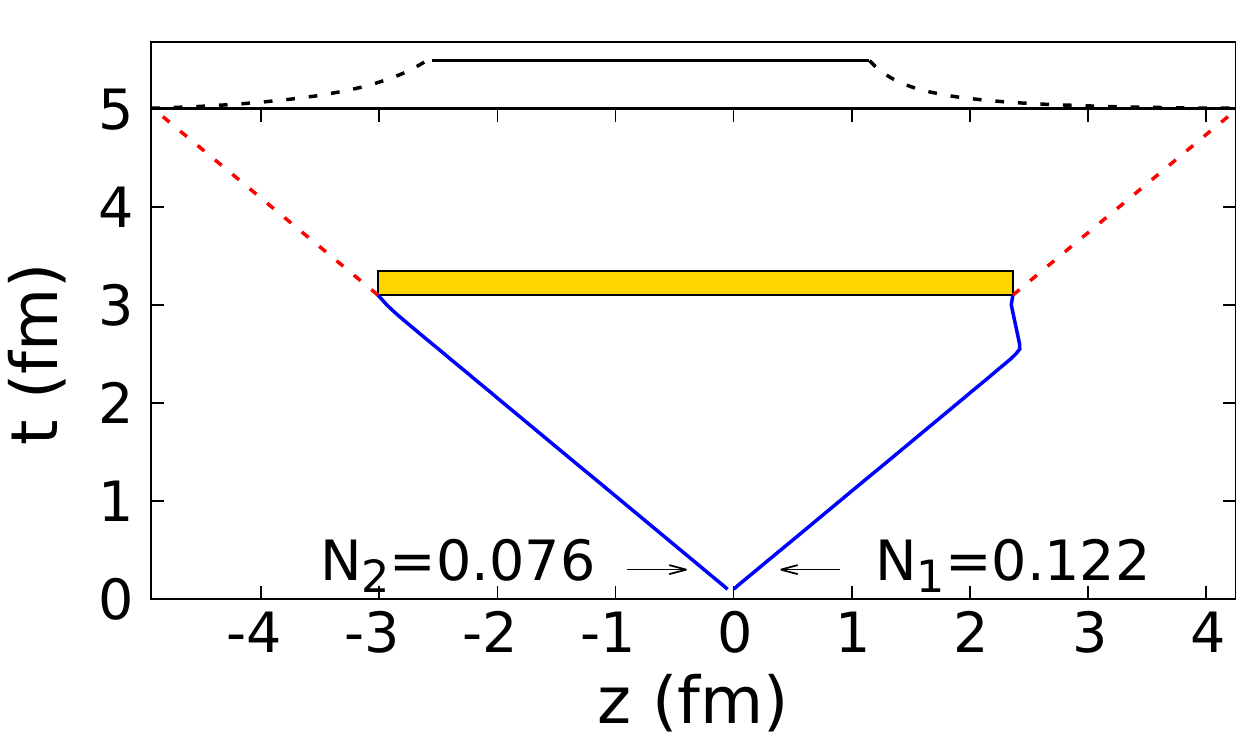}}
    \caption{(color online) Top plot: Sketch of two relativistic heavy ions just before a collision at impact parameter $\vec{b}$ (in terms of streak-streak collisions), each of which will only happen for those streaks with the same transverse coordinates $\left\lbrace x_i,y_i \right\rbrace$. Bottom plot: Example of the space-time evolution of two colliding streaks with given $N_1$ and $N_2$ number of nucleons. At some $t_{streak}=$Max$\{ t_{i,turn}\}$ we form a homogeneous final streak, sketched in the figure, which starts expansion into the vacuum  with velocity of light according to the analytical solution \cite{Rischke:1995ir,NPA7121672002}. Finally, at $t=t_{fin}=5$ fm, our final streak has an energy density profile shown at the top of the plot.
    }
    \label{SSCPT}
  \end{center}
\end{figure}

A first step in this model is the creation of a grid in the plane transverse to the beam ($[xy]$-plane). The collision between two nuclei is then described as a set of independent streak-streak collisions corresponding to the same transverse coordinates $\left\lbrace x_i,y_i \right\rbrace$. In the top plot of Fig. \ref{SSCPT} such a collision is sketched in the reaction plane. The length of each colliding streak is calculated assuming a uniform nuclear matter distribution. Those streaks corresponding to the target, right nucleus, are labeled by $l_1$ and those corresponding to the projectile, left nucleus, are labeled by $l_2$. 
The space-time evolution of the streaks is governed by the chromo-electric field generated by the color exchange of colliding partons, and by the energy-momentum and baryon charge conservation laws. The bottom plot of Fig. \ref{SSCPT} illustrates such an evolution corresponding to two peripheral colliding streaks. At $t=0$ both streaks come into contact. The kinetic energy of partons is so high that it is assumed that these can go through the opposite slab of matter without stopping. Only when they have completely passed through each other the chromo-electric flux tube or color string is created (extending up to the target and projectile streak ends), which will slow down and stop the matter. This field is assumed to be uniform as a string, or it is better to say several parallel strings $\equiv$ string rope, because its string tension, $\sigma$, is much higher than that for ordinary hadronic strings.  In the ESRM \cite{MMM,PRC6401492001,NPA7121672002} this uniform field strength was calculated in the following way: 
\be
\sigma =
A\left( \frac{\varepsilon_0}{M_n}\right)^2n_0\sqrt{l_1l_2}  =
A\left( \frac{\varepsilon_0}{M_n}\right)\frac{\sqrt{N_1N_2}}{\Delta x \Delta y},
\label{Oldsigma}
\ee
where $M_n$ is the nucleon mass, $\varepsilon_0$ is the initial energy per nucleon in the Center-of-Mass frame (=Lab frame for RHIC), $N_1$ and $N_2$ are the baryon charges of the corresponding colliding streaks, and $\Delta x\Delta y$ is the cross section of the streaks. The typical values of dimensionless parameter $A$ are around 0.06-0.08 (for $\sigma$ measured in GeV/fm). 
The typical values of $\sigma$ are 6-15 GeV/fm for $\varepsilon_0$ $=$ 100 GeV, and an order of magnitude higher for $\varepsilon_0$ $=$ 1.38 TeV.

The key point of the ESRM is the exact conservation of the energy and momentum, in contrast, for example, with the Bjorken scenario. At the moment when the colliding streaks have just passed through each other and have created the chromo-electric field with string tension, $\sigma$, the partons from the colliding streaks still move with the initial rapidities $y_0$ and $-y_0$ (in CM frame). Then they are slowly loosing their velocity/rapidity, since part of their kinetic energy is converted into an energy of the field, which is stretched in between. The equations which govern such a motion can be found in the original ESRM publications  \cite{PRC6401492001,NPA7121672002}, and we do not want to include all those here, but qualitatively the evolution we look like follows. 
Colliding streaks are gradually decelerated, at some moment, $t=t_{i,turn}$, they stop (the smaller one stops earlier) and start to move backward being gradually accelerated (the string rope length is decreasing – the field energy is converted into kinetic energy of the partons). And if we don’t make any additional assumption these streaks will reach $y_0$ and $-y_0$ rapidities again (in the opposite directions), will go through each other again, will create the same field $\sigma$ again and the oscillation will repeat, i.e we will observe the Yo-Yo-like motion known from the string theory.     

This scenario of the contraction of the reaction volume is unlikely to occur due to string fragmentations and string-string interactions. In the ESRM it is assumed that at $t_{streak}=$Max$\{ t_{i,turn}\}$ a uniform streak of length $\Delta l_f$, defined by the motion of the outer edges of the colliding streaks, is formed. The uniform energy, $e_f$, and baryon, $n_f$, densities of this final streak as well as its  unique rapidity, $y_f$, are calculated from the energy, momentum and baryon charge conservation laws, and the corresponding Equation of State (EoS). We assume for the QGP a Stefan-Boltzmann gas in the Bag EoS, i.e. $p=e/3-4B/3$, where $p$ is the gas pressure density, $e$ is the energy density and $B$ is the Bag constant (more details will be shown in section \ref{mu_T_S}).

Once this final streak is formed it starts to expand into the vacuum with the velocity of light according to the analytical solution \cite{Rischke:1995ir,NPA7121672002}. 

The bottom plot of Fig. \ref{SSCPT} shows the trajectory of leading partons from $t=0$ up to $t=t_{streak}$, blue solid lines, and the further expansion, red dashed lines. 
The analytical solution for the expansion of the homogeneous final streak into the vacuum \cite{Rischke:1995ir,NPA7121672002} allows us to know the energy and baryon density distributions and the local rapidity along the streak. This is valid until the expansion waves coming from the edges with the velocity of sound did not cross in the middle of the streak. 

Thus, considering the overall reaction volume, we can present the initial state, i.e. energy density, baryon density and flow distributions of the fireball at some time moment $t=t_{fin}$. This time, $t_{fin}$ is well chosen  if at this moment the final streaks have already been formed and started their expansion for most of the transverse coordinates, and, at the same time, none of them has reached already the moment when the analytical expanding solutions are not applicable \cite{NPA7121672002}. Please note that by assumption we neglect the transverse expansion in the model until  $t=t_{fin}$; thus, at this moment all the cells have only longitudinal velocity. 

\begin{figure}[ht]
	\begin{center}
		\resizebox{0.98\columnwidth}{!}
		{\includegraphics{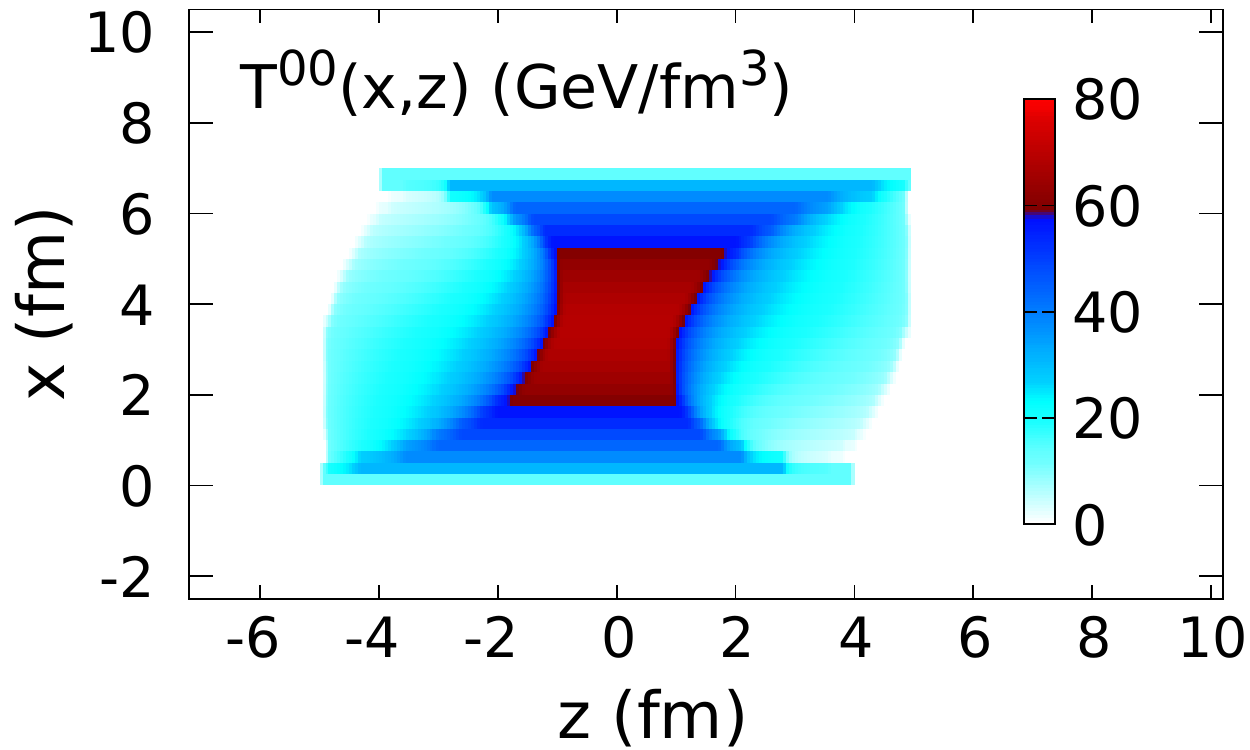}}
		\caption{ (color online)
			Energy density distribution in the reaction plane ([xz]-plane) obtained from the ESRM. The results correspond to symmetric Au+Au collisions at $\sqrt{S_{NN}}=$ 200 GeV, $b=(R_{Au}+R_{Au})/2$, $A=0.0784$, $t_{fin}=$ 5 fm.}
		\label{ESRMEDRP}
	\end{center}
\end{figure}

\begin{figure}[ht]     
	\begin{center}
		\resizebox{0.98\columnwidth}{!}
		{\includegraphics{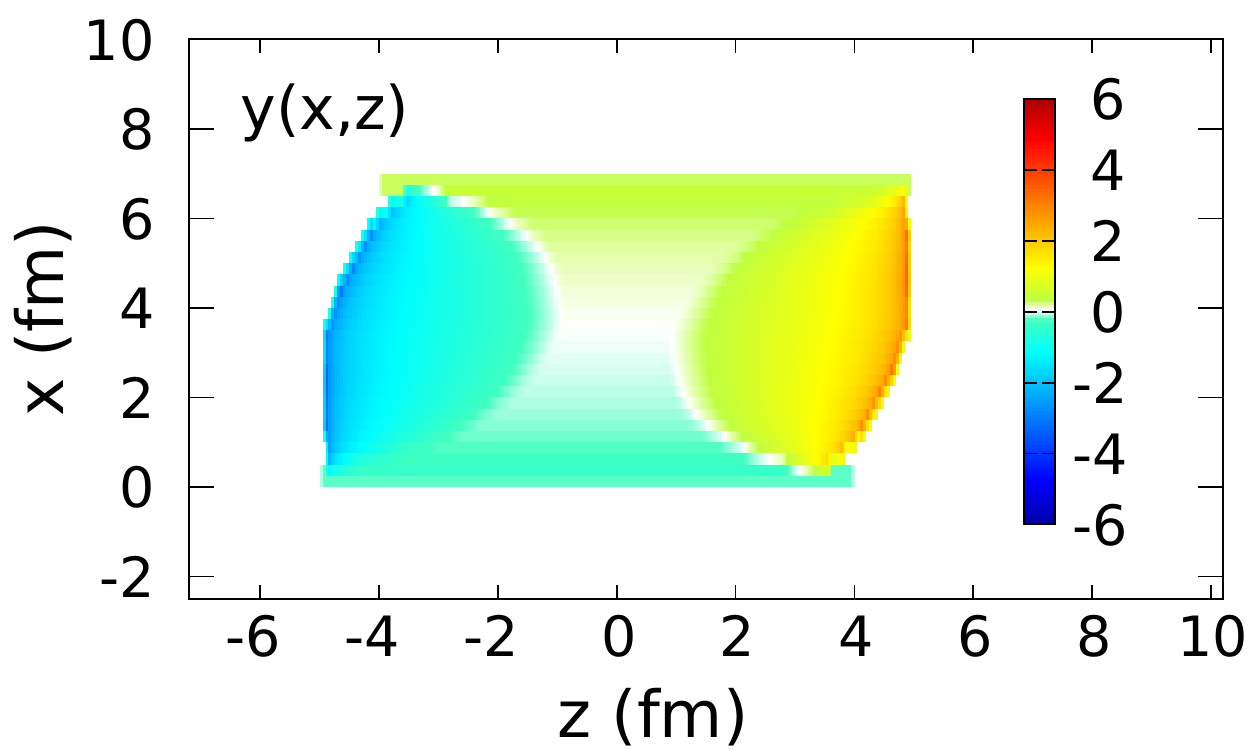}}
		\caption{ (color online)
			Rapidity distribution ($y$) in the reaction plane ([xz]-plane) for the same calculation as in Fig. \ref{ESRMEDRP}.
}
		\label{ESRMVDRP}
	\end{center}
\end{figure}

In Figs. \ref{ESRMEDRP} and \ref{ESRMVDRP} we show the energy density and rapidity distributions in the reaction plane obtained from the ESRM with final expanding streaks for symmetric Au+Au collisions at initial energy  $\varepsilon_0=$100 GeV per nucleon (i.e. $\sqrt{S_{NN}}=$ 200 GeV) for impact parameter $b_0=0.5$. In general case in our notation the module of the impact parameter $\vec{b}$ is given by 
\be 
|\vec{b}|=b_0\cdot(R_1+R_2)\,.
\label{CCON}
\ee

From the energy density distribution one can see that, at non-zero impact parameter, the system forms a type of tilted disk, and, thus, the direction of fastest expansion, generated by the strongest pressure gradient, will deviate from both the beam direction and the transverse flow direction, giving rise to a new flow component called \textit{third flow} component or \textit{antiflow} \cite{Csernai:1999nf}. 

The rapidity distribution shows that while in the central zone of the collision partons move rather slowly, those at the ends of the streaks move with a much higher velocity.
Obviously the initial flow distribution, shown in Fig. \ref{ESRMVDRP}, will generate a high vorticity; this quantity will be discussed in more detail in section \ref{IV.vort}.

\section{GENERALIZED EFFECTIVE STRING ROPE MODEL}\label{SECIII}

In this section we present the Generalized Effective String Rope Model, in which the fluctuations in the initial state of relativistic heavy ion collisions are taken into account following the Glauber Monte Carlo  approach \cite{GMHENC}.

\subsection{The Glauber Monte Carlo approach}\label{GMCA}

In the Glauber Monte Carlo approach fluctuations are introduced  randomly distributing positions of the nucleons, which allows us to obtain different configurations of those for each colliding nucleus in each collision. Therefore, the number of participant nucleons will now fluctuate event-by-event leading to fluctuations in the different physics quantities such as baryon charge, energy, the total momentum and consequently the central rapidity.

\subsubsection{Nucleon random distribution}\label{NRD}

A first step in the implementation of fluctuations in the frame of the Glauber Monte Carlo approach consists in random distribution of nucleons within the nucleus. In this work we have used different nuclear matter density distributions depending on the mass number of the colliding nuclei. For example, for a Au nucleus we have used a Woods-Saxon distribution while for a Pb nucleus we have used a 2-parameter Fermi distribution (2pF). As an illustration of this procedure, we'll use a Au: First we randomly generate four numbers $\left\{x,y,z,\delta\right\}$, where $\left\{x,y,z \right\}$ are Cartesian coordinates and $\delta$ is a random number, which takes the values in the range 0 $\leq$ $\delta$ $\leq$ 1. If the following condition is satisfied
\be
\delta \leq \rho_{WS}(x,y,z) = \frac{\rho_0}{1+exp \!\left \{ \!\left( \sqrt{x^2+y^2+z^2}-R\right)/a \right \}},
\label{CCON}
\ee
then we take the numbers $\left\{x,y,z\right\}$ as the coordinates of the center of a given nucleon, otherwise, we generate a new set of random numbers until this condition is satisfied. We repeat this process until all nucleons be randomly distributed. In this way we ensure that, in the average, our nucleons are distributed according to a Woods-Saxon (WS) nuclear matter density distribution. The skin depth $a$ and the nuclear radius parameter $R$ have been taken from Ref. \cite{GMHENC} corresponding to a $^{197}$Au nucleus, namely $a_{Au}=0.535$ fm and $R_{Au}=6.38$ fm. 
Please note that in the ESRM colliding Au nuclei have been considered as a homogeneous spheres with the radius of  $7$ fm, and thus using the above WS distribution we change a bit the geometry of the collision; and now impact parameter $b_0=0.5$, i.e. $b=0.5(R_{Au}+R_{Au})=R_{Au}=6.38$ fm, doesn't mean the same as in original ESRM calculations \cite{PRC6401492001,NPA7121672002}.

\begin{figure}[ht]     
	\begin{center}
		\resizebox{1.1\columnwidth}{!}
		{\includegraphics{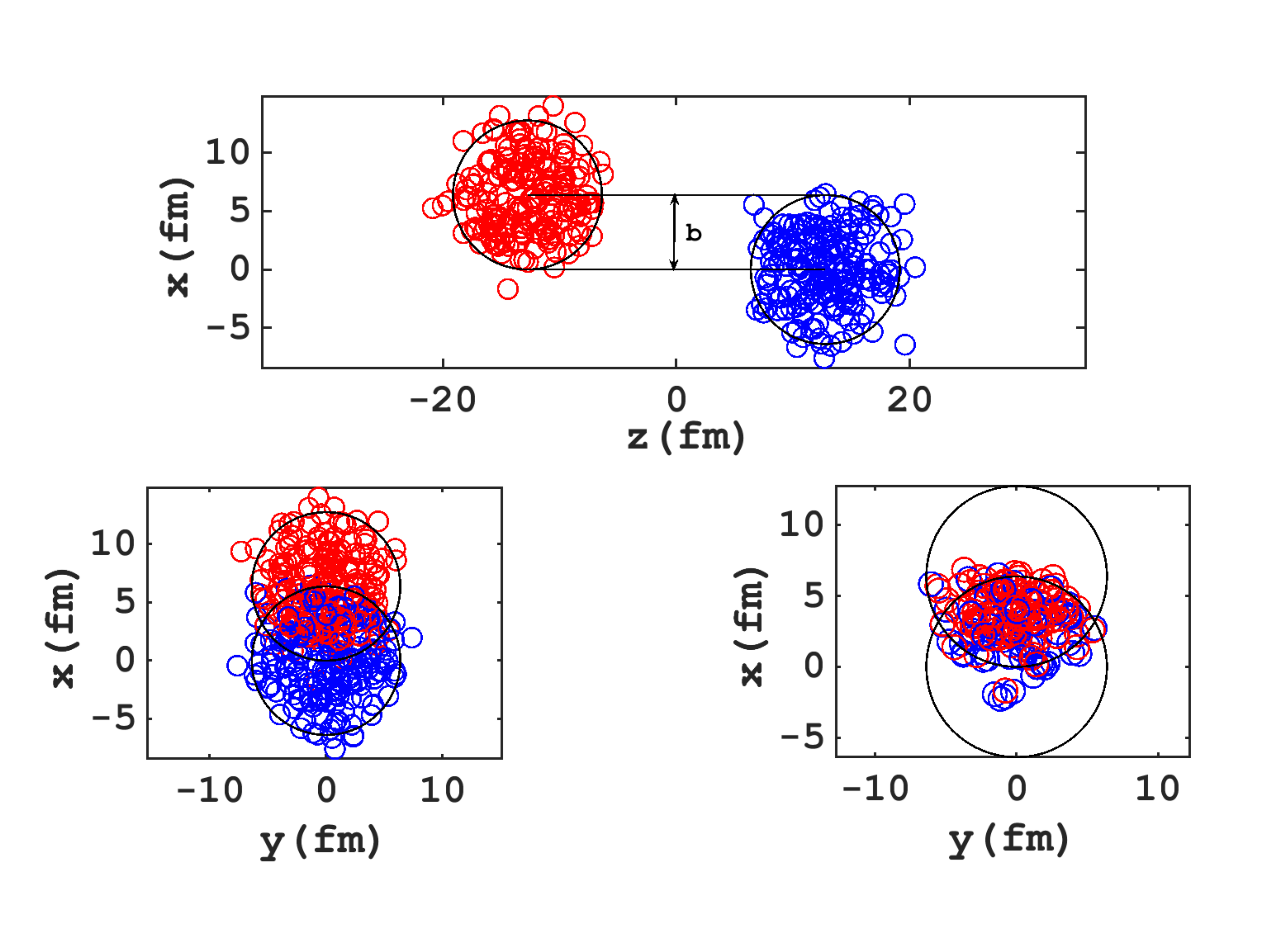}}
		\caption{ (color online)
			 Fluctuations in the initial state following the Glauber Monte Carlo approach. Top plot represents a random nucleon distribution in the reaction plane ([xz]-plane). Bottom plots represent the same distribution in the transverse plane ([xy]-plane). In the right bottom plot only the participant nucleons are shown.
		}
		\label{3DNRD}
	\end{center}
\end{figure}

The second step is to consider each nucleon as a sphere of radius $R_N=0.842$ fm \cite{PDG}, in which the nucleon baryon charge is homogeously distributed. In Fig. \ref{3DNRD} is shown an example of two given configurations of nucleons corresponding to two $^{197}$Au nuclei just before a collision at impact parameter $b_0=0.5$. These distributions are shown in the reaction plane, top plot, and in the transverse plane, bottom plot. As can be seen, the \textquotedblleft almond-shape\textquotedblright of the overlapping region between both nuclei is not observed on event-by-event basis. Instead, this will fluctuate around the average geometry producing any order geometric deformations. The almond-shape of the overlapping region can be restored averaging over many events.

\subsubsection{Nucleonic matter discretization}\label{NMD}

Once the nucleons have been randomly distributed, we generate a two-dimensional grid in the transverse plane, in which the cell-size is taken to be: $\Delta x\Delta y = (R_{Au}/10)(R_{Au}/10)$, and we calculate the baryon charge in each cell. However, there is no an easy analytical way to perform this calculation. So instead we will follow a Monte Carlo approach and consider each nucleon as a set of fictitious particles randomly distributed within an homogeneous sphere of $R_N$ radius. This method allows us to obtain the baryon charge in each transverse cell in an easy and fast way.

Each of these fictitious particles are associated with a certain baryon charge whose value depends on the total number of these. In our case, we have used 1000 fictitious particles per nucleon, and thus the  baryon charge associated to each particle is $10^{-3}$. The total baryon charge in each transverse cell is given by the sum of the charges of all fictitious particles within it.

For a given nucleon, we generate a set of three random numbers $\left\{x,y,z\right\}$ within the intervals; $x_i-R_N \leq x \leq x_i+R_N$, $y_i-R_N \leq y \leq y_i+R_N$ and $z_i-R_N \leq z \leq z_i+R_N$, where  $\left\{x_i,y_i,z_i\right\}$ are the coordinates of the center of the $i$th nucleon. If the set of random numbers $\left\{x,y,z\right\}$ satisfied the relation
\be
	(x-x_i)^2+(y-y_i)^2+(z-z_i)^2\leq R_N,
\ee
we take these numbers as the coordinates of a given fictitious particle, otherwise, we generate a new set of random numbers until this condition is satisfied.

\begin{figure}[h!]  
	\begin{center}
		\resizebox{1.1\columnwidth}{!}
		{\includegraphics[width=\linewidth]{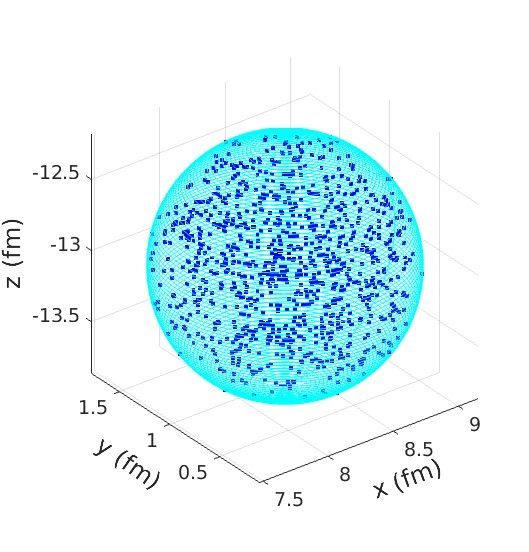}}
		\\
		\resizebox{1.0\columnwidth}{!}
		{\includegraphics[width=\linewidth,angle =0]{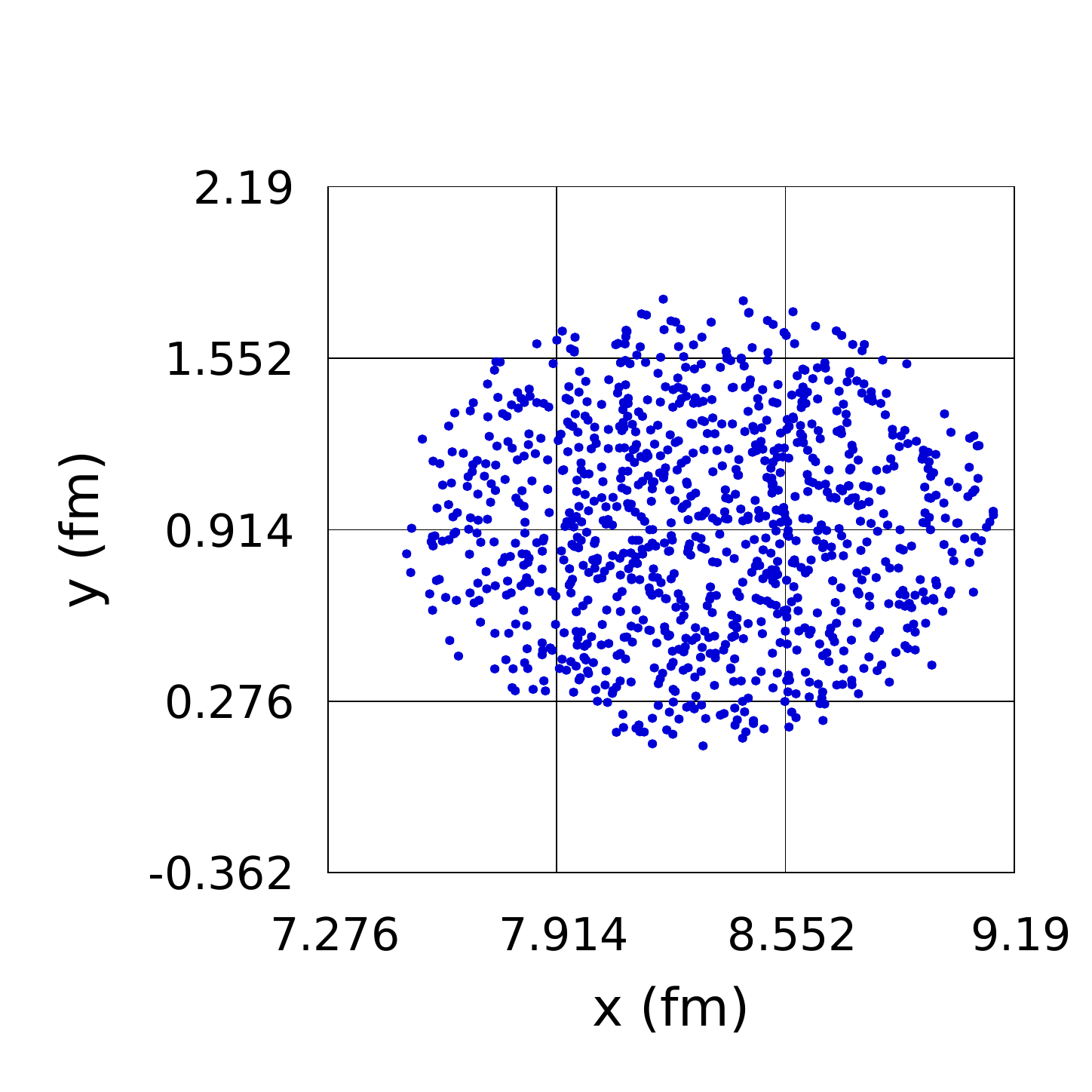}}
		
		\caption{ (color online)
			Hereby we illustrate the method used to compute the total baryon charge in each transverse cell. 3D random distribution of fictitious particles within a given nucleon is shown in the top plot, and the same distribution projected on the transverse plane ([xy]-plane) is represented in the bottom plot. 
		}
		\label{FPD}
	\end{center}
\end{figure}

In Fig. \ref{FPD} an example of such a distribution for 1000 fictitious particles is shown for a given nucleon. The top plot illustrates the fictitious particle distribution in 3 dimensions, while in the bottom plot the same distribution projected in the transverse plane is represented. The size of each cell of the grid has been chosen to be equal to the one used in our calculations, i.e. $\Delta x=\Delta y = R_{Au}/10=0.638$ fm. 

\subsection{Implementation of fluctuations in the ESRM}

\begin{figure}[h!]  
	\begin{center}
		\resizebox{1.00\columnwidth}{!}
		{\includegraphics[width=\linewidth]{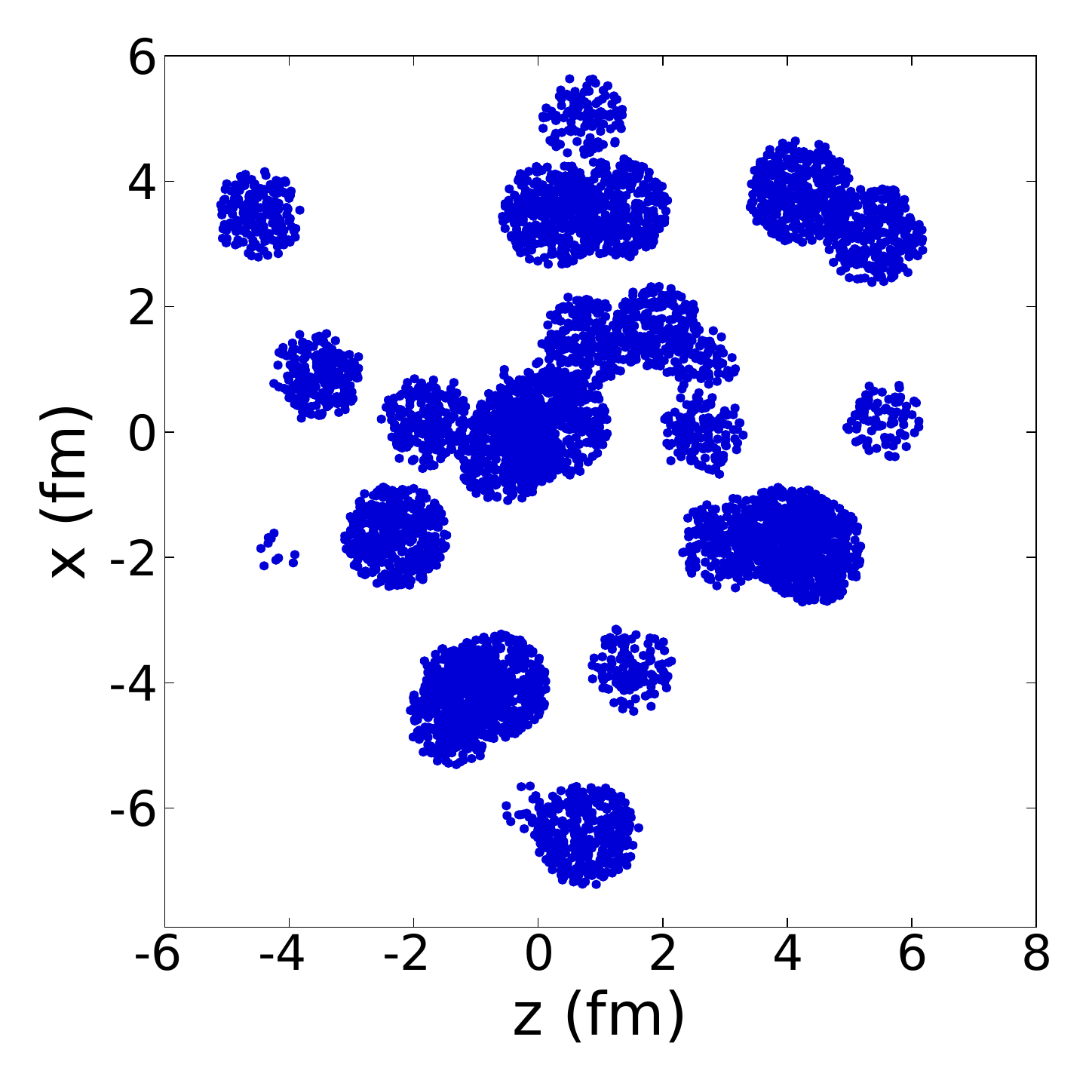}}
		\\
		\resizebox{0.75\columnwidth}{!}
		{\includegraphics[width=\linewidth,angle =90]{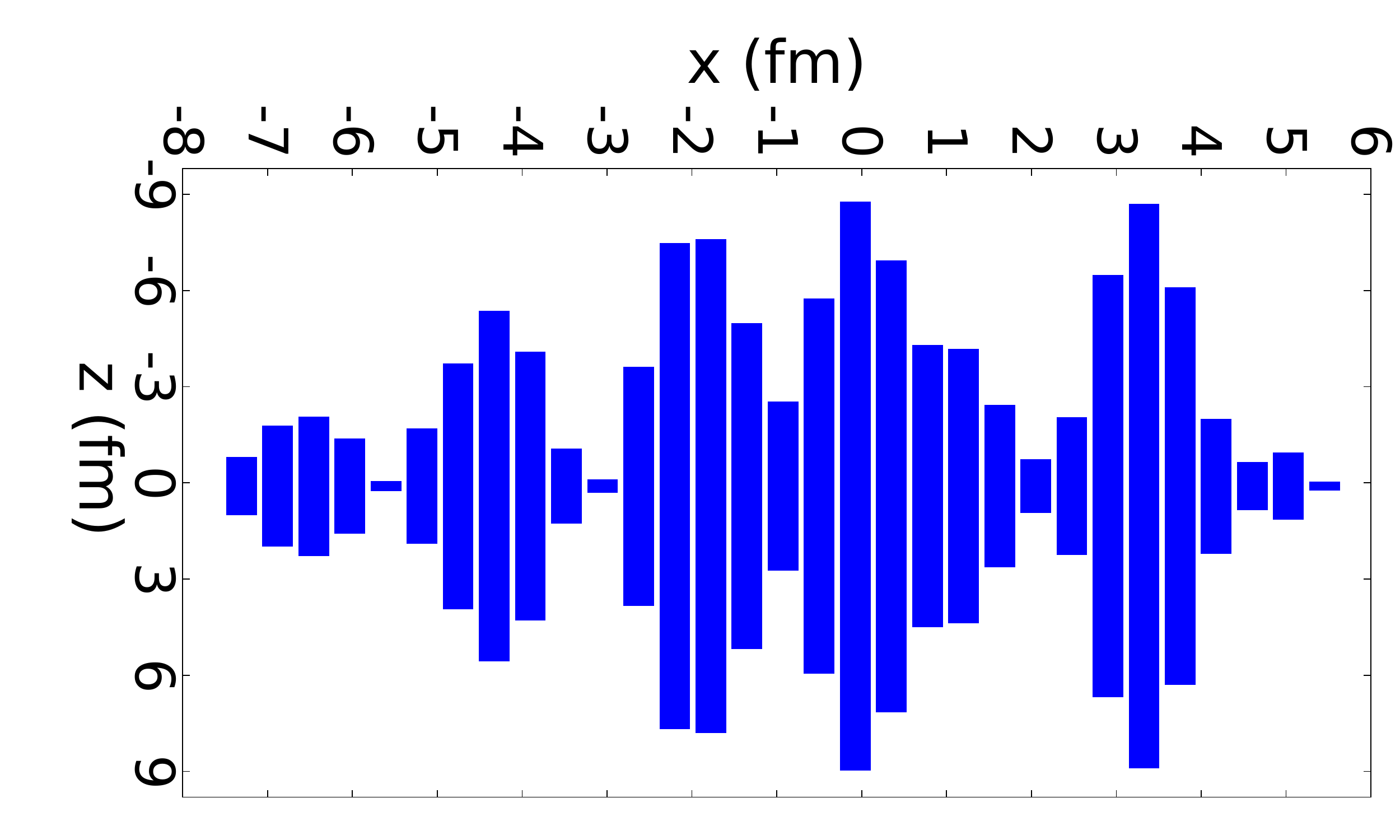}}
		
		\caption{ (color online)
			Top plot: random baryon charge distribution in the reaction plane ([xz]-plane) for a nucleus at rest. Bottom plot: longitudinal streak lengths obtained from the top plot distribution using the Eq. (\ref{L_in}).
		}
		\label{PFRP}
	\end{center}
\end{figure}

In the ESRM main equations are related to the lengths, $l_1$ and $l_2$, of the colliding streaks \cite{PRC6401492001,NPA7121672002}. Now we have the baryon charge in each transverse cell, $N_1$ or $N_2$, and its distribution along the $z$-direction. In the top plot of Fig. \ref{PFRP} an example of the baryon charge distribution in the reaction plane for a nucleus at rest is given. In the Glauber Monte Carlo approach the nucleons are homogeneous spheres of the radius $R_N$, randomly distributed according to the corresponding WS distribution function. Thus, the reaction plane will cut some of these nucleon spheres, what means that we should see them as circles with radius $0< r\le R_N$. Since the randomly distributed nucleons can overlap – some of these circles may overlap. And since we simulate a nucleon as a set of 1000 ficticious particles, homogeneously distributed within the sphere of radius $R_N$, in the reaction plane we see the groups of these particles homogeneously distributed within the corresponding circles. 

Most impotantly, we can see that, due to the random position of nucleons,  the actual length of the streak, $[z_{min}, z_{max}]$, has nothing to do with its baryon content: there will be “long” steaks with matter only at the edges and zero in the middle, and there will be “short” streaks with a very high baryon content due to overlap of several nucleons in this region. However, since the colliding streaks have to go through each other, their real lengths do not matter; what matters is their baryon content, because the number of parton-parton collisions will be proportional to $N_1 \cdot N_2$. In this way the number of the individual hadronic streaks, which will form the “string rope” after interpenetration and thus define the string tension $\sigma$, will also depend on the same product, not on the actual lengths of colliding streaks. Please also note that the colliding nuclei  will be Lorentz contracted, what will  make for ultra-relativistic reaction all these streak lengths rather small anyway.    

On the other hand, in the ESRM with homogeneous nuclei the baryon content was directly proportional to the streak length and all the formulas for the further streak evolution were written in term of $l_1$ and $l_2$  \cite{PRC6401492001,NPA7121672002}. Therefore, in order to be able to use the ESRM core for streak-streak collision without modifications, and since the actual length of the streaks is not so important anyway, we will define effective streak lengths in the following way. For each streak, containing baryon charge  $N$, we will assume that this baryon charge is homogeneously distributed in a volume $\Delta x\, \Delta y\, l$, where $\Delta x\, \Delta y$ is the transverse area of each cell and $l$ is the length of the streak. Thus, effective streak lengths are given by 
\be
	l_{i,j}^{\alpha} = \frac{N_{i,j}^{\alpha}}{\rho_0 \Delta x\Delta y}\,
	\label{L_in}
\ee
where $\rho_0$ is the normal nuclear matter density and $N_{i,j}^{\alpha}$ is the baryon charge in the transverse cell ($i,j$) corresponding to the $\alpha$th nucleus. In the bottom plot of Fig. \ref{PFRP} the length of each streak, obtained from the total baryon charge distribution of the top plot, is illustrated. Due to fluctuations, the largest streaks are not necessarily located in the middle region of the nucleus but these can be now in any place, even at the top and bottom  extremes, leading to fluctuations around the average geometry. Thus, the spherical symmetry assumed in a $^{197}$Au nucleus at rest is no longer true on event-by-event basis. However, averaging over many events the geometry can be restored. 

Describing a collision of two nuclei which are moving with relativistic energies it is necessary to take into account the Lorentz contraction, and, thus, Eq. (\ref{L_in}) is generalized to
\be
l_{i,j}^{\alpha} = \frac{N_{i,j}^{\alpha}}{\rho_0\gamma\Delta x\Delta y}.
\label{L}
\ee
Fluctuations on the number of colliding baryons, which is equivalent, according to Eq. (\ref{L}), to fluctuations in the lengths  of colliding streaks,  could lead to formation of holes and cells with low baryon charge in the central zone of the overlapping region between both colliding nuclei. 
The string tension, $\sigma$, of the chromo-electric field formed from the collision of two of such streaks will be very small. 
On the other hand, to maintain the average, in almost every event there appear some streaks (one or more) which generate a rather high string tension, and consequently will form the final streak in a short time and, as it was discussed above in section \ref{ESRM},  these will limit the maximal time when we should stop our model and fix the produced initial state. 
 
If the old definition of string tension is used, see Eq. (\ref{Oldsigma}), then, first of all, we have to reduce the final time, $t_{fin}$, from $5$ fm, to $4.5$ fm, for $A=0.65$, or even less for higher $A$. At the same time we observe that there was not enough time to start the final streak expansion for most of the transverse cells. 

\begin{figure}[ht]
	\begin{center}
		\resizebox{1.00\columnwidth}{!}
		{\includegraphics{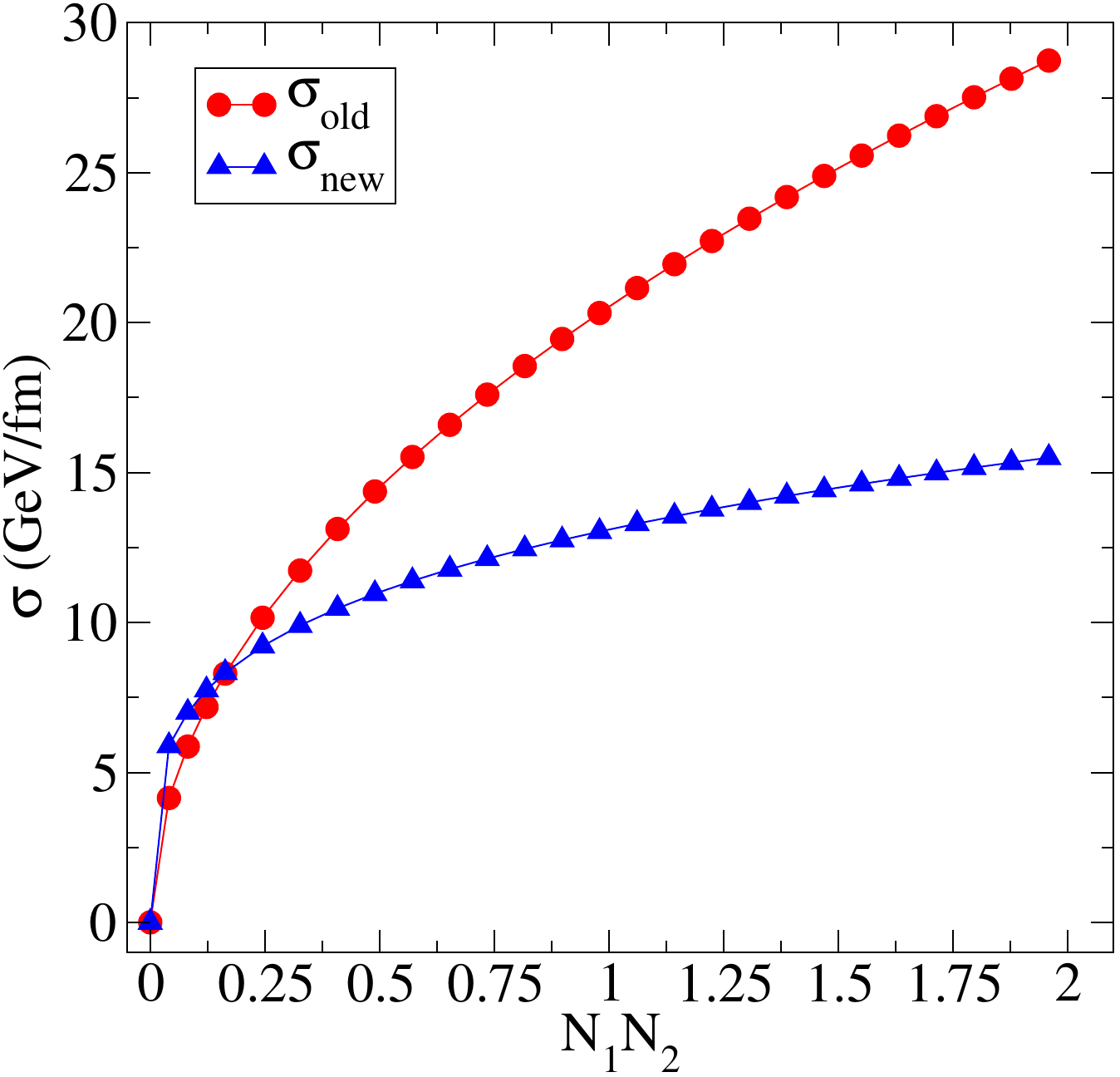}}
		\caption{ (color online)
			String tension, $\sigma$, as a function of the product of the baryon charge ($N_1$ and $N_2$) inside two colliding streaks. Red points correspond to the parametrization of the string tension used in the ESRM (Eq. (\ref{Oldsigma})), while blue triangles correspond to that used in the present work (Eq. (\ref{Newsigma})).
		}
		\label{STNe1Ne2}
	\end{center}
\end{figure}

 To ensure that most of the final streaks have been formed and have started their expansion a modification of $\sigma$ is necessary. In the present work we have used the following parametrization for $\sigma$:
\be
\sigma =
A\left( \frac{\varepsilon_0}{M_n}\right)\frac{(N_1N_2)^{1/4}}{\sqrt{\Delta x \Delta y}},
\label{Newsigma}
\ee
where $A$ is now a dimensional parameter measured in GeV; in the calculations we used $A = 0.05/\sqrt{\Delta x\Delta y} = 0.0784$ GeV. 
This new expression for $\sigma$ presents a lower dependence on the product $N_1 \cdot N_2$, 
and, thus, the difference between lower and higher string tension will be much less. In consequence, the final streak formation times for different transverse coordinates will be much more homogeneous.  In Fig. \ref{STNe1Ne2} both definitions of $\sigma$ as a function of the product of the baryon charges inside the colliding streaks ($N_1$ and $N_2$) are presented. The expression used in the ESRM has been indicated by $\sigma_{old}$  while that used in the present work is indicated by $\sigma_{new}$. In both cases, the same values have been used to the parameters present in these expressions: $A=0.0784$ (in GeV for the new definition of $\sigma$ (see Eq. \ref{Newsigma})), $\varepsilon_0$ $=$ 100 GeV and $\Delta x\Delta y$ $=$ 0.407 fm$^2$. 

\section{RESULTS}\label{SECIV}

In this section we present the results obtained from the GESRM corresponding to symmetric Au+Au collisions at RHIC energies and Pb+Pb collisions at LHC energies, and asymmetric A+Au head on collisions, being A the mass number of a given nucleus. We compare results obtained on event-by-event basis with those coming from averaging over $N$ events, where $N$ will vary from 10 to 500.000 events. The latter should be qualitatively comparable with the ESRM results, but we note that even if we average over a very large  number of events we will not reproduce the ESRM case, and this has to do not only with 
 the new definition of $\sigma$ (Eq. (\ref{Newsigma})) and WS distribution of nucleons in Au. Actually these modifications would only generate a small quantitative, but not qualitative difference.
We will show that the initial state fluctuations lead to principle differences in the initial state not only for a single event, but also if we perform an averaging over many events. 

\begin{figure}[ht]
	\begin{center}
		\resizebox{1.0\columnwidth}{!}
		{\includegraphics[width=\linewidth]{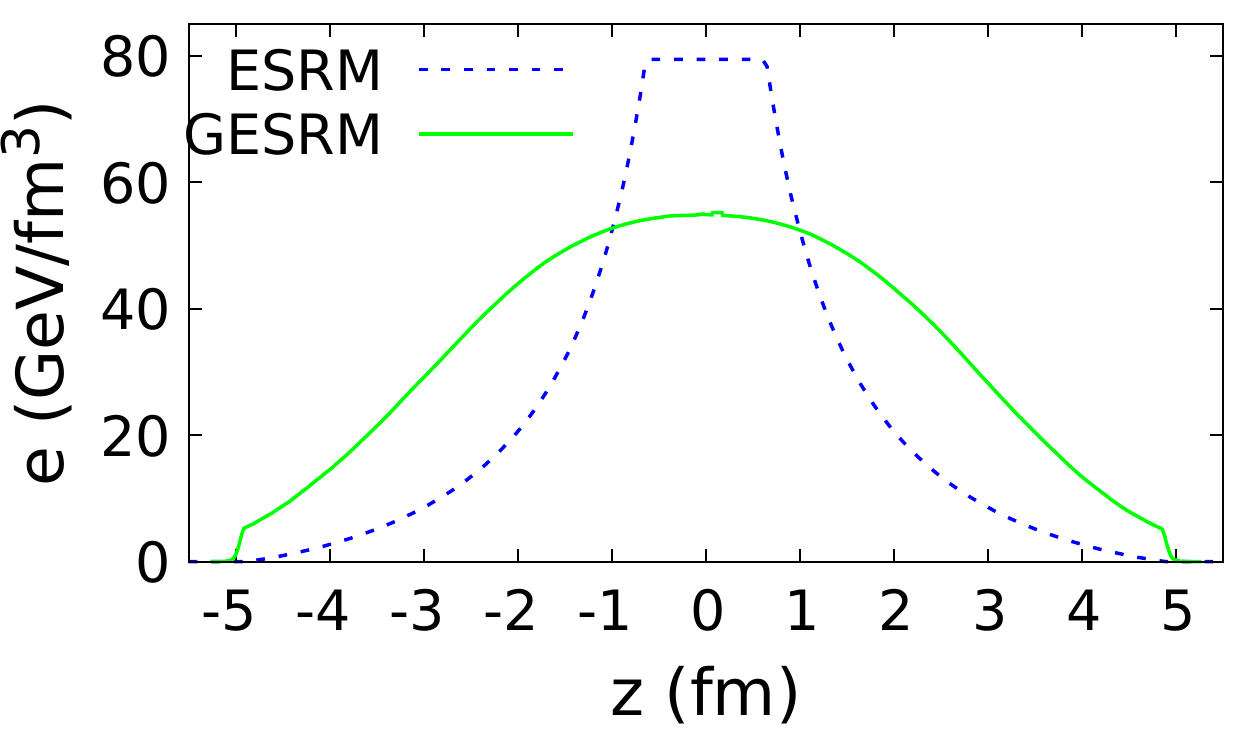}}
		\caption{ (color online)
			Energy density profiles of two given final expanding streaks corresponding to symmetric Au+Au collisions at impact parameter $b= (R_{Au}+R_{Au})/2$ and $\sqrt{S_{NN}}=$200 GeV, obtained from the ESRM, blue dashed line, and from the GESRM averaging over $N = 10000$ events, green solid line. In both cases the same values of parameters $A$ in definitions of the string tension $\sigma$ (Eqs. (\ref{Oldsigma}) and (\ref{Newsigma})) have been used.}
		\label{ENERDENSPCS}
	\end{center}
\end{figure}

As an illustration let us consider the energy density profile of a given final expanding streak corresponding to symmetric Au+Au collisions at impact parameter $b= (R_{Au}+R_{Au})/2$, $\sqrt{S_{NN}}=$200 GeV and $t=t_{fin}$. The result of the ESRM is well known and shown by a blue dashed line in Fig. \ref{ENERDENSPCS}. In the generalized model the situation will change:
the number of colliding nucleons (from both sides) will now fluctuate; this will lead to different string tensions $\sigma$ and, correspondingly, to different final streak length in each event, as well as to a different rapidity of the final streaks $y_f$ (which is defined by the momentum conservation). If we average over many such events then, according to Central Limiting Theorem, we can expect to see some Guassian-like shape.  This is exactly what we see in Fig. \ref{ENERDENSPCS}, where green solid line corresponds to an averaging over $N = 10000$ events.

\begin{figure}[ht!]  
	\begin{center}
		\resizebox{1.0\columnwidth}{!}
		{\includegraphics[width=\linewidth]{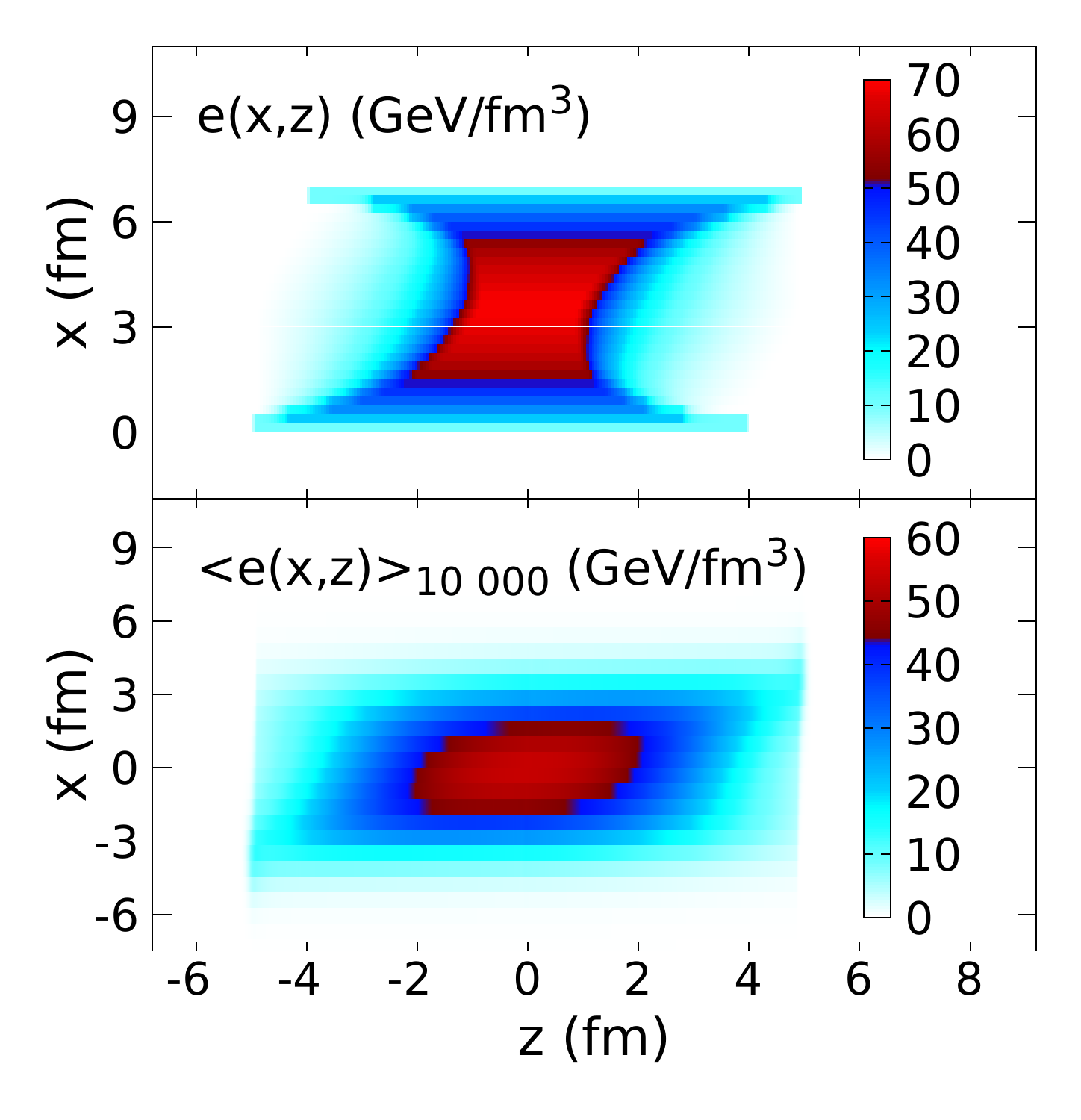}}
		\caption{ (color online)
			Energy density distributions in the reaction plane ([xz]-plane) obtained from the ESRM, top plot, and from the GESRM averaging over $N = 10000$ events, bottom plot. The results correspond to symmetric Au+Au collisions at $\sqrt{S_{NN}}=$ 200 GeV,\linebreak $b= (R_{Au}+R_{Au})/2$, $t_{fin}=$ 5 fm.}
		\label{ENERDENSOLDNEW}
	\end{center}
\end{figure}

Generalizing this discussion for the whole reaction volume, we show in Fig. \ref{ENERDENSOLDNEW} the energy density distributions in the reaction plane, obtained from the ESRM, top plot, and from the GESRM averaging over $N =$ 10000 events, bottom plot, for Au+Au collisions $\sqrt{S_{NN}}=$200 GeV for impact parameter $b= (R_{Au}+R_{Au})/2$. In both cases, we used the same values of parameters $A=0.0784$ (in GeV for the new definition of $\sigma$ (see Eq. \ref{Newsigma})). We can see that qualitatively these distributions are similar, for example both are titled, but the averaged GESRM initial state is a bit wider and less peaked (=more smoothed) in the middle, just like it is illustrated for one final streak in Fig. \ref{ENERDENSPCS}. 

\subsection{Reaction volume and number of participant nucleons}
\label {reacVol}

\begin{figure}[h!]  
\begin{center}
\resizebox{1.00\columnwidth}{!}
{\includegraphics[width=\linewidth]{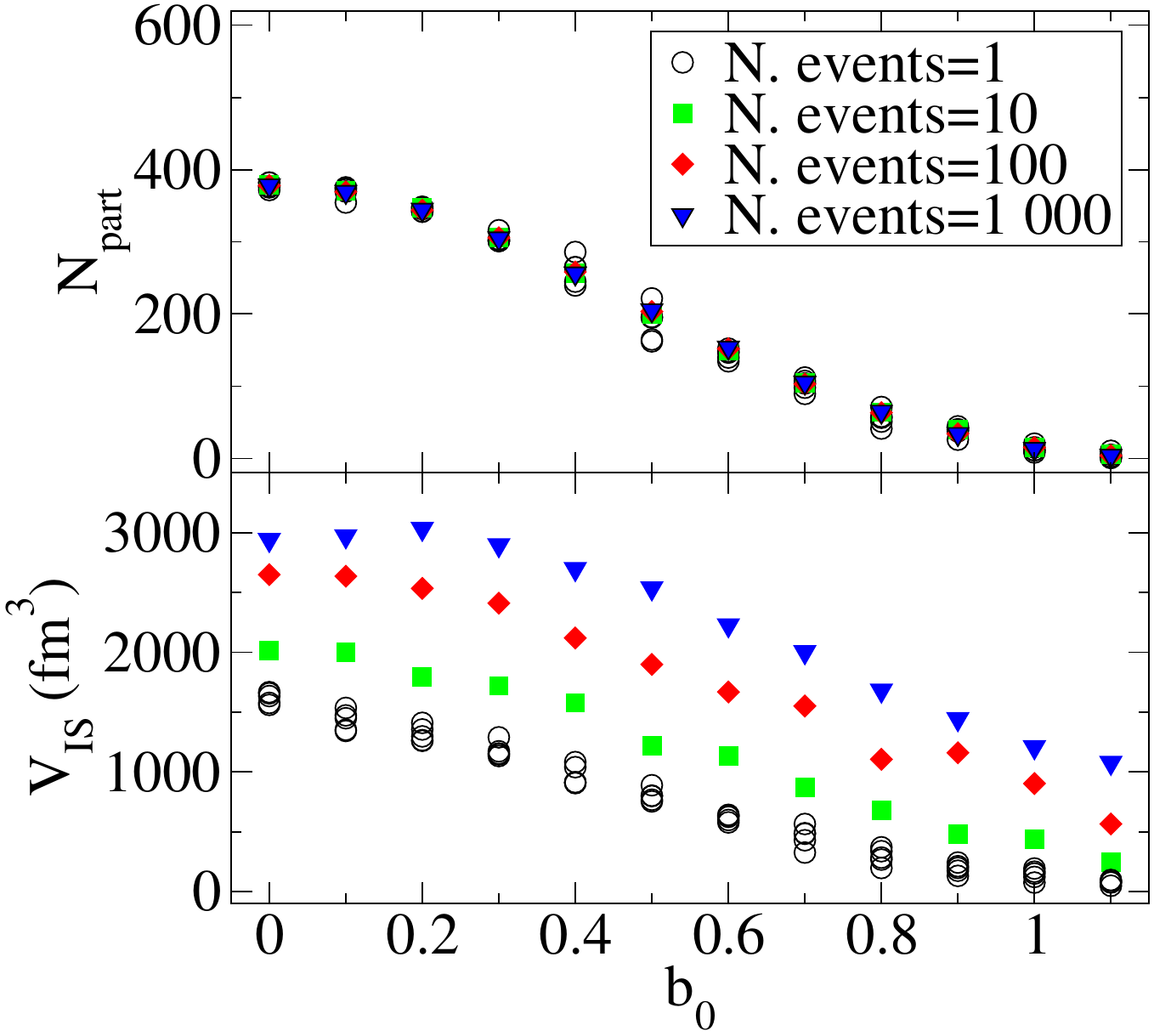}}
\caption{ (color online)
Number of participant nucleons ($N_{part}$), top plot, and initial state volume ($V_{IS}$), bottom plot, as a function of the impact parameter ($b_0$) for different number of events: 1, 10, 100 and 1000. The results correspond to symmetric Au+Au collisions at $\sqrt{S_{NN}}=$ 200 GeV , $t_{fin}=$ 5 fm.
}
\label{RVBN}
\end{center}
\end{figure}

We consider the reaction volume as the region of space occupied by our expanding system at $t_{fin} = 5$ fm. Unlike the ESRM, where the reaction volume will be fixed for a given impact parameter, in the GESRM, due to the random distribution of nucleons, this will fluctuate event-by-event. The bottom plot of Fig. \ref{RVBN} is representing the initial state volume ($V_{IS}$) as a function of the impact parameter for $N=1,10,100$ and $1000$ events, correponding to symmetric Au+Au collisions at $\sqrt{S_{NN}}=$200 GeV. For a single event, most of the non-empty cells are located in the overlapping region and only a few are outside (see Fig. \ref{3DNRD}). However, when we average over many events, the contribution of the outside region becomes more and more significant giving rise to a considerable increment in the reaction volume. Please note that this is just a pure effect of the geometric fluctuations; this doesn't mean that the number of participants grows, since those peripheral cells have very small densities. 

This behavior of the number of participant nucleons ($N_{part}$) is clearly seen in the top plot of Fig. \ref{RVBN}. 
The $N_{part}$ is calculated using the baryon charge density $n_i$ inside each non-empty cell as follows: 
\be
	N_{part} = V_{cell}\sum_{i=1}^{N_{cell}} n_i\gamma_i,
\ee
where $V_{cell}=\Delta x\Delta y\Delta z$, $N_{cell}$ is the total number of non-empty cells and $\gamma_i$ is the Lorentz factor of the cell $i$. We can see that for a single event the fluctuations of $N_{part}$ increase when we increase impact parameter.
On the other hand, with increasing number of events the number of participant nucleons for averaged initial state rapidly converges to the average value, which obviously depends on the given impact parameter. Please note that in the generalized model for central collisions ($b=0$) the average number of the participants is not $N_{part}=197+197=394$, but, due to initial state fluctuations, it is a bit smaller: $<N_{part}(b=0)>\simeq 378$.   
 
\begin{figure}[h!]  
	\begin{center}
		\resizebox{1.0\linewidth}{!}
		{\includegraphics[width=\linewidth]{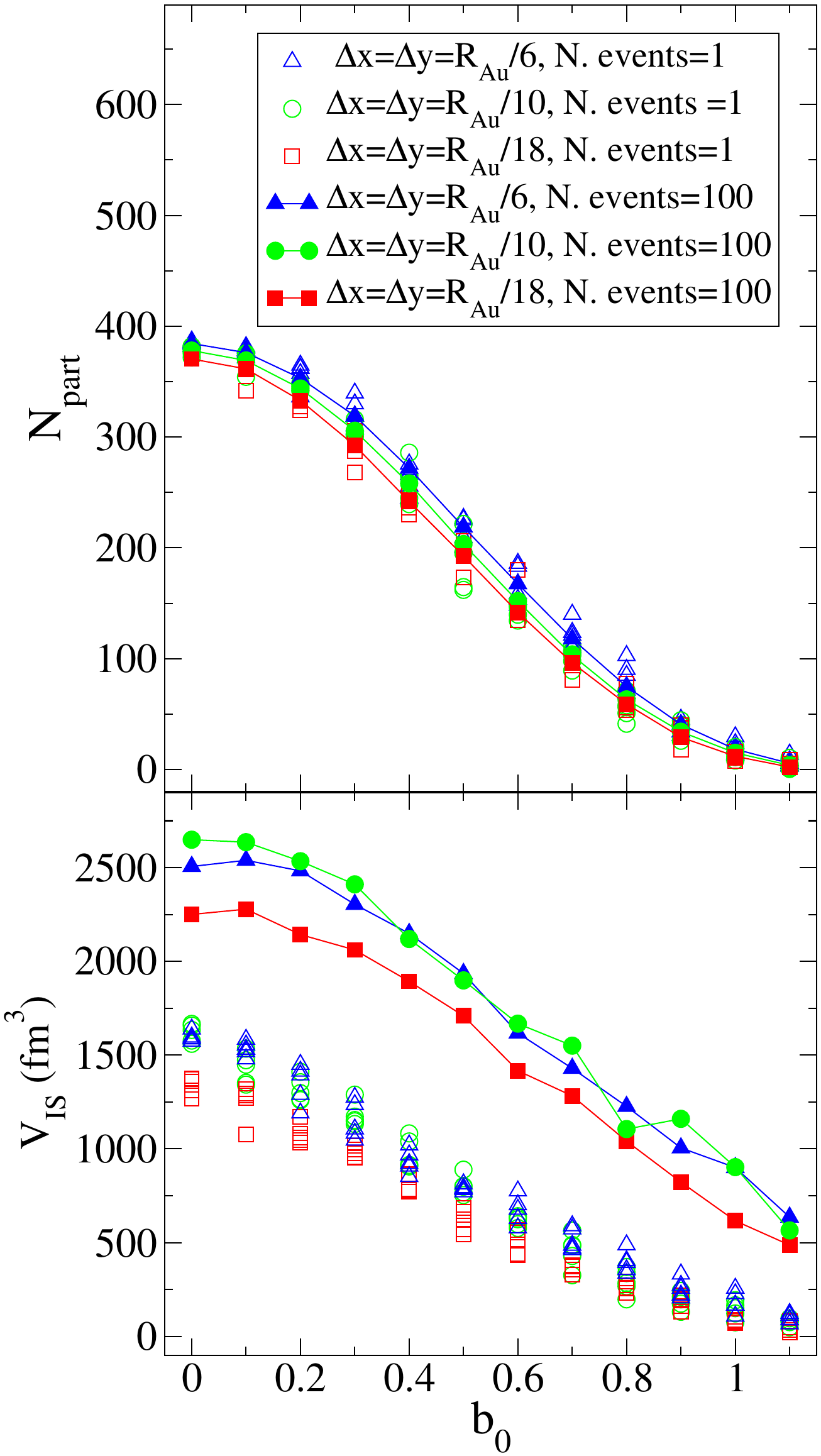}}
		\caption{ (color online)
			Number of participant nucleons ($N_{part}$), top plot, and initial state volume ($V_{IS}$), bottom plot, as a function of the impact parameter ($b_0$) for different cell sizes. The simulation is performed for symmetric Au+Au collisions at $\sqrt{S_{NN}}=$ 200 GeV , $t_{fin}=$ 5 fm, as in Fig. \ref{RVBN}, for a single event and averaging over 100 events.
		}
		\label{RVBN-delta}
	\end{center}
\end{figure}

In order to check the stability of our calculation we performed simulations with different cell sizes, namely $\Delta x=\Delta y = R_{Au}/6$;  $R_{Au}/10$ and $R_{Au}/18$. The corresponding results are shown in Fig. \ref{RVBN-delta}. Both the number of participants, top plot, and the reaction volume, bottom plot, grows a bit for a bigger cell size, what can be expected. We also note that this difference is reduced for the most peripheral collisions.   

\begin{figure*}[ht!]  
	\begin{center}
		{\includegraphics[width=\textwidth]{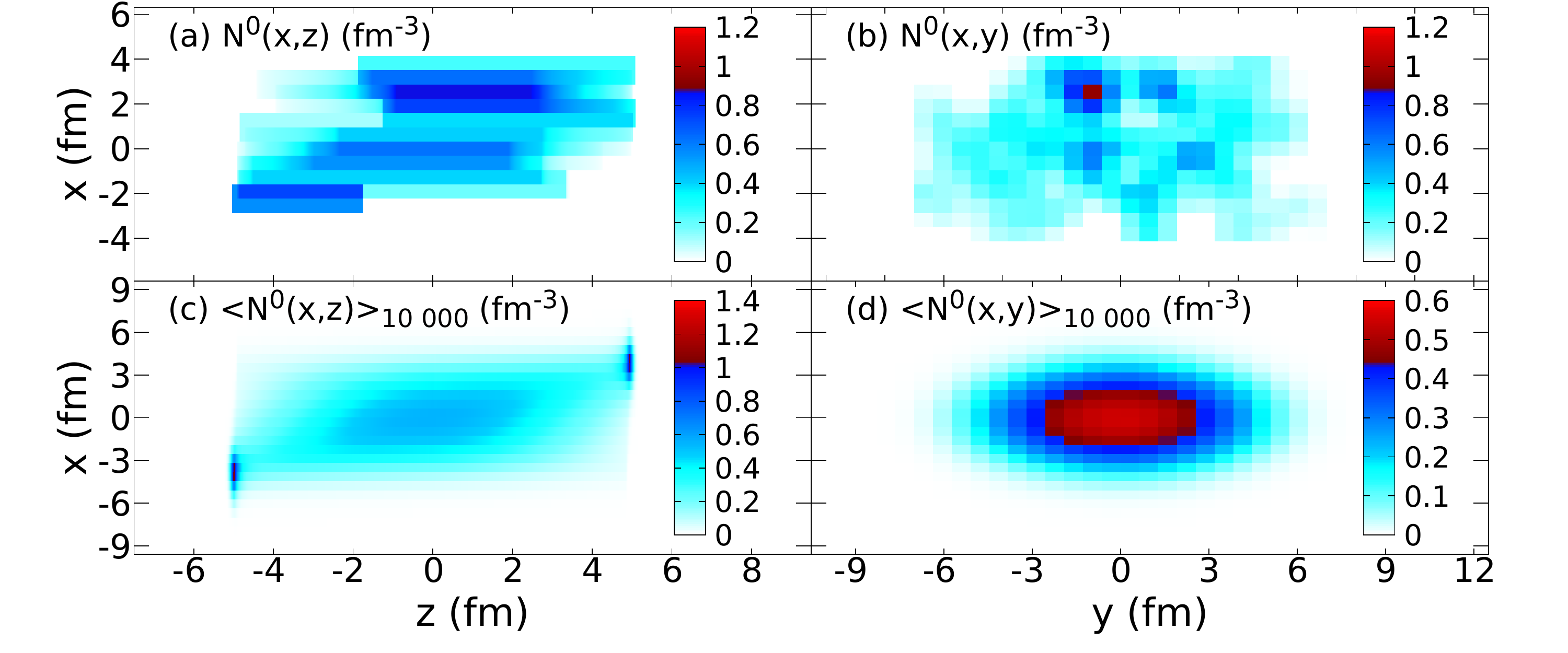}}
		\caption{ (color online)
			Initial state baryon charge density distributions, $N^0$, for symmetric Au+Au collisions at $\sqrt{S_{NN}}=$ 200 GeV, $b=(R_{Au}+R_{Au})/2$, $t_{fin}=$ 5 fm. The top plots ((a) and (b)) represent $N^0$ for a single event and the bottom ones ((c) and (d)) - the average over $N=$10000 events. The left plots ((a) and (c))correspond to $N^0$ in the reaction plane ([xz]-plane), while the right ones ((b) and (d)) show $N^0$ in the transverse plane ([xy]-plane).}
		\label{BCD}
	\end{center}
\end{figure*} 

\begin{figure*}[ht!] 
	\begin{center}
		\resizebox{1.00\textwidth}{!}
		{\includegraphics[width=\textwidth]{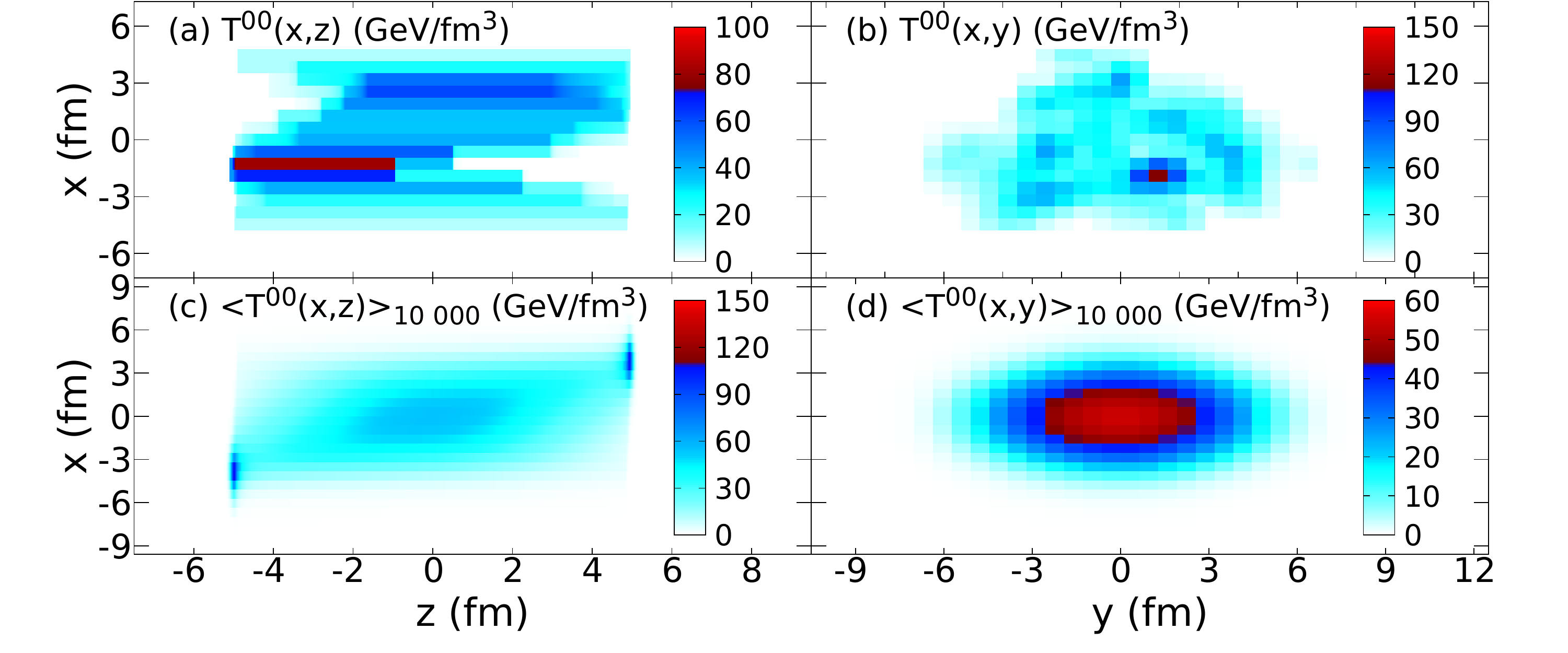}}
		\caption{ (color online)
			Initial state energy density distributions, $T^{00}$, for symmetric Au+Au collisions at $\sqrt{S_{NN}}=$ 200 GeV, $b=(R_{Au}+R_{Au})/2$, $t_{fin}=$ 5 fm. The top plots ((a) and (b)) represent $T^{00}$ for a single event and the bottom ones ((c) and (d)) - the average over $N=$10000 events. The left plots ((a) and (c)) correspond to $T^{00}$ in the reaction plane ([xz]-plane), while the right ones ((b) and (d)) show $T^{00}$ in the transverse plane ([xy]-plane).}   
		\label{ED}
	\end{center}
\end{figure*}  

\subsection{Baryon charge and energy density distributions}

The baryon charge and energy density distributions at $t=t_{fin}$ are the main output of the GESRM, which should be used as an input for further hydrodynamical evolution.  

\begin{figure*}[ht] 
	\begin{center}
		\resizebox{1.00\textwidth}{!}
		{\includegraphics[width=\textwidth]{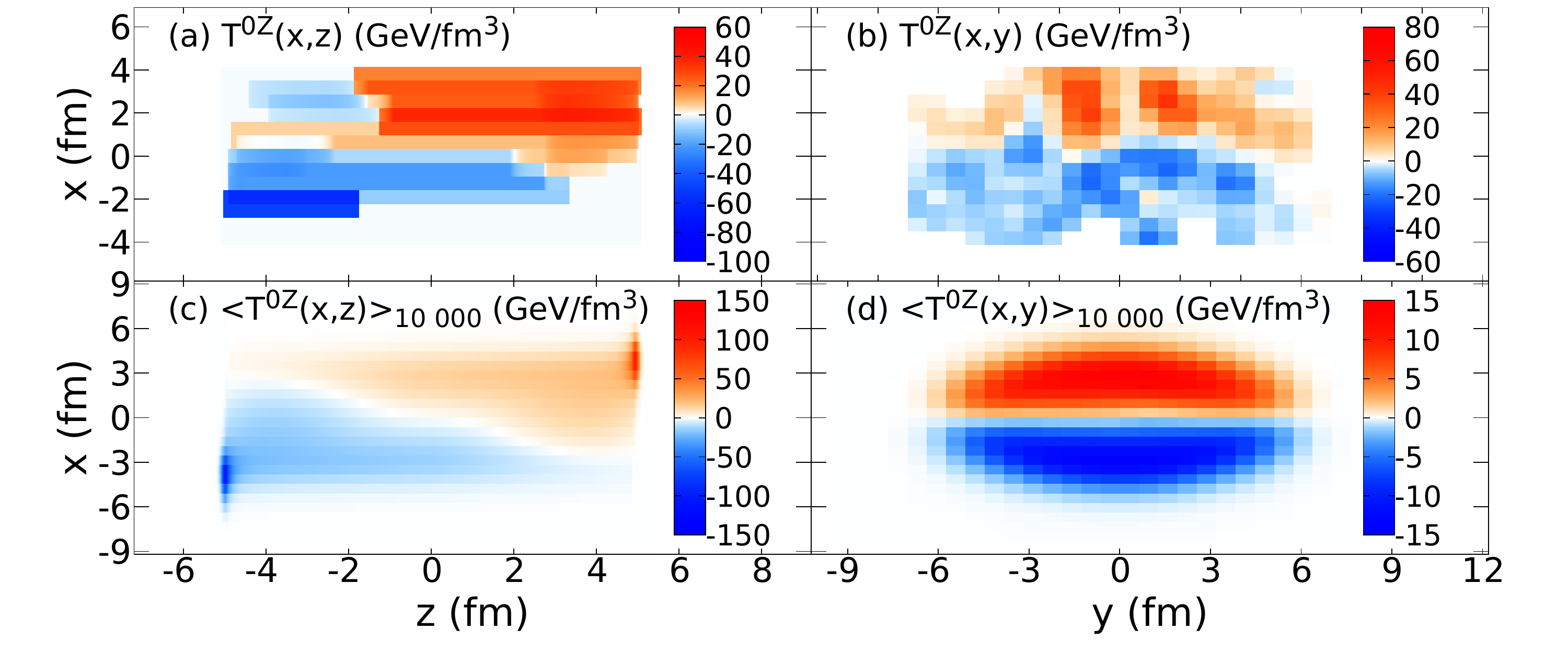}}
		\caption{ (color online)
			Initial state longitudinal momentum density distributions, $T^{0Z}$, for symmetric Au+Au collisions at $\sqrt{S_{NN}}=$ 200 GeV, $b=(R_{Au}+R_{Au})/2$, $t_{fin}=$ 5 fm. The top plots ((a) and (b)) represent $T^{0Z}$ for a single event and the bottom ones ((c) and (d)) - the average over $N=$10000 events. The left plots ((a) and (c)) correspond to $T^{0Z}$ in the reaction plane ([xz]-plane), while the right ones ((b) and (d)) show $T^{0Z}$ in the transverse plane ([xy]-plane).} 
		\label{Pz}
	\end{center}
\end{figure*}  
\begin{figure*}[ht]  
	\begin{center}
		\resizebox{1.00\textwidth}{!}
		{\includegraphics[width=\textwidth]{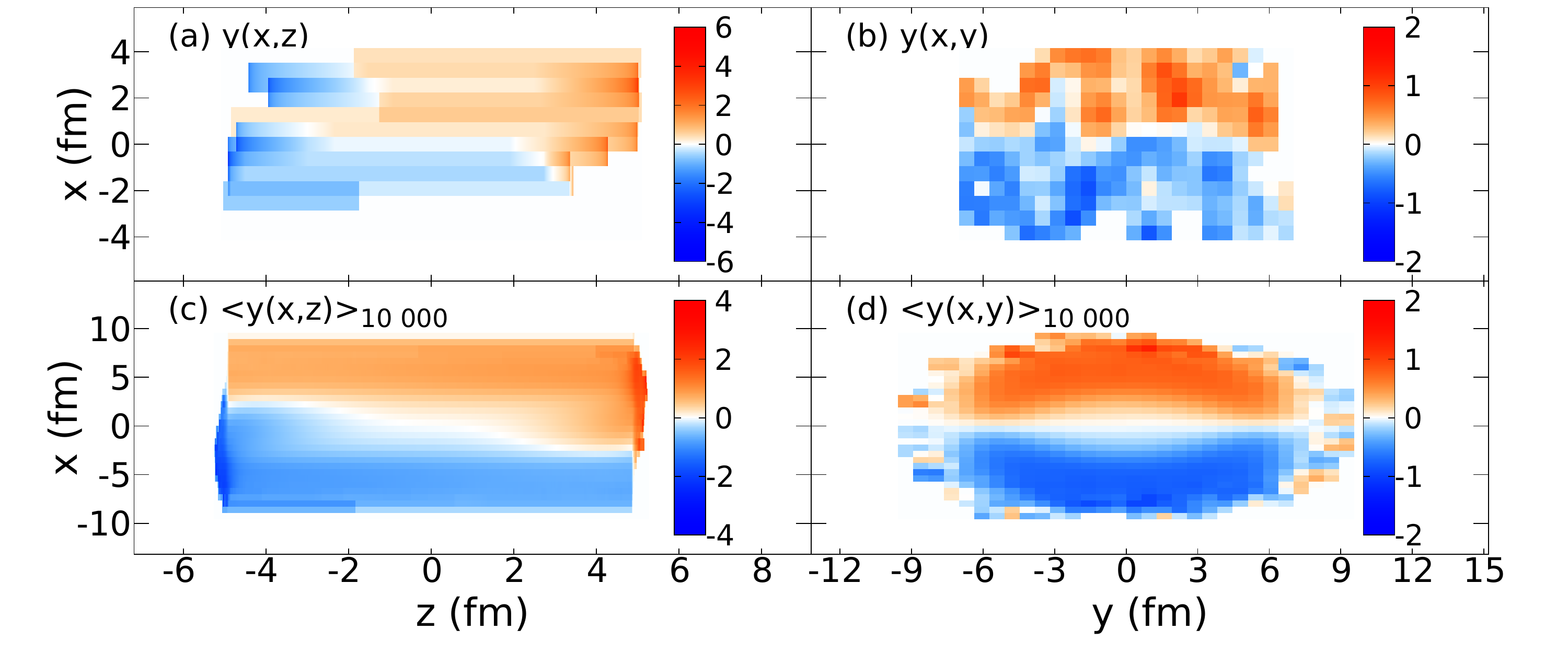}}
		\caption{ (color online)
			Initial state rapidity distributions, $y$, for symmetric Au+Au collisions at $\sqrt{S_{NN}}=$ 200 GeV, $b=(R_{Au}+R_{Au})/2$, $t_{fin}=$ 5 fm. The top plots ((a) and (b)) represent $y$ for a single event and the bottom ones ((c) and (d)) - the average over $N=$10000 events. The left plots ((a) and (c)) correspond to $y$ in the reaction plane ([xz]-plane), while the right ones ((b) and (d)) show $y$ in the transverse plane ([xy]-plane).}   
		\label{RAPIDITY}
	\end{center}
\end{figure*}  

In Figs. \ref{BCD} and \ref{ED} we represent the baryon charge and the energy density distributions for a single event with fluctuations, top plots, and averaging over $N=10000$ events, bottom plots, for symmetric Au+Au collisions at $\sqrt{S_{NN}}=$ 200 GeV, $b= (R_{Au}+R_{Au})/2$, $t_{fin}=$ 5 fm. These are shown in the reaction plane, left plots, and in the transverse plane, right plots. Similarly to the results obtained from the ESRM, when we average over many events the baryon charge and energy density distributions show a type of tilted disk in the reaction plane. This is not clearly seen on event-by-event basis because of fluctuations. With respect to the transverse plane, we see that for N=10000 events the overlapping region has an \textquotedblleft almond shape\textquotedblright giving rise to a strong elliptic flow.  However, on event-by-event basis the geometry of the collision will fluctuate around the average geometry, giving rise to all possible geometric deformations: elliptical $\varepsilon_2$, triangular $\varepsilon_3$, quadrupole $\varepsilon_4$ and other harmonics, which will generate momentum anisotropies quantified by flow harmonics $v_n$ \cite{Mazeliauskas2017} as discussed in Eq. (\ref{MAD}). 

\begin{figure*}[ht] 
	\centering
	\resizebox{1.01\textwidth}{!}
	{\includegraphics[width=\textwidth]{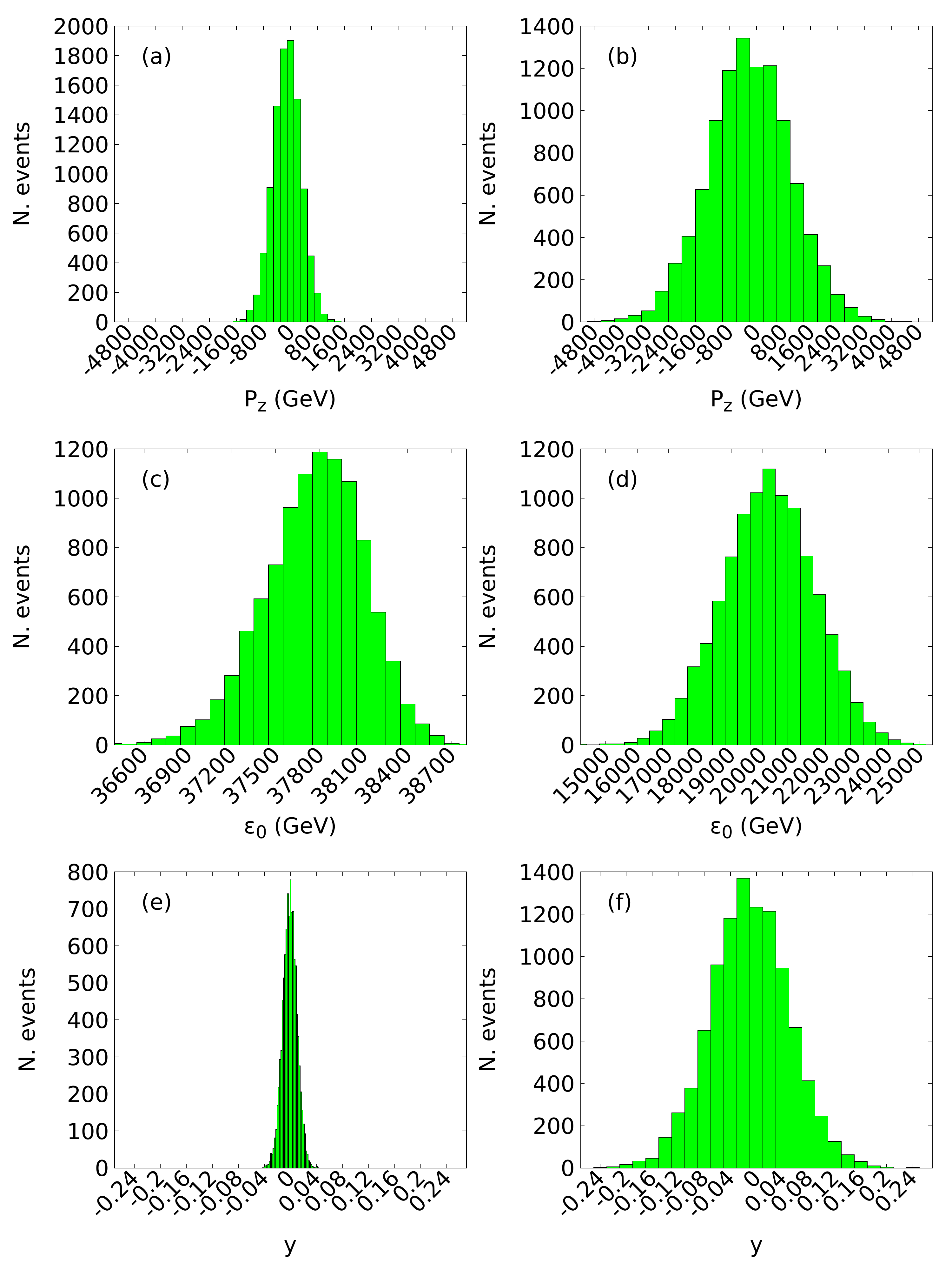}}
	\caption{ (color online) 
		Distributions of the single event results for the total longitudinal linear momentum, top plots ((a) and (b)), total energy, middle plots ((c) and (d)), and central rapidity, bottom plots ((e) and (f)), at impact parameters $b=0$ fm, left plots ((a), (c) and (e)), and $b=(R_{Au}+R_{Au})/2$ fm, right plots ((b), (d) and (f)), for symmetric Au+Au collisions at $\sqrt{S_{NN}}=$200 GeV, $t_{fin}=$ 5 fm. The total number of results presented at each histogram is 10000.
	}
	\label{ENERPzRAPFLUCT}
\end{figure*} 

In Figs. \ref{Pz} and \ref{RAPIDITY} we show the longitudinal linear momentum, $T^{0Z}$,  and the rapidity distributions, $y$, respectively, for the initial states. As we can see, when we average over $N=10000$ events the fluid cells in the central zone of the collision are approximately at rest because the colliding streaks are symmetric in this region. This is not observed on event by event basis because of fluctuations of the streak lengths. 
There is a remarkable difference between the distributions obtained averaging over many events, bottom plots of Fig. \ref{RAPIDITY}, and those obtained by the ESRM without fluctuations, see Fig. \ref{ESRMVDRP}. In the latter the streak ends move in opposite directions while in the former both ends move in the same direction, except in the central region of the collision.

From Central Limiting Theorem it is expected that the distribution of any fluctuating physics quantity should have a Gaussian-shape if we look for large enough number of events. This is illustrated in Fig. \ref{ENERPzRAPFLUCT}, where the distributions of the results from different single events for the total longitudinal linear momentum of the collisions (top plots), its total energy (middle plots), and center of mass rapidity of the system (bottom plots) are shown for symmetric Au+Au collisions at impact parameters $b=0$ fm (left plots), and $b=(R_{Au}+R_{Au})/2$ fm (right plots). At each histogram of Fig.  \ref{ENERPzRAPFLUCT}  we present 10000 results.

For the $b=0$ case, as it is expected for symmetric collisions, the Gaussians corresponding to the total linear momentum and to the center-of-mass rapidity are centered around $0$. In the case of the energy, the Gaussian is centered around the average energy of the system corresponding to zero impact parameter, and it can be calculated following the result of Fig. \ref{RVBN}:  $E=\varepsilon_0 <N_{part}(b=0)>\simeq 37800$ GeV. 

\begin{figure}[h!] 
	\begin{center}
		\resizebox{1.01\columnwidth}{!}
		{\includegraphics[width=\linewidth]{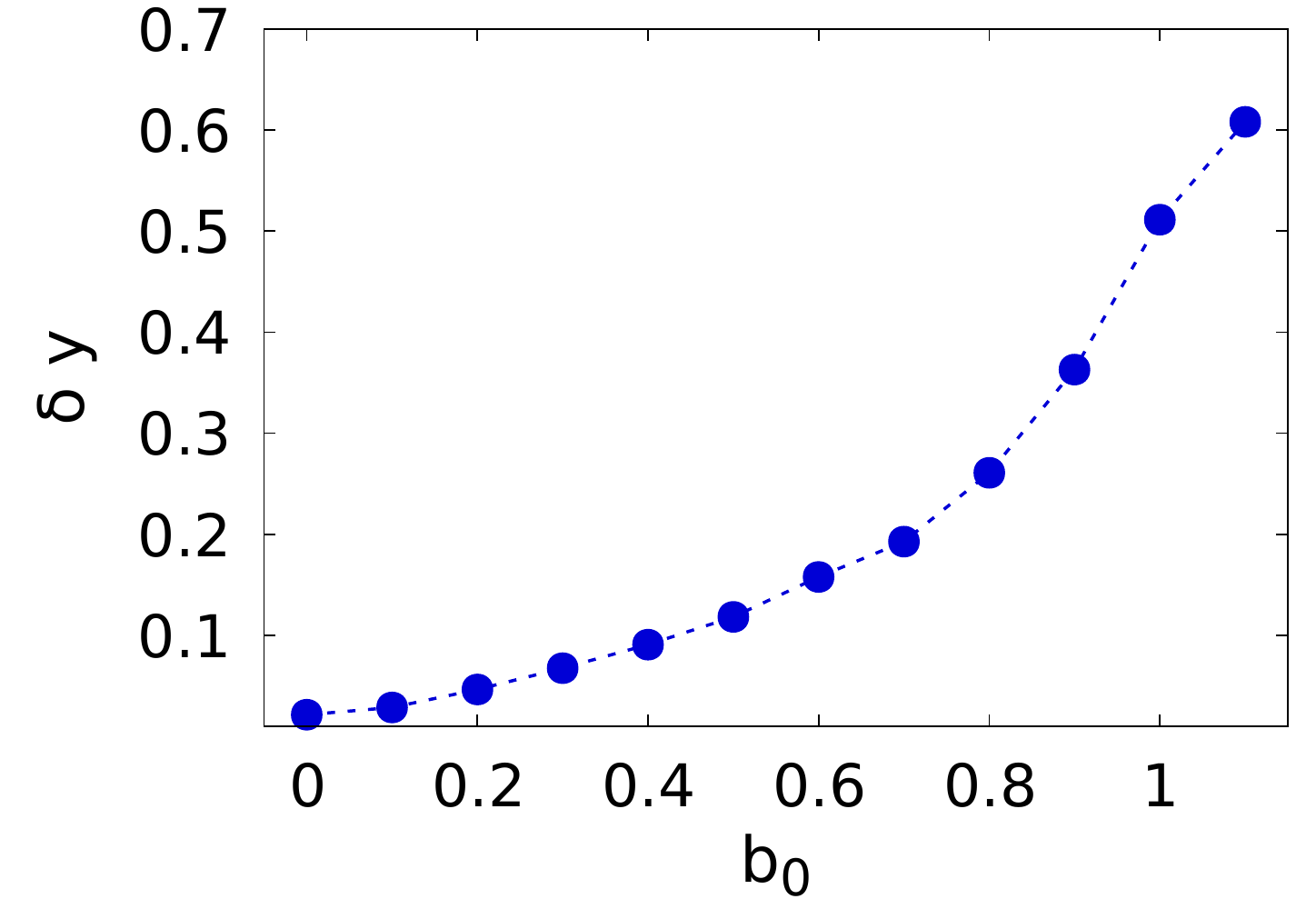}}
		\caption{ (color online)
			Center of mass rapidity fluctuations as a function of the impact parameter for symmetric Au+Au collisions at $\sqrt{S_{NN}}=$ 200 GeV, $t_{fin}=$ 5 fm.
		}
		\label{RAPFLUCT}
	\end{center}
\end{figure}

It is interesting to compare the widths of the corresponding Gaussian distributions; these clearly grow with impact parameter. This is not a surprise since we have already seen that the relative fluctuations are stronger for smaller systems. To be more quantitative, in 
Fig. \ref{RAPFLUCT} we present the Gaussian widths corresponding to the center-of-mass rapidity ($\delta y$) as a function of the impact parameter. 
The results of Fig. \ref{RAPFLUCT} show that in Ref. \cite{PRC840249142011} the authors have overestimated the center-of-mass rapidity fluctuations, taking  $\delta y=1$ and $2$, even for very peripheral events.

\subsection{From RHIC to LHC energies}

\begin{figure*}[ht]  
	\begin{center}
		\resizebox{1.00\textwidth}{!}
		{\includegraphics[width=\textwidth]{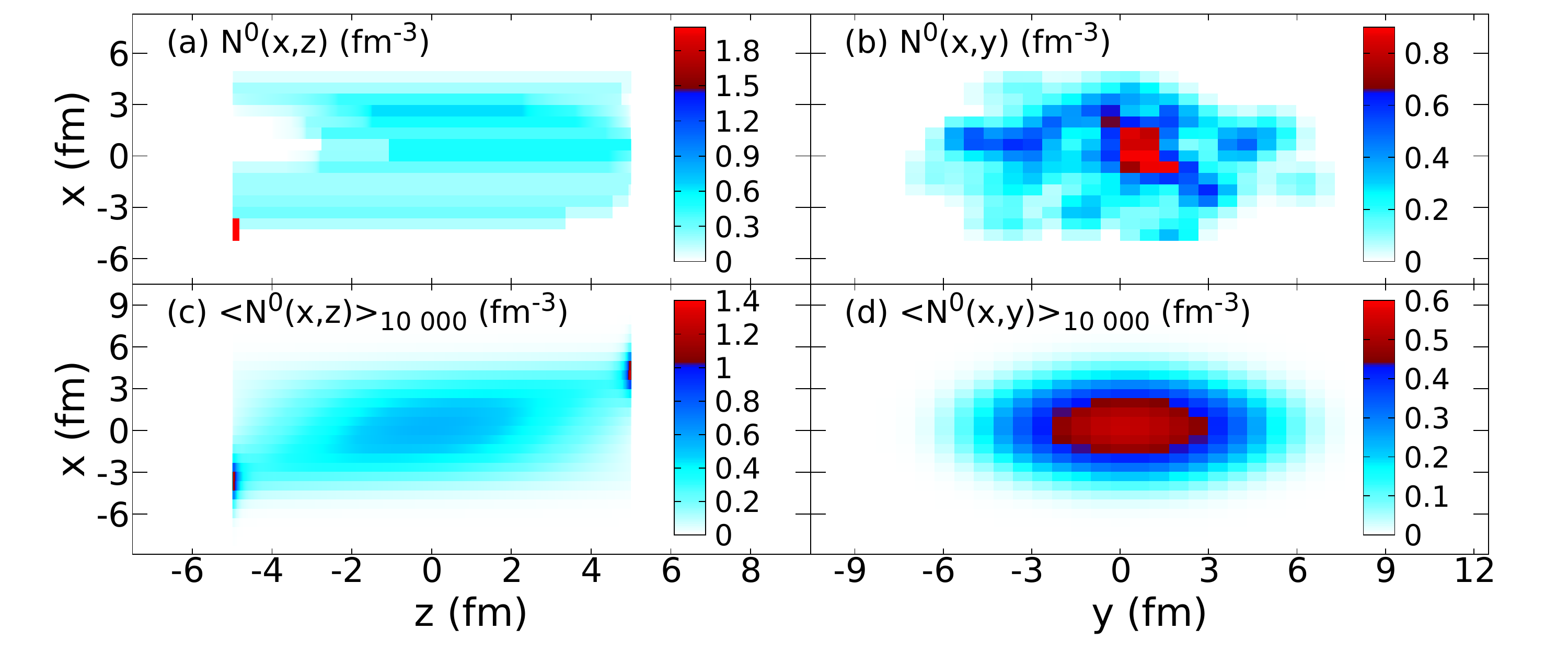}}
		\caption{ (color online)
			Initial state baryon charge density distributions, $N^0$, for symmetric Pb+Pb collisions at $\sqrt{S_{NN}}=$ 2.76 TeV, $b=(R_{Pb}+R_{Pb})/2$, $t_{fin}=$ 5 fm. The top plots ((a) and (b)) represent $N^0$ for a single event and the bottom ones ((c) and (d)) - the average over $N=$10000 events. The left plots ((a) and (c)) correspond to $N^0$ in the reaction plane ([xz]-plane), while the right ones ((b) and (d)) show $N^0$ in the transverse plane ([xy]-plane).}
		\label{BCD-LHC}
	\end{center}
\end{figure*}

\begin{figure*}[ht] 
	\begin{center}
		\resizebox{1.00\textwidth}{!}
		{\includegraphics[width=\textwidth]{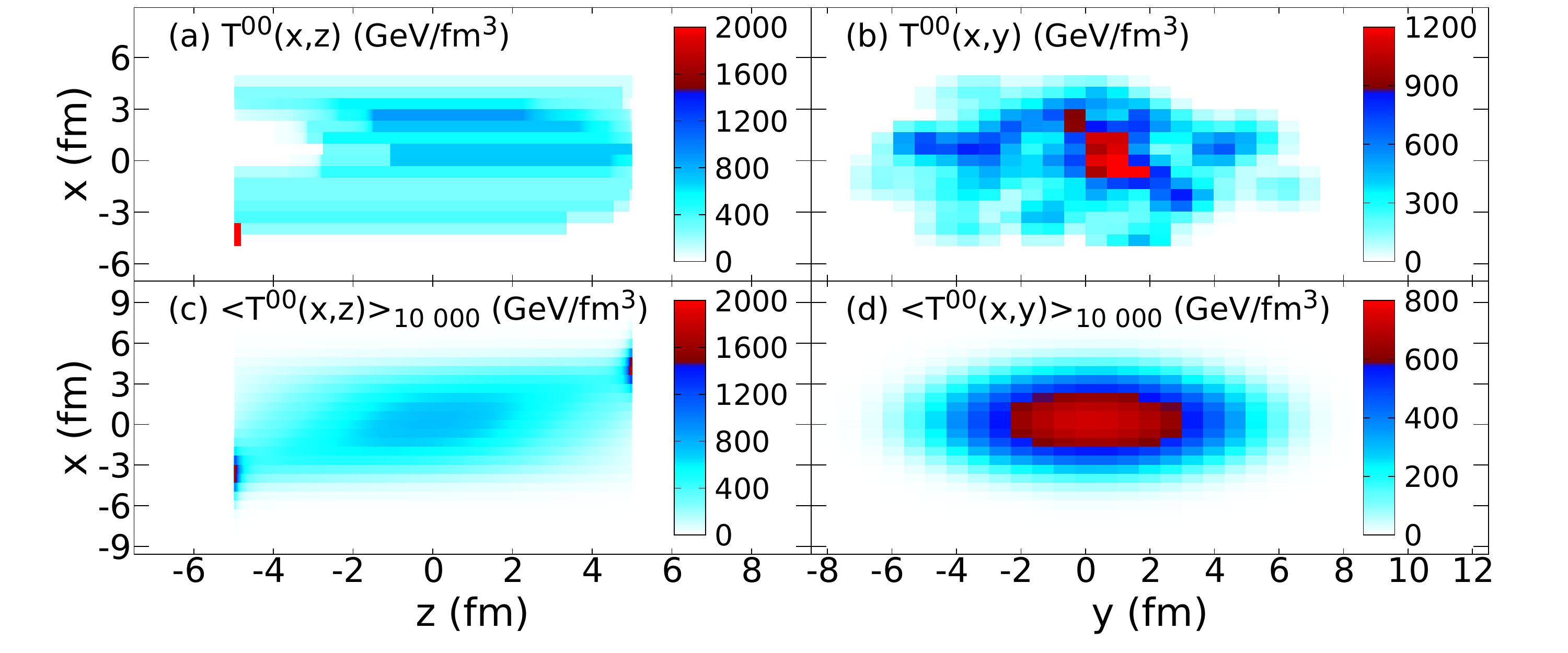}}
		\caption{ (color online)
			Initial state energy density distributions, $T^{00}$, for symmetric Pb+Pb collisions at $\sqrt{S_{NN}}=$ 2.76 TeV, $b=(R_{Pb}+R_{Pb})/2$, $t_{fin}=$ 5 fm. The top plots ((a) and (b)) represent $T^{00}$ for a single event and the bottom ones ((c) and (d)) - the average over $N=$10000 events. The left plots ((a) and (c)) correspond to $T^{00}$ in the reaction plane ([xz]-plane), while the right ones ((b) and (d)) show $T^{00}$ in the transverse plane ([xy]-plane).}   
		\label{ED-LHC}
	\end{center}
\end{figure*}

As it was mentioned in the introduction we expect the GESRM model to be applicable for collision energies higher than a few dozen of GeV per nucleon, let us say $\sqrt{S_{NN}}\gtrsim 50$ GeV, i.e. at RHIC and LHC.  As an illustration let us perform the similar study as before for the ALICE at LHC energy; the model parameters stay the same: $A=0.0784$ GeV, $t_{fin}=5$ fm. In Figs. \ref{BCD-LHC} and \ref{ED-LHC} the baryon charge density, $N^{00}$, and energy density, $T^{00}$, distributions are presented for a single event with fluctuations, top plots, and averaging over $N=10000$ events, bottom plots, for symmetric Pb+Pb collisions at $\sqrt{S_{NN}}=$ 2.76 TeV for $b=(R_{Pb}+R_{Pb})/2$. 
We can see that although the the initial energy density is substantially higher than that for the RHIC energy, compare Figs. \ref{ED-LHC} and \ref{ED},  the baryon density distribution remains rather similar, Figs \ref{BCD-LHC} and \ref{BCD}, what is known property of the ESRM.

\begin{figure*}[ht]
	\begin{center}
		\resizebox{1.01\textwidth}{!}
		{\includegraphics[width=\textwidth]{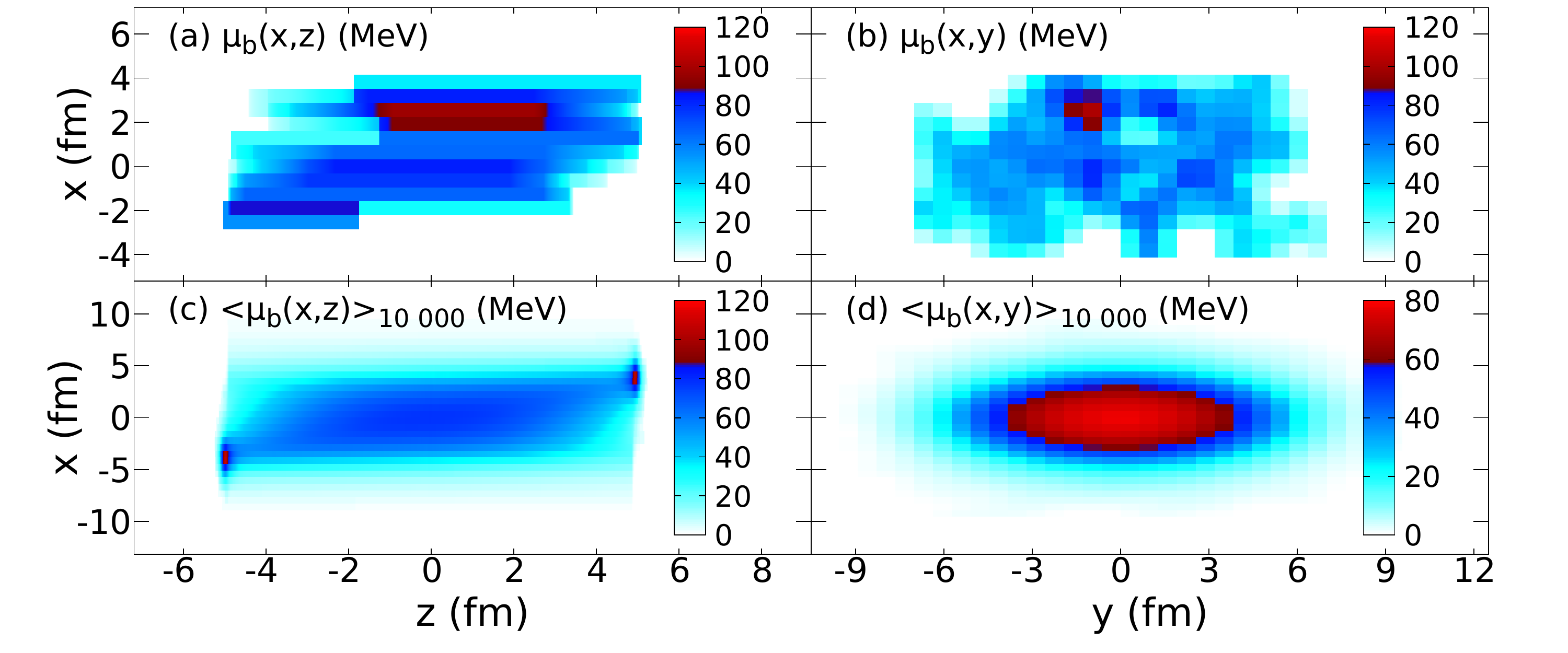}}
		\caption{ (color online)
			Initial state baryon chemical potential distributions, $\mu_b$, for symmetric Au+Au collisions at $\sqrt{S_{NN}}=$ 200 GeV, $b=(R_{Au}+R_{Au})/2$, $t_{fin}=$ 5 fm. The top plots ((a) and (b)) represent $\mu_b$ for a single event and the bottom ones ((c) and (d)) - the average over $N=$10000 events. The left plots ((a) and (c)) correspond to $\mu_b$ in the reaction plane ([xz]-plane), while the right ones ((b) and (d)) show $\mu_b$ in the transverse plane ([xy]-plane).} 
		\label{CHEMPOT}
	\end{center}
\end{figure*}  
\begin{figure*}[ht]  
	\begin{center}
		\resizebox{1.01\textwidth}{!}
		{\includegraphics[width=\textwidth]{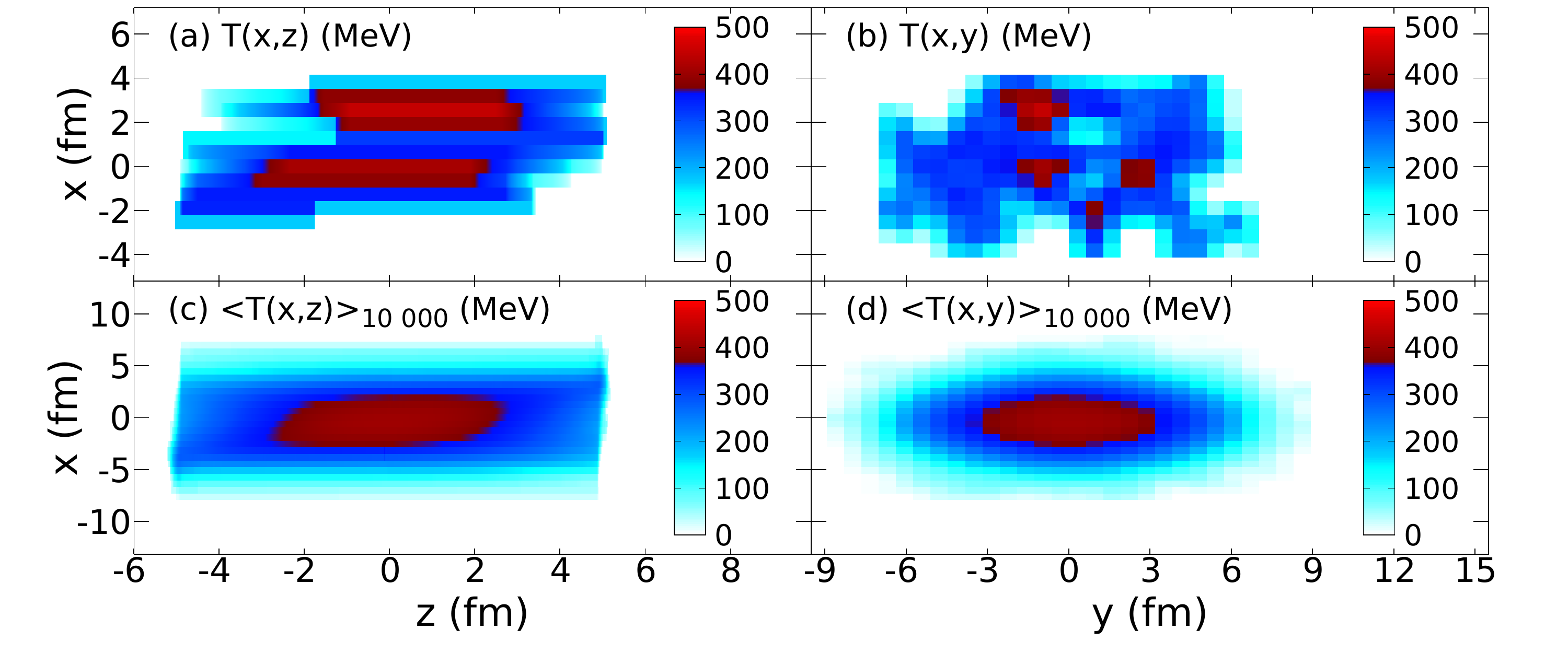}}
		\caption{ (color online)
			Initial state temperature distributions, $T$, for symmetric Au+Au collisions at $\sqrt{S_{NN}}=$ 200 GeV, $b=(R_{Au}+R_{Au})/2$, $t_{fin}=$ 5 fm. The top plots ((a) and (b)) represent $T$ for a single event and the bottom ones ((c) and (d)) - the average over $N=$10000 events. The left plots ((a) and (c)) correspond to $T$ in the reaction plane ([xz]-plane), while the right ones ((b) and (d)) show $T$ in the transverse plane ([xy]-plane).} 
		\label{TEMPERATURE}
	\end{center}
\end{figure*}  
\begin{figure*}[ht]  
	\begin{center}
		\resizebox{1.01\textwidth}{!}
		{\includegraphics[width=\textwidth]{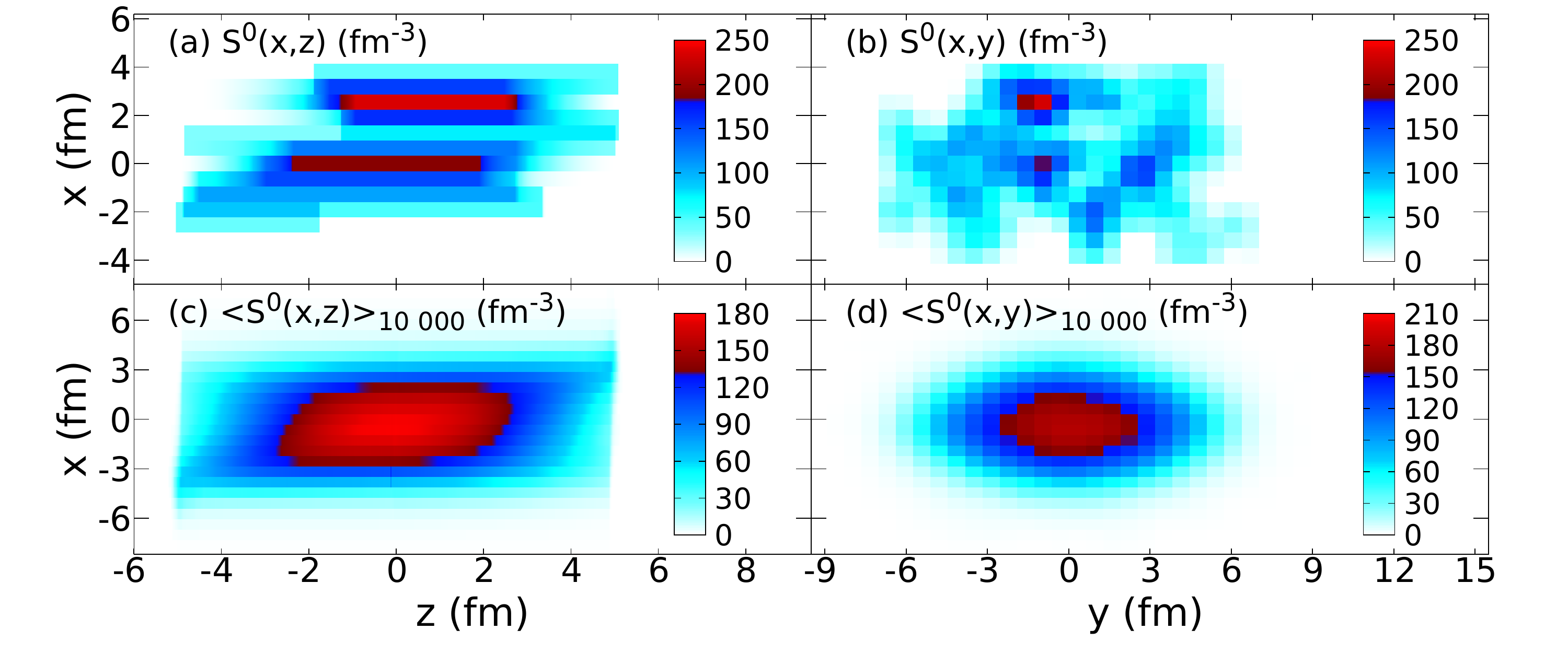}}
		\caption{ (color online)
			Initial state entropy density distributions, $s$, for symmetric Au+Au collisions at $\sqrt{S_{NN}}=$ 200 GeV, $b=(R_{Au}+R_{Au})/2$, $t_{fin}=$ 5 fm. The top plots ((a) and (b)) represent $s$ for a single event and the bottom ones ((c) and (d)) - the average over $N=$10000 events. The left plots ((a) and (c)) correspond to $s$ in the reaction plane ([xz]-plane), while the right ones ((b) and (d)) show $s$ in the transverse plane ([xy]-plane).} 
		\label{ENTDENS}
	\end{center}
\end{figure*}  

\subsection{Chemical potential, Temperature and Entropy}
\label{mu_T_S}

One can obtain temperature and baryon chemical potential from the energy and baryon charge density distributions, calculated in our model, applying the corresponding EoS. For an ideal gas of $N_f$ flavor quarks, with their respective $N_c=3$ colors, and ($N_c^2$-1) gluons, we assume the EoS given by the Stefan-Boltzmann expressions:
\be
\begin{aligned}
 e_{SB}(T,\mu_q) = 
\frac{\pi^2}{15}\left(N_c^2-1+\frac{7N_cN_f}{4}\right)T^4+ \\
\frac{N_cN_f}{2}\left(T^2\mu_q^2+\frac{\mu_q^4}{2\pi^2}\right), 
\label{SBEdens}
\end{aligned}
\ee
\be
P_{SB}(T,\mu_q) = 
\frac{1}{3}e_{SB}(T,\mu_q),
\label{SBP}
\ee
\be
n_{SB}(T,\mu_q) = 
\frac{N_cN_f}{9\pi^2}\left(\mu_q^3+\pi^2T^2\mu_q\right),
\label{SBBC}
\ee
where $T$ and $\mu_q$ are the quark temperature and chemical potential ($\mu_b = 3\mu_q$), and $n_{SB}=n_b$ is the baryon charge density in the quark phase. From Eq. (\ref{SBBC}) it is possible to write the temperature as a function of $n_{SB}$ and $\mu_b$, and replace it in Eq. (\ref{SBEdens}) in order to obtain $\mu_b$. For $N_c=N_f=3$ we get:
\be
-\frac{2}{243\pi^2}\mu_b^6+\frac{4}{9}n_b\mu_b^3-e_{SB}\mu_b^2+\frac{57\pi^2}{4}n_b^2=0.
\label{T_eq}
\ee
Please note that in the MIT Bag Model EoS, used for our calculations, the energy density and pressure of the parton gas differ from $e_{SB}$ and $P_{SB}$ by the Bag constant $B$:
\be
	e=e_{SB}+B\,,\quad P=P_{SB}-B\,.
\ee
In our calculations we have used $B=0.330$ GeV/fm$^{3}$.

If the pressure is high enough then $e_{SB}=e-B$ and we can proceed solving Eq. (\ref{T_eq}). However, if\linebreak $P_{SB}=e_{SB}/3<B$ we might run into the problem of negative pressure. 
Following the prescription given in Ref. \cite{PLB6922772010}, we assign in such a situation $P=0$ and $e_{SB}=3B\left(\frac{e}{4B}\right)^{4/3}$.

Once we have calculated the baryon chemical potential and the temperature, we can easily obtain the entropy density, $s$, from the following thermodynamical relation:
\be
Ts = e+P-\mu_b n_b.
\label{FLT}
\ee
In Figs. \ref{CHEMPOT}, \ref{TEMPERATURE} and \ref{ENTDENS} the baryon chemical potential, $\mu_b$, the temperature, $T$, and the entropy density, $s$, distributions are represented respectively, for symmetric Au+Au collisions at $\sqrt{S_{NN}}=$ 200 GeV, $b=(R_{Au}+R_{Au})/2$, $t_{fin}=5$ fm. As we can see, the temperatures reached in the middle region of the collision are very high - up to 450-500 MeV.

\begin{figure}[h!]  
	\begin{center}
		\resizebox{1.0\columnwidth}{!}
		{\includegraphics[width=\linewidth]{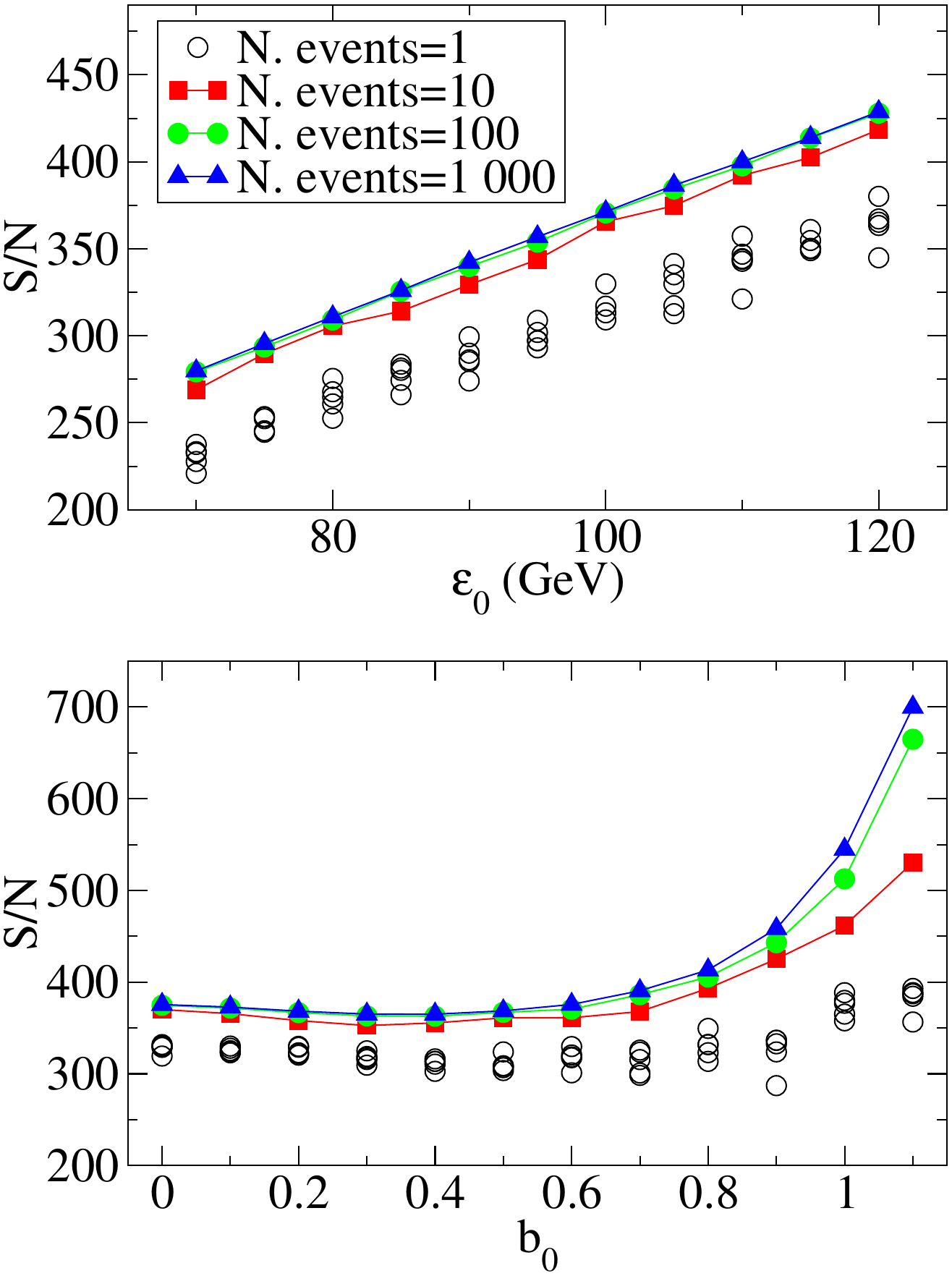}}
		\caption{ (color online)
			Top plot: entropy  per nucleon ($S/N$) as a function of the initial energy per nucleon ($\varepsilon_0$) for symmetric Au+Au collisions at $b=(R_{Au}+R_{Au})/2$, $t_{fin}=$5 fm. Bottom plot: $S/N$ as a function of the impact parameter for symmetric Au+Au collisions at $\sqrt{S_{NN}}=200$ GeV, $t=$5 fm.}
		\label{SPN}
	\end{center}
\end{figure}

From the entropy density one can calculate the total entropy per baryon charge, $S/N$:
\be
\frac{S}{N}=\frac{\sum_i^{N_{cell}}s_i\gamma_i}
{\sum_i^{N_{cell}}n_i\gamma_i}=\frac{S}{N_{part}}\,.
\label{SpNeq}
\ee
In Fig. \ref{SPN} we present the entropy per baryon charge in our initial state, simulated for symmetric Au+Au collisions, as a function of the initial energy per nucleon, $\varepsilon_0$, at impact parameter $b=(R_{Au}+R_{Au})/2$, top plot, and  as a function of the impact parameter, $b_0$, at $\sqrt{S_{NN}}=200$ GeV, bottom plot.

In the top plot we can see that, first of all,  $S/N$  increases linearly with $\varepsilon_0$. This linear rise, observed for a given impact parameter, i.e., given average $<N_{part}>$, is directly related to the rise of entropy with average temperature of the initial state in more energetic collisions.   Secondly, we note that $S/N$  saturates with the number of events, although not very rapidly - the saturation is reached around N events = 100, but the saturated value is much higher than $S/N$ in a single event! 

\begin{figure}[h!]  
	\begin{center}
		\resizebox{1.0\columnwidth}{!}
		{\includegraphics[width=\linewidth]{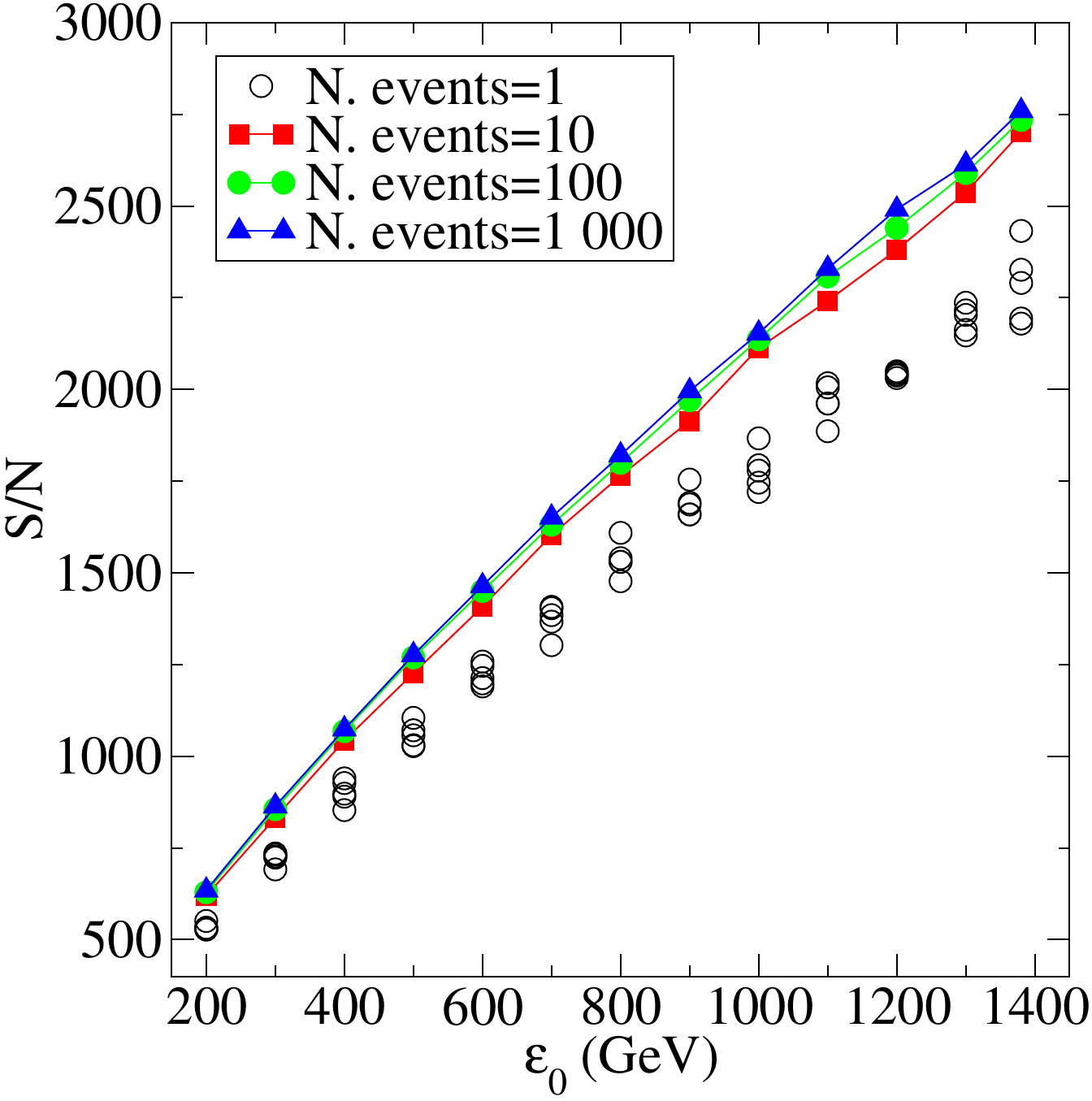}}
		\caption{ (color online)
			Entropy  per nucleon ($S/N$) as a function of the initial energy per nucleon, varying from RHIC to LHC, at $b_0=0.5$, $t_{fin}=$5 fm.}
		\label{SPN-LHC}
	\end{center}
\end{figure}

In Fig. \ref{SPN-LHC}  the entropy per baryon charge, $S/N$, is shown for energies from RHIC to LHC for Au+Au collisions at impact parameter $b=(R_{Au}+R_{Au})/2$, $t_{fin}=5$ fm. And we can see that both above discussed tendencies are correct even in much vwider range of energies. Firstly, the entropy per baryon charge strongly increases with  $\varepsilon_0$, although at LHC energies the entropy per baryon charge growth starts to deviate from linear rise behavior. This deviation is more noticable for single events. 
Secondly, the grater the initial energy per baryon charge is, the greater is the difference between values obtained averaging over different number of events and those obtained in single events. And thirdly, the single event fluctuations increase with the energy of the collision; this was actually not clear from the top plot of Fig. \ref{SPN}, but becomes notisable at larger scale. 

In Ref. \cite{PLB6922772010} the value of $S/N\simeq 225$ for the initial state calculated in ESRM model for Au+Au collision at $\varepsilon_0=65$ GeV per nucleon for central collision, $b_0=0$.  This value is compatible with our new results in the generalized model for a single event. However, for the initial state averaged over many events the $S/N$ is much bigger. Thus, although the averaged  initial state looks like smoothed over corresponding initial state obtained in ESRM, Fig. \ref{ENERDENSOLDNEW}, it will lead in a subsequent hydrodynamic evolution to rather different results due to much higher initial entropy.

In the top plot of Fig. \ref{RVBN} we have seen that for a given impact parameter, $N_{part}$ saturates very quickly with number of the events. 
Furthermore, the bottom plot of Fig. \ref{RVBN} shows how the reaction volume increases with the number of events for given impact parameter. We think that this is the effect, that explains the rise of saturated value of $S/N$ with respect to that in a single event.
 Since $N_{part}$ saturates rapidly the observed fluctuations in $S/N$ are related to fluctuations of the entropy of initial state. To study these fluctuations in more details we present these for different system sizes in
the bottom plot of Fig. \ref{SPN}.

In the bottom plot of Fig. \ref{SPN} one can distinguish two regions: the first one corresponds to $0\leq b_0 < 0.6$, where the $S/N$ is practically independent on $b_0$, and the second one with $b_0 \geq 0.6$, where a strong dependence on $b_0$ and on the number of events is observed. The more peripheral the collision is, the greater is the entropy generated by our model per baryon charge. Please note that also the  fluctuations of $S/N$ for a single event strongly increase with impact parameter. 

To understand these results we have to recall the basic thermodynamic relation: 
\be
dS = \frac{dU}{T}+\frac{P}{T}dV=\varepsilon_0 \frac{dN}{T}+\frac{P}{T}dV\,.
\label{dS}
\ee
To estimate the average variation of $S/N$ we first calculate the corresponding differential  
\be
d\left( \frac{S}{N}\right)=\frac{dS\ N - dN\ S}{N^2}=\frac{dS}{N}-\frac{S}{N}\frac{dN}{N}
\label{dSN}
\ee
and using Eq. (\ref{dS}) we finally obtain the following relation
\be
d\left( \frac{S}{N}\right)=
\left(\frac{\varepsilon_0}{T}
-\frac{S}{N}\right)
\frac{d N}{N}+
\frac{P}{T} \frac{d V}{N}
\label{dSN2}
\ee
From Fig. \ref{RVBN} we know that the fluctuations of relative volume and of $N_{part}$  increase for smaller systems, i.e. for higher impact parameters, and thus, following Eq. (\ref{dSN2}), the $S/N$ fluctuations in a single event will increase with $b_0$, as it is seen in the bottom plot of Fig. \ref{SPN}. 
However, why the $S/N$ value itself grows with impact parameter for peripheral collisions?

\begin{figure}[h!]  
	\begin{center}
		\resizebox{1.01\columnwidth}{!}
		{\includegraphics[width=\linewidth]{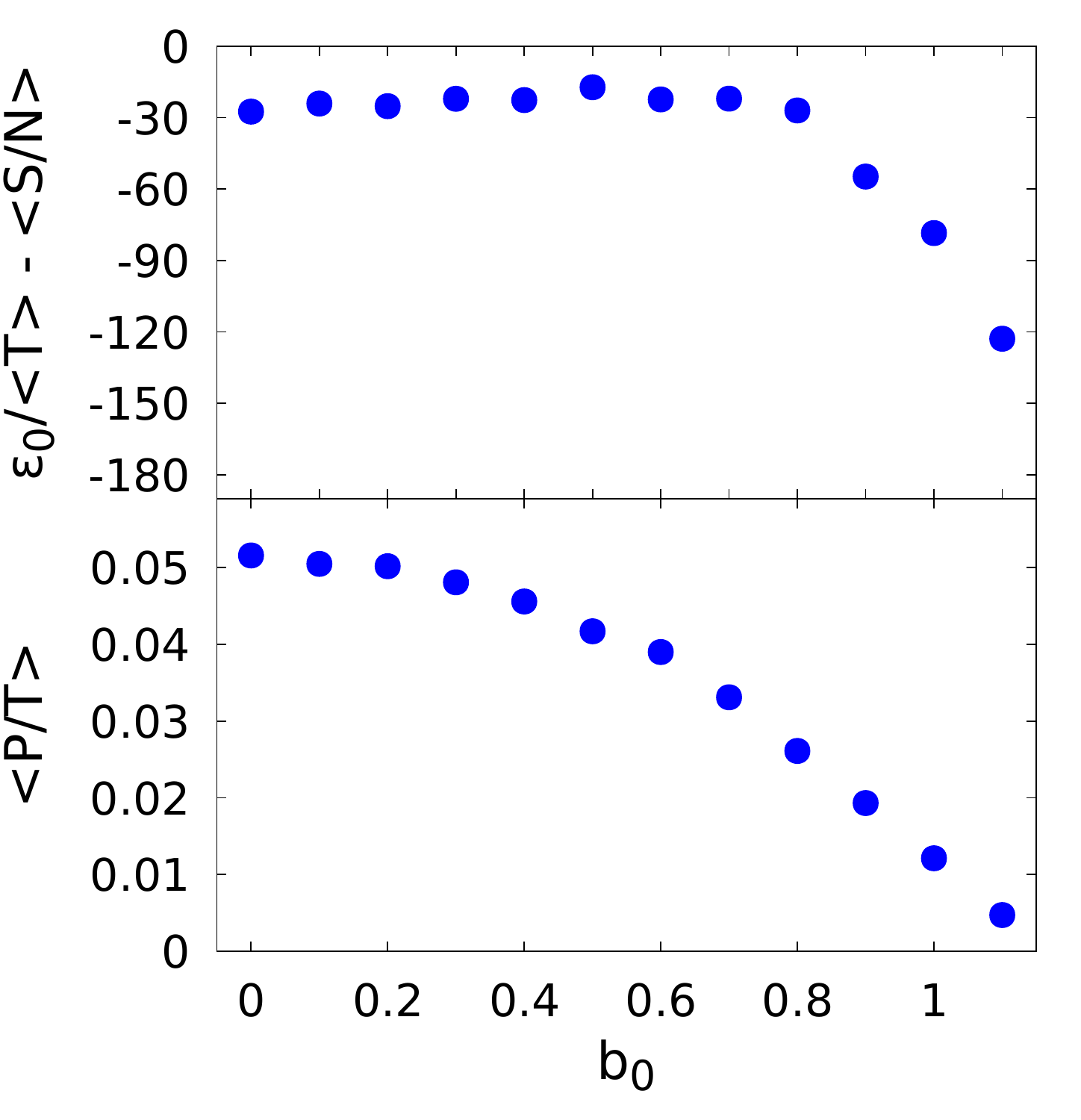}}
		\caption{ (color online)
			The coefficients of the Eq. (\ref{deltaSN}) as functions of impact parameter for symmetric Au+Au collisions at $\sqrt{S_{NN}}=$ 200 GeV $t_{fin}=$ 5 fm. The top plot shows the values of $\frac{\varepsilon_0}{\left\langle T\right\rangle}
			-\left\langle\frac{S}{N}\right\rangle$, while the bottom one shows the evolution of $\left\langle\frac{P}{T}\right\rangle$. The averaging is done over 10000 events. }
		\label{coeff}
	\end{center}
\end{figure}  

As impact parameter grows, $\Delta b_0>0$, the number of participants and the reaction volume decrease, $\Delta N<0$ and $\Delta V<0$, and the average change of $S/N$ can be estimated from Eq. (\ref{dSN2}):
\be
\left\langle \Delta\left( \frac{S}{N}\right)\right\rangle=
\left(\frac{\varepsilon_0}{\left\langle T\right\rangle}
-\left\langle\frac{S}{N}\right\rangle\right)
\left\langle\frac{\Delta N}{N}\right\rangle+
\left\langle\frac{P}{T}\right\rangle \left\langle\frac{\Delta V}{N}\right\rangle\,.
\label{deltaSN}
\ee
Since $\frac{P}{T} > 0$ the second term always generates a decrease of $S/N$. On the other hand the first term can be both positive or negative, and for our simulations it is positive. As we can see in the top plot of  Fig. \ref{coeff} the first coefficient  
$
\frac{100\ {\rm GeV} }{\left\langle T\right\rangle}
-\left\langle\frac{S}{N}\right\rangle <0
$ for all the values of the impact parameter, and, thus, the first term is always positive. However, in the first region ($0\leq b_0 < 0.6$) this increase of $S/N$ is compensated by the second negative term, see the bottom plot of Fig. \ref{coeff}. As we go to more and more peripheral collisions the average temperature slowly drops down while $S/N$ slowly grows maintaining the first term practically constant, while the value of the second term decreases, since $\frac{P}{T}\sim T^3$. 
Thus, at some moment ($b_0 \simeq 0.6$) the $S/N$ starts to grow and then it grows faster and faster, just as we observe in Fig. \ref{SPN}, since according to Eq. (\ref{deltaSN}) this is an auto-reinforced process. The bigger $S/N$ is, the bigger is the first positive term and, thus, the bigger is the increase of $S/N$.

\begin{figure}[h!]  
	\begin{center}
		\resizebox{1.01\columnwidth}{!}
		{\includegraphics[width=\linewidth]{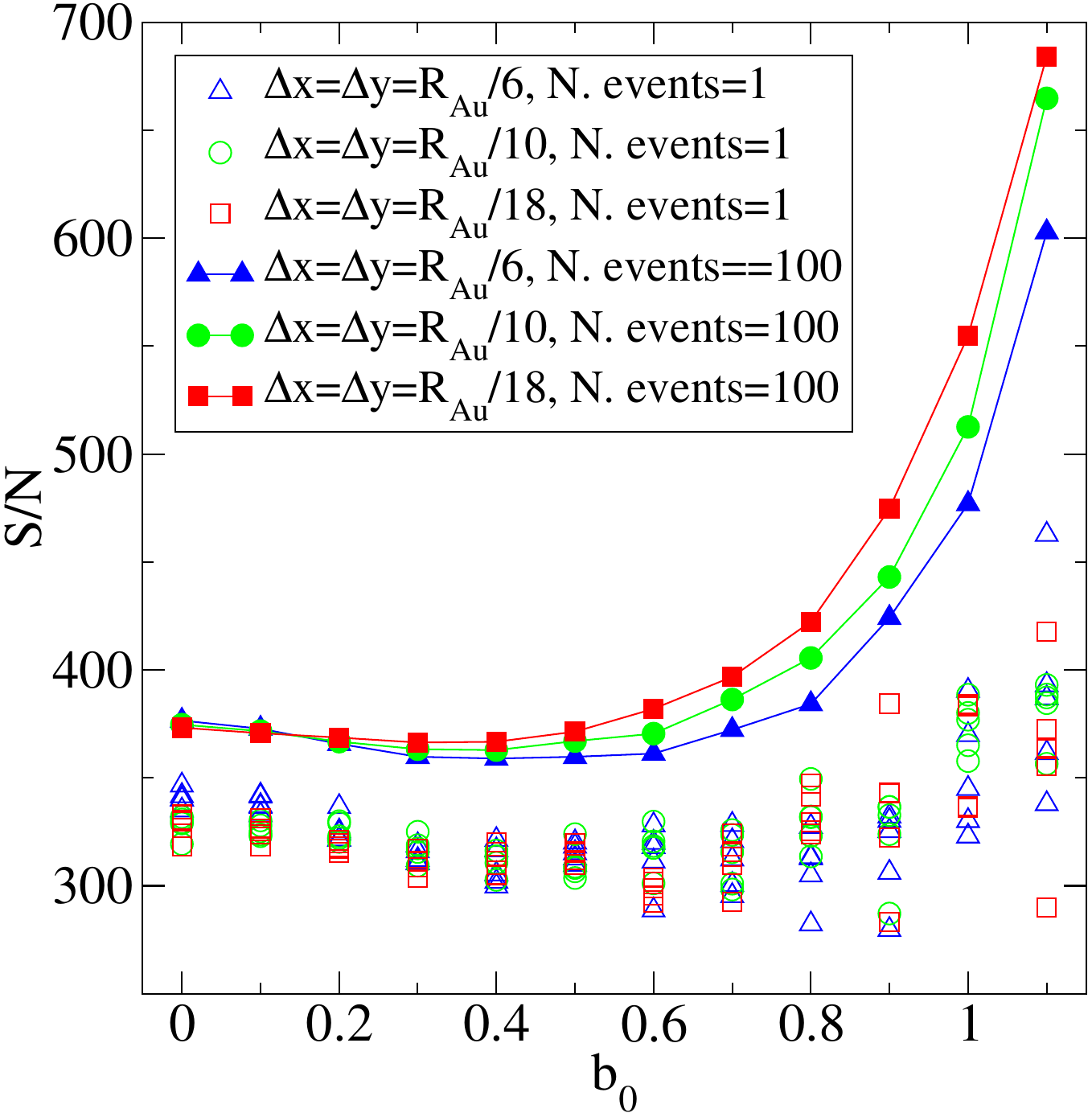}}
		\caption{ (color online)
			 $S/N$ as a function of the impact parameter for symmetric Au+Au collisions at $\sqrt{S_{NN}}=200$ GeV, $t_{fin}=$5 fm, for different cell sizes. }
		\label{SPN-cell}
	\end{center}
\end{figure}

We have also checked whether the $S/N$ production is sensitive to the cell size. Fig. \ref{SPN-cell} shows that it has a rather little sensitivity. The most central events ($b_0 \le 0.5$) are practically independent to the cell size, and for more peripheral collisions we observe a small increase of  $S/N$ with decrease of cell size. Taking into account that from  Fig. \ref{RVBN-delta} we know that $N_{part}$ for very peripheral collisions ($b_0 \ge 0.9$) is insensitive to the cell size, Fig. \ref{SPN-cell} tells us that the entropy production in the peripheral collisions grows a bit if one considers smaller cells.  

We can also see in Fig. \ref{SPN-cell} that the important effect, seen earlier in the bottom plot of Fig. \ref{SPN}, that entropy per nucleon production for the initial state averaged over many events strongly increases with respect to that for a single event, does not depend on the cell size.

\subsection{Vorticity}\label{IV.vort}

The vorticity field is a quantitative measure of the local circulation of a fluid. It is calculated in each point of the fluid and can be shown via  the so-called \textit{vortex lines}. 
A vortex line is a line whose tangent is everywhere parallel to the local vorticity vector.
 Vortex lines cannot cross each other, which implies that vortex lines cannot emerge or terminate anywhere of the fluid, but must keep on going until they reach the boundaries of the flow. 
 
 In classical physics for incompressible, perfect fluids vorticity exhibits an impressive conservation law: the conservation of circulation \cite{BL2011}. The relativistic fluid dynamical calculations indicate \cite{LPC-PRC-2012} that typical flow patterns and instabilities may occur here also. Thus, their studies can provide insight into the properties of the QGP.

In peripheral heavy-ion collisions the initial angular momentum of the system generates a strong shear and vorticity in the flow \cite{LPC2013}, which may lead to rotation \cite{PRC840249142011} and even Kelvin Helmholtz instability \cite{LPC-PRC-2012} in the reaction plane for low-viscosity QGP. At later stages of the reaction  particles produced in the vortical matter are expected to be polarized.

\subsubsection{Classical vorticity}

Mathematically, the classical vorticity $\vec{\omega}$ is defined as the curl of the flow velocity $\vec{v}$:
\be
\vec{\omega}\equiv\frac{1}{2}\nabla\times\vec{v}.
\label{classvort} 
\ee
Thus, for example, in the reaction plane the vorticity will be 
\begin{equation}
\omega_y\equiv\omega_{xz}\equiv-\omega_{zx}\equiv\frac{1}{2}\left(\partial_zv_x-\partial_xv_z\right),
\label{classvortRP}
\end{equation}
where the $x$,$y$,$z$ components of the 3-velocity $\vec{v}$ are denoted  respectively. In this definition of vorticity we have already included the factor $\frac{1}{2}$ for the symmetrization to have the same magnitude of vorticity as for symmetrized volume divergence or expansion rate \cite{DW2014}.

\begin{figure}[ht] 
\begin{center}
\resizebox{1.\columnwidth}{!}
{\includegraphics{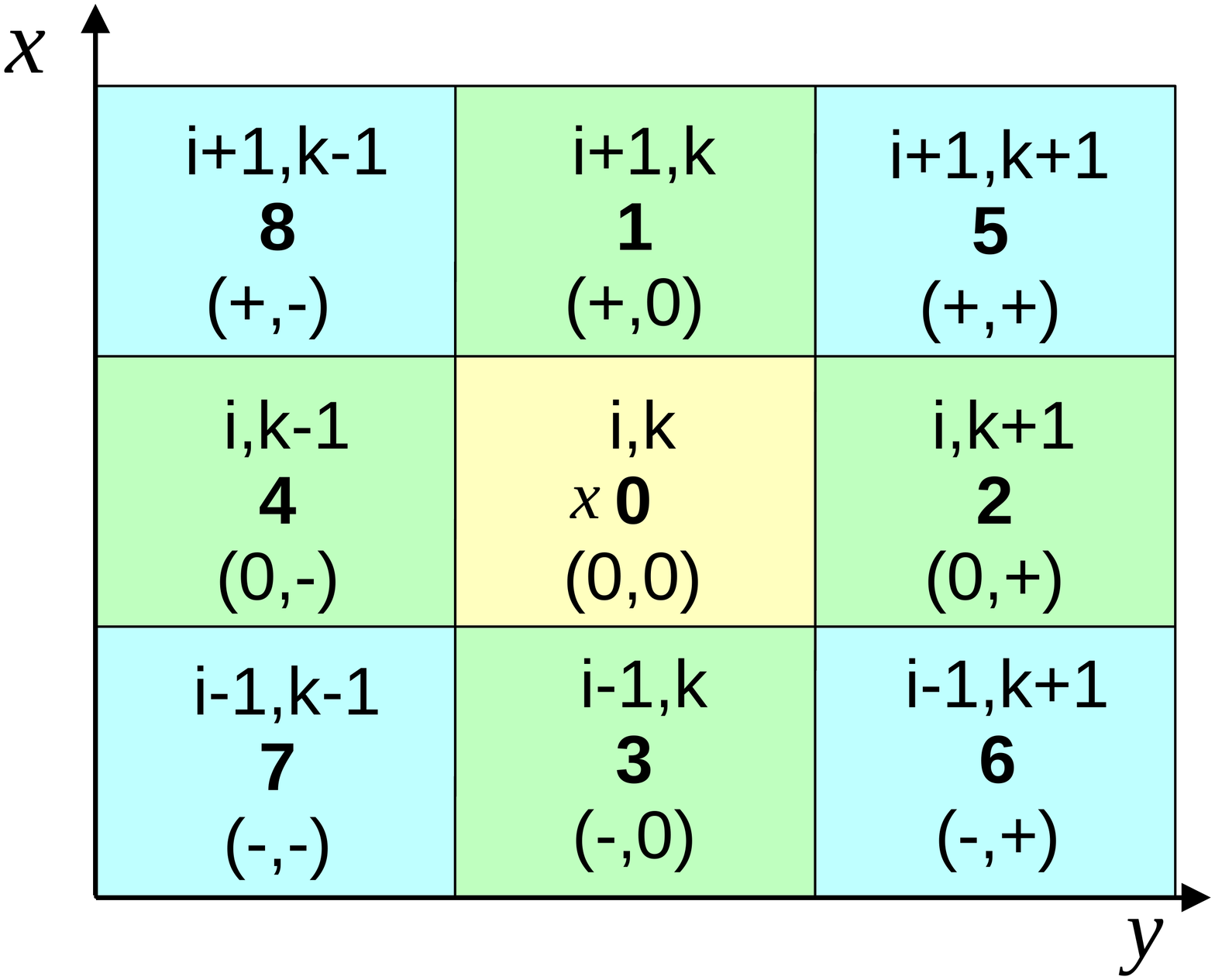}}
\caption{ (color online)
 Sketched of a fluid cell at $i$,$k$ in the reaction plane ([xz]-plane) with its eight neighbors. The central cell, $0$, is labeled as $(0,0)$; while the nearest four side neighbors indicated by 1,2,3,4 are labeled as $(+,0),\ (0,+),\ (-,0),\ (0,-)$; and the four corner neighbors indicated by 5,6,7,8 are labeled as $(+,+),\ (-,+),\ (-,-),\ (+,-)$. Figure adapted from Ref. \cite{DW2014}.
}
\label{GVC}
\end{center}
\end{figure}

To calculate the y-component of vorticity, $\omega_y$, in each cell of the reaction plane we proceed as follows:
First we label each cell by the indexes $i$,$k$ corresponding to $x$,$z$ axes, respectively. For a given layer $y$, we have a contribution from the side points 1,2,3,4 and the corner neighboring points 5,6,7,8 (see Fig. \ref{GVC}).

\begin{figure*}[ht] 
	\begin{center}
		\resizebox{1.01\textwidth}{!}
		{\includegraphics[width=\textwidth]{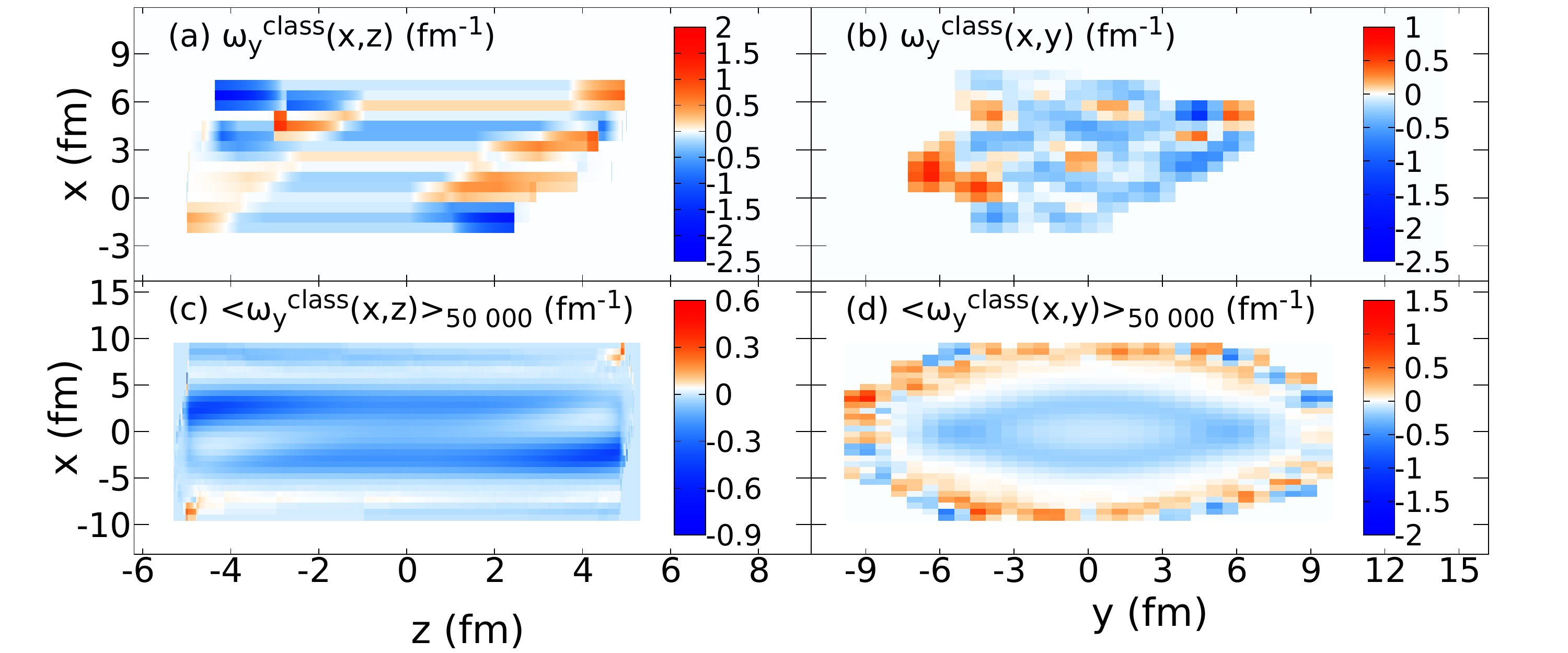}}
		\caption{ (color online)
			The y-component of the classical vorticity is shown for symmetric Au+Au collisions at $\sqrt{S_{NN}}=$ 200 GeV, $b=(R_{Au}+R_{Au})/2$, $t_{fin}=$ 5 fm. 
			In the top plots ((a) and (b)) $\omega_y^{class}$ is represented for a single event and in the bottom ones ((c) and (d)) for IS averaged over $N=$50000 events. The left plots ((a) and (c)) correspond to $\omega_y^{class}$ in the reaction plane ([xz]-plane) and the right ones ((b) and (d)) for that in the transverse plane ([xy]-plane).}
		\label{CV}
	\end{center}
\end{figure*} 
\begin{figure}[h!] 
	\begin{center}
		\resizebox{1.01\columnwidth}{!}
		{\includegraphics[width=\linewidth]{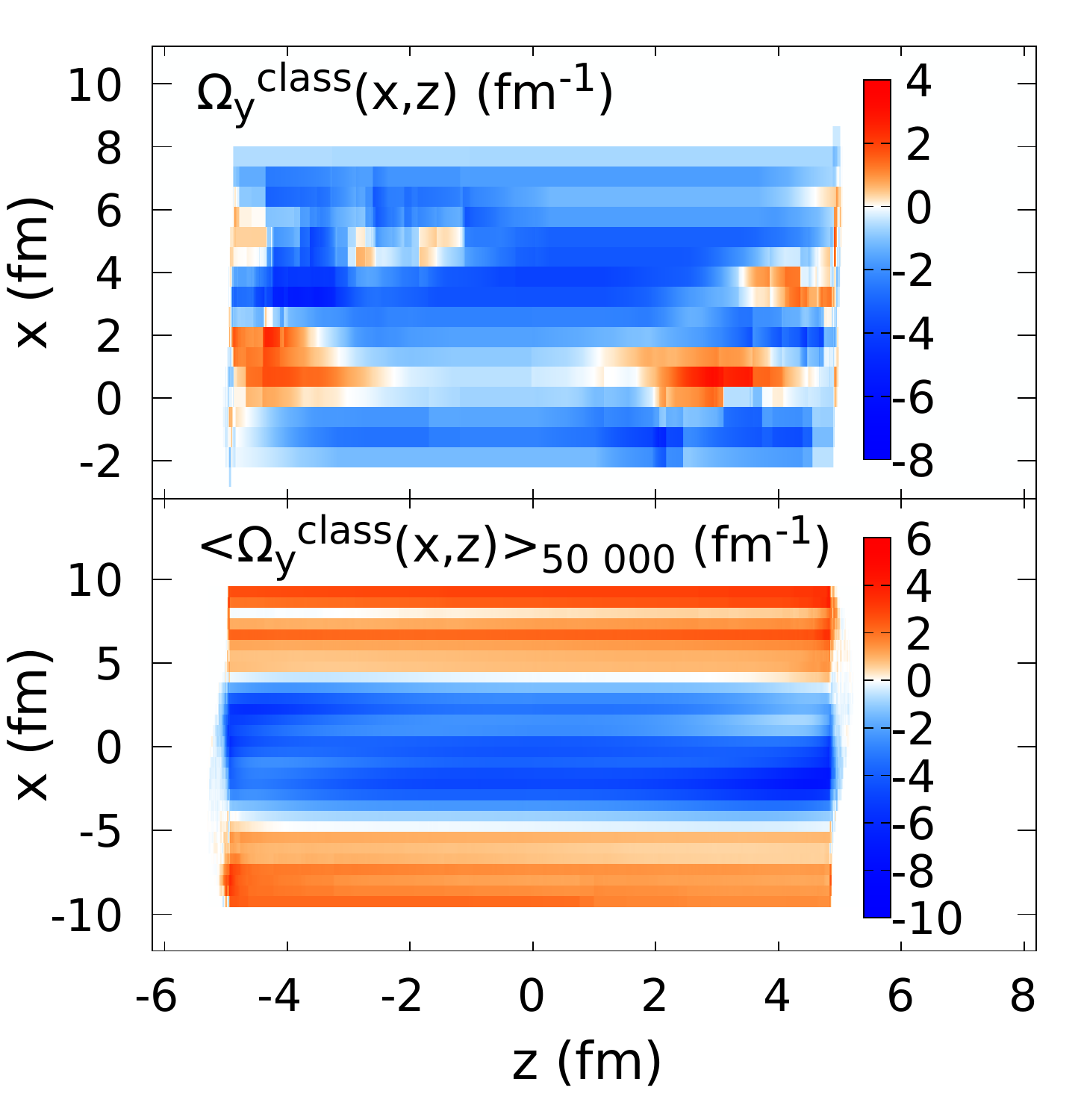}}
		\caption{ (color online)
			The y-component of the classical vorticity summed over all $y$-layers ($\Omega_y^{class}$) is shown in the reaction plane for symmetric Au+Au collisions at $\sqrt{S_{NN}}=$ 200 GeV, $b=(R_{Au}+R_{Au})/2$, $t_{fin}=$ 5 fm. 
			In the top plots $\Omega_y^{class}$ is represented for a single event and in the bottom ones for IS averaged over $N=$50000 events.
		}
		\label{TCV}
	\end{center}
\end{figure}  

Since in our model there is only expansion in the z-direction, the first term on the right side of Eq. (\ref{classvortRP}) vanishes. From this and the definition of partial derivative we can rewrite Eq. (\ref{classvortRP}) in the form
\be
\begin{aligned}
 \omega_y \equiv \omega_{xz} \equiv - \frac{v_z(x + \Delta x,y,z)-v_z(x,y,z)}{2\Delta x}.
 \label{classvortRP2}
\end{aligned} 
\ee
We now take all possible differences between neighbors cells in the x-direction, i.e., we take the differences: 1-0,5-2,2-6,0-3,4-7,8-4. So the vorticity for a given $i$,$k$ cell will be given by
\begin{multline} 
\begin{array}{ll}
\omega_y(i,k) = \frac{1}{2} \times & \\
\\
\left[ \frac{v_z^{+0} - v_z^{00}}{2\Delta x} + \frac{v_z^{00} - v_z^{-0}}{2\Delta x} + \frac{v_z^{++} - v_z^{0+}}{4\Delta x} + \frac{v_z^{0-} - v_z^{--}}{4\Delta x}\right. & \\
\\
 + \left. \frac{v_z^{+-} - v_z^{0-}}{4\Delta x} + \frac{v_z^{0+} - v_z^{-+}}{4\Delta x} \right]. 
\end{array}
 \label{CVPIC}
 \end{multline}
If all cells are filled with matter, the terms $v_z^{00}$, $v_z^{0-}$ and $v_z^{0+}$ cancel each other resulting in the following simplified expression for the classic vorticity \cite{DW2014}:
\begin{multline}
	\begin{array}{lcl}
		\omega_y(i,k) & = & \frac{1}{2} \times \left[ \frac{v_z^{+0} - v_z^{-0}}{2\Delta x} + \frac{v_z^{++} - v_z^{--}}{4\Delta x} + \frac{v_z^{+-} - v_z^{-+}}{4\Delta x} \right].
	\end{array}
\label{CVFC}
\end{multline}
This is not true for surface cells since we must delete every term where there is an empty cell, since to replace the empty cell by zero velocity would lead to a large derivative. Obviously the process is the same for all $y$-layers.

In Fig. \ref{CV} the classical vorticity distributions obtained for a single event, top plots, and averaged over $N=$50000 events, bottom plots, are shown for symmetric Au+Au collisions at impact parameter $b=(R_{Au}+R_{Au})/2$, $\sqrt{S_{NN}}=$200 GeV. $t_{fin}=$5 fm. Please note that, from the angular momentum conservation, the overall vorticity $\langle\omega_y^{class}\rangle$ should be negative, what is clearly seen for the averaged initial state. 
We can note that for $N=$50000 events vorticity takes positive values only at the  extremes of the reaction volume, as it has been seen in the other smooth initial state models, for example \cite{LPC2013,Magas:2017nuw}, while for a single event positive values can be distributed throughout all the reaction volume due to fluctuations. If we look at the average vorticity in the transverse plane, we can  see two well separated regions: a central one, where the vorticity takes small and negative values, and an outer one, where strong fluctuations are observed. However, the cells corresponding to this outer region contain small amount of matter and energy (see Figs. \ref{BCD} and \ref{ED}). So it is expected that if one studies the energy or matter weighted vorticity these strong fluctuations will vanish.

In Fig. \ref{TCV} the results obtained for the y-component of the classical vorticity summed over all $y$-layers are presented, i.e. 
\be
\Omega_y^{class} = \sum_i \omega_{y_i}^{class}\,.
\label{omega}
\ee
 The top plot shows $\Omega_y^{class}$ for a particular single event with fluctuations, while the bottom plot presents $\Omega_y$ averaged over $N=$50000 events. 
Again, in the case of $N=$50000 events, vorticity takes only positive values at the top and bottom extremes, while for a single event, positive and negative values are found distributed over all reaction plane.

\subsubsection{Relativistic vorticity}

Unlike to the classical hydrodynamics, where vorticity is defined as the curl of the velocity field $\bold{v}$, several vorticities can be defined in relativistic hydrodynamics \cite{IK-2021}: the kinematical vorticity, the kinematical transverse vorticity, the T-vorticity, the thermal vorticity,... In this work we follow the definition used in Ref. \cite{DW2014}, known as the transverse kinematical vorticity for the relativistic case. The corresponding vorticity tensor is defined as
\be
 \omega_{\nu}^{\mu} \equiv \frac{1}{2} \left( \nabla_{\nu} u^{\mu} - \nabla^{\mu} u_{\nu} \right)\,,
 \label{RVD}
\ee
where for any 4-vector $q^{\mu}$ the quantity $\nabla_{\alpha} q^{\mu} \equiv \Delta_{\alpha}^{\beta} \partial_{\beta} q^{\mu} = \Delta_{\alpha}^{\beta} q_{,\beta}^{\mu}$ and $\Delta^{\mu\nu} \equiv g^{\mu\nu} - u^{\mu}u^{\nu}$. This leads to
\begin{multline}
	\begin{array}{lcl}
		\omega_{\nu}^{\mu} & = & \frac{1}{2} \Delta^{\mu\alpha} \Delta_{\nu}^{\beta} \left( u_{\alpha,\beta} - u_{\beta,\alpha} \right)\\
		\\
		& = & \frac{1}{2} \left[ \left( \partial_{\nu} u^{\mu} - \partial^{\mu} u_{\nu} \right) + \left( u^{\mu} u^{\alpha} \partial_{\alpha} u_{\nu} -  u_{\nu} u^{\alpha} \partial_{\alpha} u^{\mu}\right)\right]\\ 
		\\
		& = & \frac{1}{2} \left[ \left( \partial_{\nu} u^{\mu} - \partial^{\mu} u_{\nu} \right) + \left( u^{\mu} \partial_{\tau} u_{\nu} -  u_{\nu} \partial_{\tau} u^{\mu}\right)\right]\,,
	\end{array}
  \label{RVD2}
\end{multline}
where $\partial_{\tau} u^{\mu} \equiv \dot{u}^{\mu} = u^{\alpha} \partial_{\alpha} u^{\mu}$ is the proper time derivative of $u^{\mu}$.

From the rapidity distributions represented in Fig. \ref{RAPIDITY} we can see that the rapidities in the central region of the collision are rather small. This indicates a strong stopping of partons in the middle region of the reaction volume and, therefore, it is expected that in this region the acceleration of the fluid elements are negligible compare to the rotation: $|\partial_{\tau}u^{\mu}| << |\partial_x u^z|$. This is also true at the extremes of the system where partons move with almost no loss of rapidity. This approximation allows us to simplify the relativistic vorticity to the following expression
\be
	\omega_{\nu}^{\mu} \approx \frac{1}{2} \left( \partial_{\nu} u^{\mu} - \partial^{\mu} u_{\nu} \right).
\ee 
Thus, for example, for vorticity development in the reaction plane we have
\be
 \omega_y^{rel}\equiv \omega_x^z = \frac{1}{2} \gamma \left( \partial_x v_z - \partial_z v_x \right) + \frac{1}{2} \left( v_z \partial_x \gamma - v_x\partial_z \gamma \right).
\label{RVFD}	
\ee
\begin{figure*}[ht]  
	\begin{center}
		\resizebox{1.01\textwidth}{!}
		{\includegraphics[width=\textwidth]{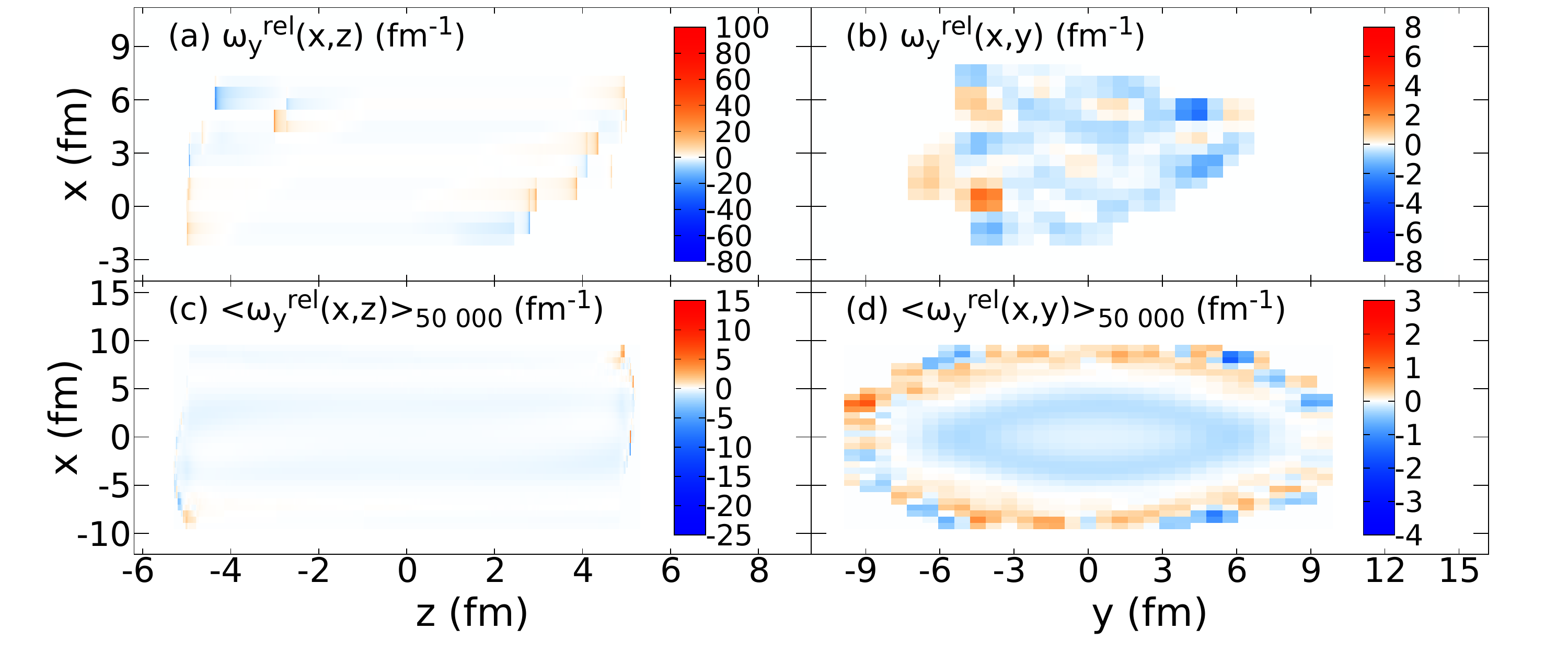}}
		\caption{ (color online)
			The y-component of the  relativistic vorticity is shown for symmetric Au+Au collisions at $\sqrt{S_{NN}}=$ 200 GeV, $b=(R_{Au}+R_{Au})/2$, $t_{fin}=$ 5 fm. In the top plots ((a) and (b)) $\omega_y^{rel}$ is represented for a single event and in the bottom ones ((c) and (d)) for IS averaged over $N=$50000 events. The left plots ((a) and (c)) correspond to $\omega_y^{rel}$ in the reaction plane ([xz]-plane) and the right ones ((b) and (d)) for that in the transverse plane ([xy]-plane).}
		\label{RVSE}
	\end{center}
\end{figure*}  
The first term is similar to the classical case, it has just to be multiplied by a factor $\gamma$, and the second (new) term has a similar structure. This results in the following expression (see Fig. \ref{GVC}): 
\begin{multline}
	\begin{array}{lcl}	
		\omega_x^z(i,k) & = & \frac{1}{8}\left[\gamma^{00}\left(2\left(\left(v_z^{+0}-v_z^{00}\right)+\left(v_z^{00}-v_z^{-0}\right)\right)\right.\right. \\
		\\
		&  + & \left.\left.\left(v_z^{++}-v_z^{00}\right) + \left(v_z^{00}-v_z^{--}\right)+\left(v_z^{+-} \right.\right.\right. \\
		\\
		& - & \left.\left.\left. v_z^{00}\right)+\left(v_z^{00}-v_z^{-+}\right)\right)/\Delta x\right]+\left[v_z^{00}\left(2\left(\right.\right.\right. \\
		\\
		&   &  \left.\left.\left. \left(\gamma^{+0}-\gamma^{00}\right)+\left(\gamma^{00}-\gamma^{-0}\right)\right)+\left(\gamma^{++}\right.\right.\right. \\
		\\
		& - & \left.\left.\left. \gamma^{00}\right)+\left(\gamma^{00}-\gamma^{--}\right)+\left(\gamma^{+-}-\gamma^{00} \right)  \right.\right. \\
		\\
		& + &  \left.\left. \left(\gamma^{00} - \gamma^{-+} \right)\right)/\Delta x\right].
		\label{eq.30}
	\end{array}
\end{multline}
If there is an empty neighbouring cell, all differences ($\gamma-\gamma$) and ($v-v$) with respect to the empty cell have to be dropped from this summation. On the other hand, if all neighbouring cells are filled the $v^{00}$ and $\gamma^{00}$ terms cancel and the expression is simplified to \cite{DW2014}
\begin{multline}
	\begin{array}{lcl}
		\omega_x^z(i,k) & = & \gamma^{00}\left[ \left(2\left(v_z^{+0}-v_z^{-0}\right) +
		\left(v_z^{++}-v_z^{--}\right)\right)/\Delta x \right.\\
		\\
		& + & \left.\left(v_z^{+-}-v_z^{-+}\right)/\Delta x\right]/8 + v_z^{00}\left[2\left(\gamma^{+0}\right.\right.\\
		\\
		& - &\left.\left.  \gamma^{-0}\right)/\Delta x+\left(\gamma^{++}-\gamma^{--}\right)/\Delta x+\left(\gamma^{+-}\right.\right.\\
		\\
		& - & \left.\left. \gamma^{-+}\right)/\Delta x\right].
	\end{array}
\label{RVNEC}
\end{multline}  

 Since for each streak the cells moving at velocities close to the speed of light are those located at the very ends of each streak, it is expected that these cells will have a much larger vorticity in the relativistic case than in the classical one. In Fig. \ref{RVSE} the y-component of the relativistic vorticity ($\omega_y^{rel}\equiv \omega_x^z$) is shown for symmetric Au+Au collisions at impact parameter $b=(R_{Au}+R_{Au})/2$, $\sqrt{S_{NN}}=$200 GeV. $t_{fin}=$5 fm. If we compare classical (Fig. \ref{CV}) and relativistic vorticity distributions in the reaction plane, we will see that the latter at the edges of the reaction volume (at large $|z|$) is one order of magnitude bigger than the former, because of the gamma factor present in the relativistic equations. On the other hand,
in the center of the reaction volume both vorticities have similar values, since in this region the gamma factor is close to unity.

Of course, the cells showing the biggest relativistic vorticity contain a very little amount of matter, and the situation might drastically change if one studies an energy or matter weighted relativistic vorticity, but we are not going to enter here in such a discussion.

\begin{figure}[h!]  
	\begin{center}
		\resizebox{1.01\columnwidth}{!}
		{\includegraphics[width=\linewidth]{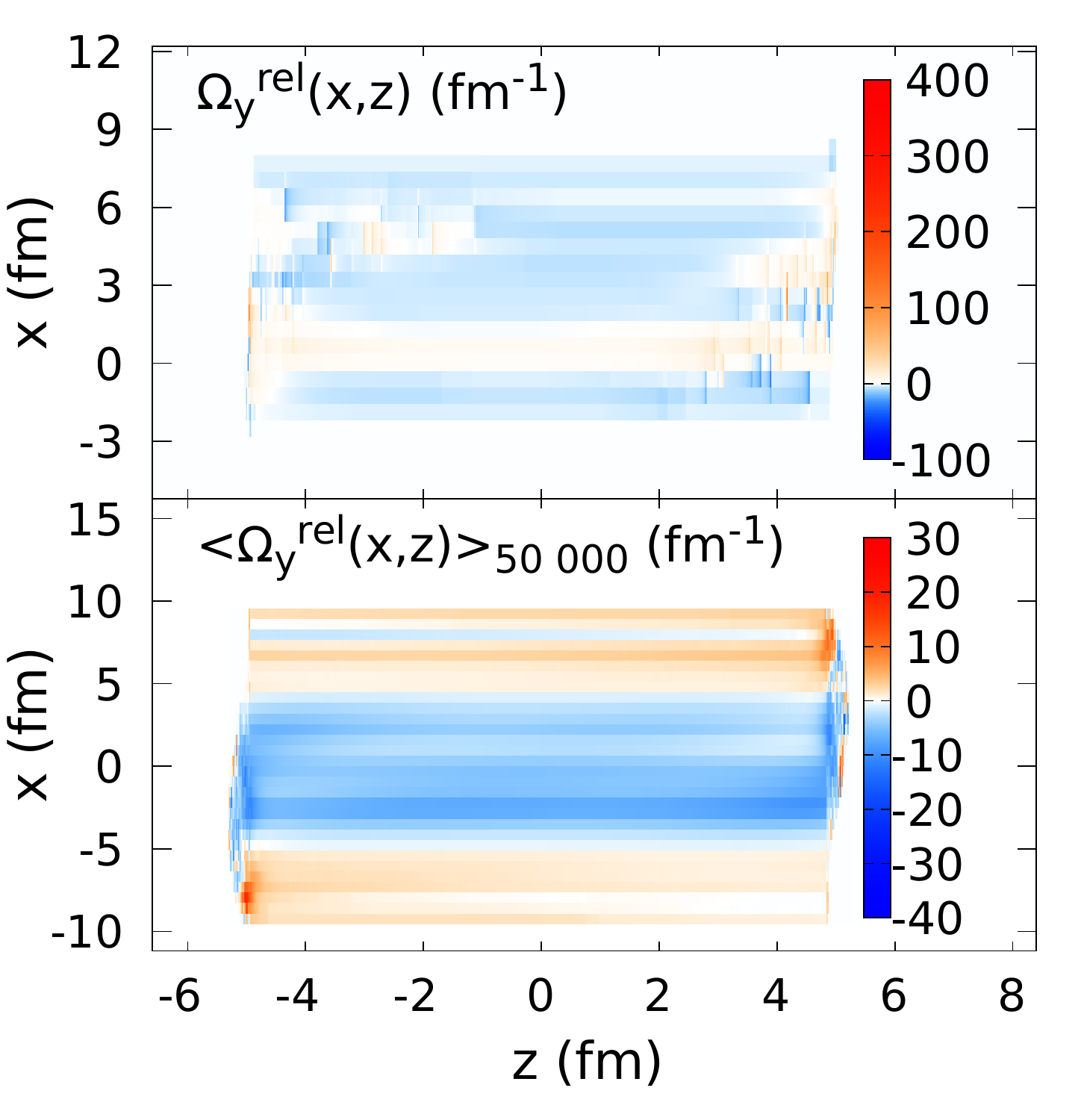}}
		\caption{ (color online)
		The y-component of the relativistic vorticity summed over all $y$-layers ($\Omega_y^{rel}$) is shown in the reaction plane for symmetric Au+Au collisions at $\sqrt{S_{NN}}=$ 200 GeV, $b=(R_{Au}+R_{Au})/2$, $t_{fin}=$ 5 fm. In the top plots $\Omega_y^{rel}$ is represented for a single event and in the bottom ones for IS averaged over $N=$50000 events.
		}
		\label{TRV}
	\end{center}
\end{figure}

In Fig. \ref{TRV} we show the relativistic vorticity summed over all y-layers ($\Omega_y^{rel}= \sum_i \omega_{y_i}^{rel}$) in the reaction plane for a single event, top plot, and averaged over $N=$50000 events, bottom plot, for symmetric Au+Au collisons at impact parameter $b=(R_{Au}+R_{Au})/2$, $\sqrt{S_{NN}}=$200 GeV. $t_{fin}=$5 fm. The comparison of Figs. \ref{TRV}  and \ref{TCV} confirms the above observations.  

\begin{figure}[h!]  
	\begin{center}
		\resizebox{1.01\columnwidth}{!}
		{\includegraphics[width=\linewidth]{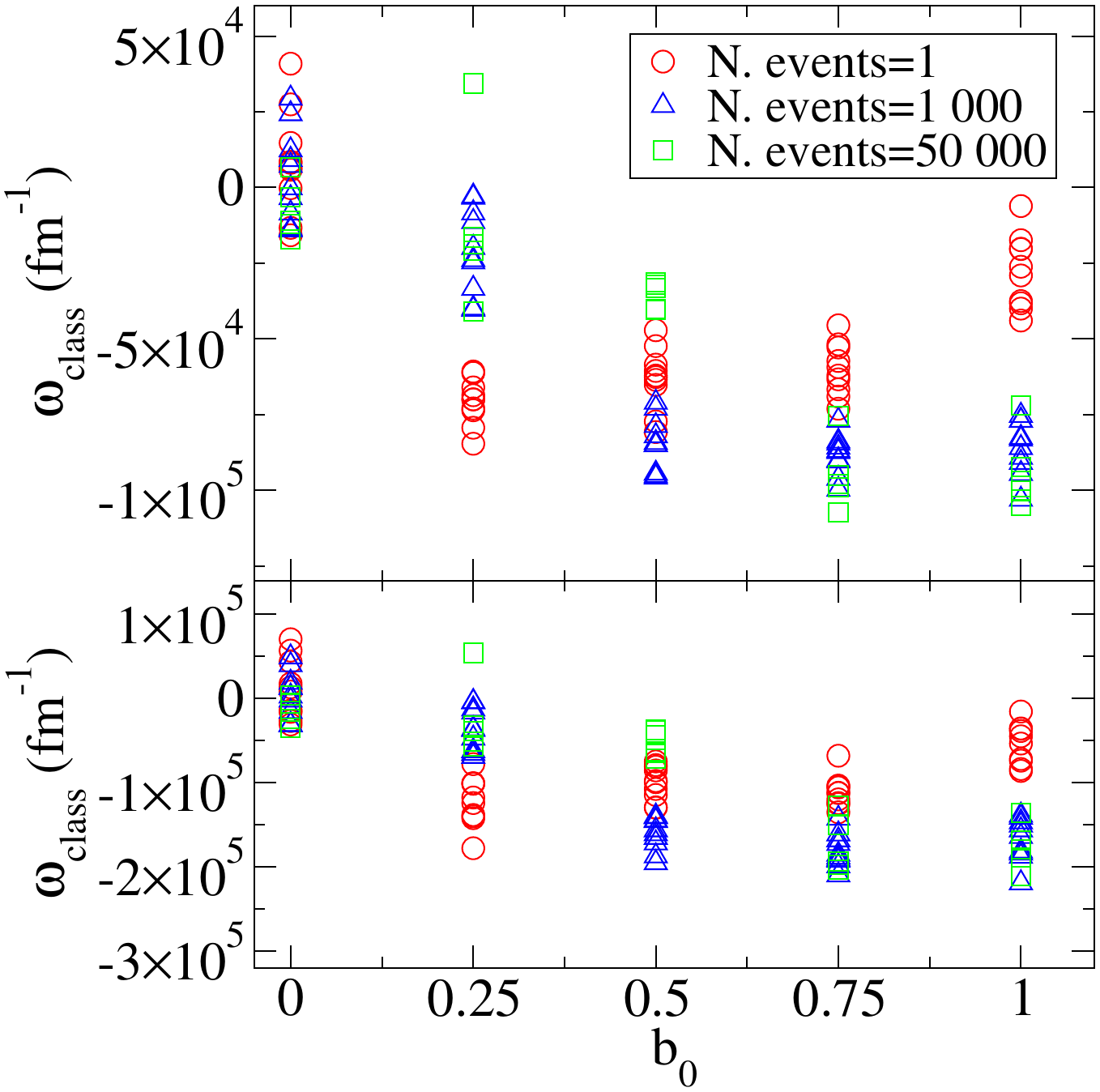}}
		\caption{ (color online)
			The y-component of the classical, top plot, and relativistic, bottom plot, total vorticities, Eq. (\ref{total_vort}),  as a function of the impact parameter for 1, 1000 and 50000 events, for symmetric Au+Au collisions at $\sqrt{S_{NN}}=$ 200 GeV, $t_{fin}=$ 5 fm.
		}
		\label{CRVb0}
	\end{center}
\end{figure}

In Fig. \ref{CRVb0} we present the classical, top plot, and relativistic, bottom plot, total vorticities, i.e. sum of the corresponding vorticities $\omega^{class/rel}_{y}$ over all non-empty cells of the grid:
\be
\omega_{class/rel}=\sum_{ijk} \omega^{class/rel}_{y, ijk}\,,
\label{total_vort}
\ee 
as a function of the impact parameter for 1, 1000 and 50000 events for symmetric Au+Au collisions at impact parameter $b=(R_{Au}+R_{Au})/2$, $\sqrt{S_{NN}}=$200 GeV. $t_{fin}=$5 fm. As we can see, the results show that the total vorticity is strongly dependent on the impact parameter, similar to the angular momentum (see for example \cite{Becattini:2007sr}), presenting a parabolic-like shape with a shallow minimum located around $b_0=$0.75, for the classical total vorticity, and $b_0=$1.0, for the relativistic total vorticity  (please note that both, relativistic and classical, total vorticities show approximately the same shape, although they differ in their magnitudes by factor 2).  

One interesting thing that can be also noticed from Fig. \ref{CRVb0} is that the total vorticities show very strong fluctuations even when we average over N=50000 events. In fact, there is a rather small difference between the spread of the points obtained for N=1000 events and those obtained for N=50000 events, what suggests a very slow "saturation" of the total vorticity with the number of events. 

This is actually an effect that can be expected. Making an initial state averaged over some number of event, what we actually average is the baryon and energy content of each cell, assigning to it $<N^0>$, $<N^z>$, $<T^{00}>$ and $<T^{0z}>$. These quantities, directly related to the corresponding conservation laws, relatively quickly saturate to their medium values determined by the impact parameter. The quantities directly related to them, for example, flow velocity $<v_z>=<N^z>/<N^0>$, show a similar behaviour
and also saturate rather rapidly. However the vorticity has to do not with the flow velocity, but with its derivative, which naturally will need much more statistics to demonstrate the saturation effect.

\begin{figure}[ht] 
	\begin{center}
		\resizebox{1.\columnwidth}{!}
		{\includegraphics[width=\columnwidth]{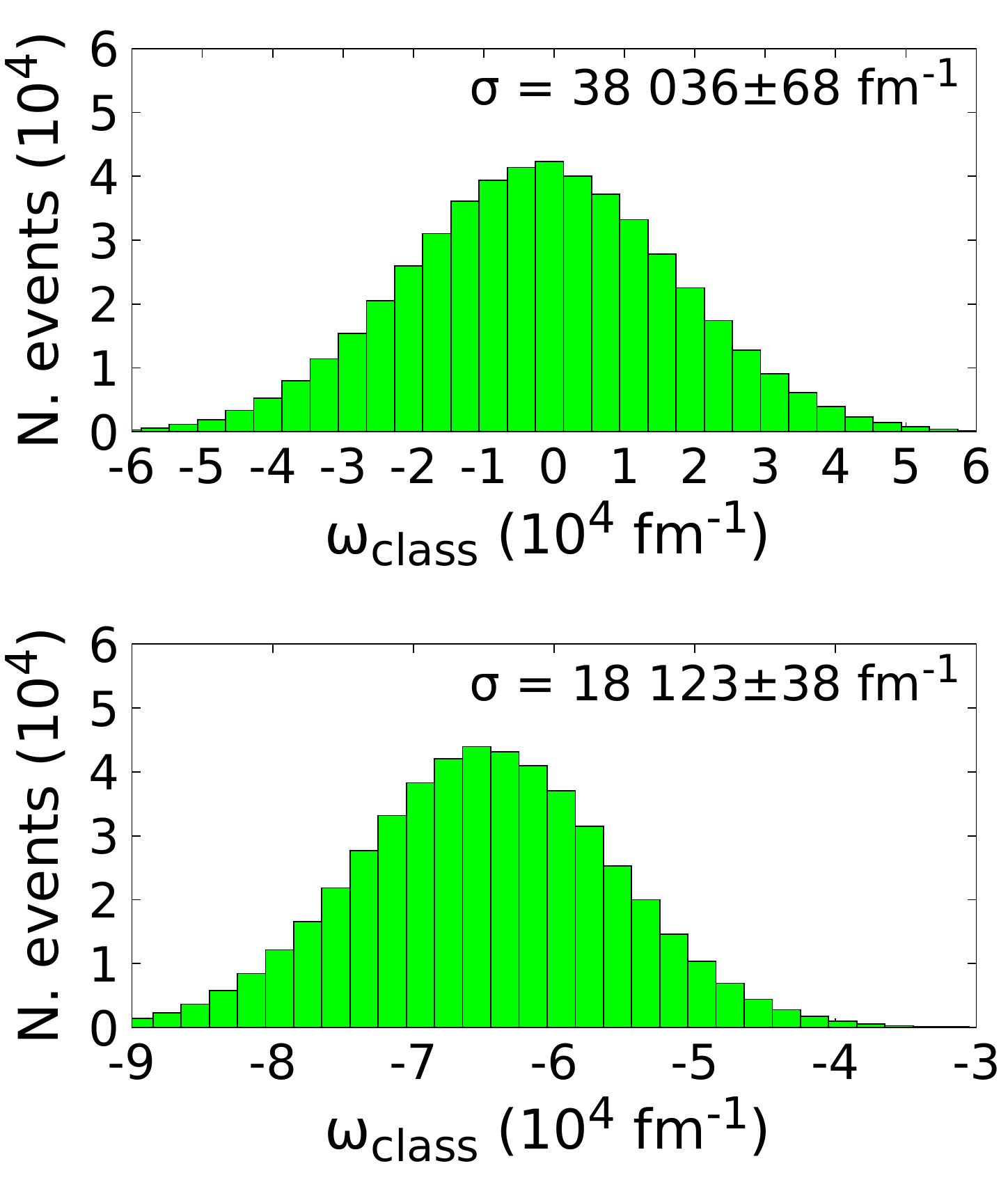}}
		\caption{ (color online)
			Distributions of the classical vorticity single event results for symmetric Au+Au collisions at$\sqrt{S_{NN}}=$200 GeV, $b=0$ (top plot) and $b=(R_{Au}+R_{Au})/2$ (bottom plot), $t_{fin}=$ 5 fm. The total number of results presented at each histogram is 500000. See text for the discussion.
		}
		\label{TCVHIST}
	\end{center}
\end{figure}

To understand better such large fluctuations of vorticity, we show in Fig. \ref{TCVHIST} distributions of the classical vorticity single event results for symmetric Au+Au collisions at $\sqrt{S_{NN}}=$200 GeV, $t_{fin}=$ 5 fm for $b=0$ (top plot) and $b=(R_{Au}+R_{Au})/2$ (bottom plot). The total number of results presented at each histogram is 500.000. 
As we see, for such a large number of events we obtained a Gaussian-shaped curve centered at $\omega_{class}$ $=$ 0 fm$^{-1}$, for $b_0=0$, and $\omega_{class}$ $\approx$ -65000 fm$^{-1}$, for $b_0=1/2$, as it should be according to Central Limiting Theorem. However we can also note that the standard deviation in the case  $b_0=0$ is surprisingly high - $\sigma =$ 38036$\pm$68 fm$^{-1}$. For $b_0=0.5$ it is approximately two times smaller - $\sigma =$ 18123$\pm$38 fm$^{-1}$, but still the distribution is rather wide, if we compare it, for example, with those of Fig. \ref{ENERPzRAPFLUCT}. Out of this we can guess that in order to get some experimental results related to the vorticity with good accuracy a rather high statistics may be needed. 

Looking at Fig. \ref{CRVb0} we can also note that the total vorticity for an averaged state may be different from the vorticity in a single event. This can be seen despite the strong fluctuations. This is the same effect as the one seen already for the entropy per baryon production, and, in our opinion, it has the same reason. From Figs. \ref{RVSE} and \ref{CV} we have learned that the most peripheral cells have the highest vorticity, and not always negative! In the initial state averaged over many events the reaction volume grows, see Fig. \ref{RVBN-delta} (bottom plot), due to an increment of peripheral cells with a little amount of matter, but with rather high vorticity, what leads to the observed results. On the other hand, since these cells contain a little amount of matter then their contribution, for example, to $\Lambda$ hyperon polarization results will be very little, since the probability that such a heavy baryon will be produced in one of these very peripheral cells is very low.

\begin{figure}[ht!]  
	\begin{center}
		\resizebox{1.01\columnwidth}{!}
		{\includegraphics[width=\linewidth]{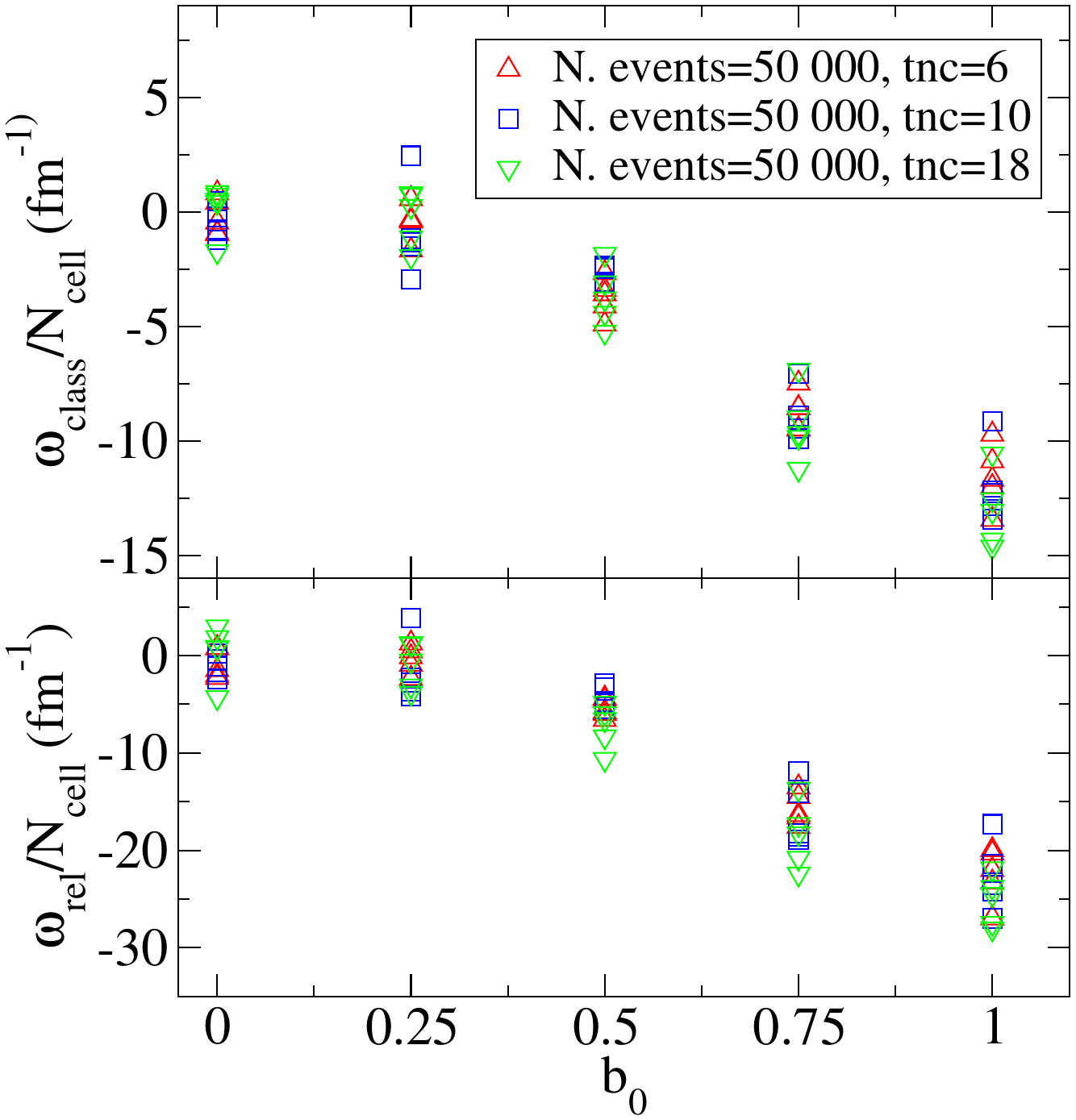}}
		\caption{ (color online)
			The y-component of the classical, top plot, and relativistic, bottom plot, total vorticities, Eq. (\ref{total_vort}), normalized to total number of non-empty cells  as a function of the impact parameter for different cell sizes for symmetric Au+Au collisions at $\sqrt{S_{NN}}=$ 200 GeV, $t_{fin}=$ 5 fm.
		}
		\label{CRVb0-delta}
	\end{center}
\end{figure}

Finally, we decided to test how sensitive are the obtained vorticity values to the cell size of our model.   In Fig. \ref{CRVb0-delta} we show the classical, top plot, and relativistic, bottom plot, vorticities  summed over all non-empty cells of the grid and normalized to total number of cells (which will obviously vary with the cell size), as a function of the impact parameter for different cell sizes. Despite the fluctuations we can  clearly see that the generated vorticity, averaged over 50000 events, is not sensitive to the size of numerical grid (not more than any other derivative). 

\subsection{Initial state for asymmetric A+Au collisions}

In order to see more clearly which features of the initial state are sensitive to the geometry of the collision and which depend only on the number of participant nucleons, we compare peripheral Au+Au with asymmetric A+Au head on collisions for different A from 2 to 180. The idea is to compare initial states generated in these geometrically rather different situation but with the same average number of participants. The corresponding initial state calculations are performed at $\sqrt{S_{NN}}=200$ GeV and $t_{fin}=$5 fm; the average number of participants is compared for $N_{\mathrm{events}}=100$. 

\begin{table}
	\centering
	\begin{tabular}{|c|c|c|c|c|}
		\hline
		A & 7 & 10 & 13 & 16 \\
		\hline
		$\alpha$ (fm) & 0.327 & 0.837 & 1.403 & 1.544 \\
		\hline
		$a$ (fm) & 1.770 & 1.710 & 1.635 & 1.833  \\
		\hline
		$\sqrt{\langle r^2 \rangle}$ (fm) & 2.39 & 2.45 & 2.44 & 2.718 \\
		\hline
	\end{tabular}
	\caption{Parameters used in the Harmonic Oscillator (HO) distribution (see Eq. (\ref{HOdist})) for nuclei with A=7, 10, 13, 16. The values are taken from Ref. \cite{ADNDT1987}.}
	\label{HOD-parameters}
\end{table}

To parameterize the nucleon density distribution of the nuclei with A nucleons we have used: 

- for A$\ge$100, a WS distribution with parameter $a_{WS}=0.535$ fm and radii calculated from equation $r=1.1 A^{1/3}$ fm.

- for 20$\leq$A$\leq$100, a WS distribution with parameter $a_{WS}=0.535$ fm and radii calculated from equation $r=1.2 A^{1/3}$ fm.

- for 6$\leq$A$\le$20, an Harmonic Oscillator (HO) distribution 
\be
\rho(r) = \rho_0\left(1+\alpha\left(r/a\right)^2\right)exp\left(-\left(r/a\right)^2\right),
\label{HOdist}
\ee  
where  the parameters $\alpha$ and $a$ are presented in Tab. \ref{HOD-parameters}. 

- for A$=$2, 3, 4  we have used an homogeneous distribution with radii $R_2=2.095$ fm, $R_3=1.976$ fm, $R_4=1.696$ fm correspondingly, taken from Ref. \cite{ADNDT1987}. 
\begin{figure}[ht!]  
	\begin{center}
		\resizebox{1.01\columnwidth}{!}
		{\includegraphics[width=\linewidth]{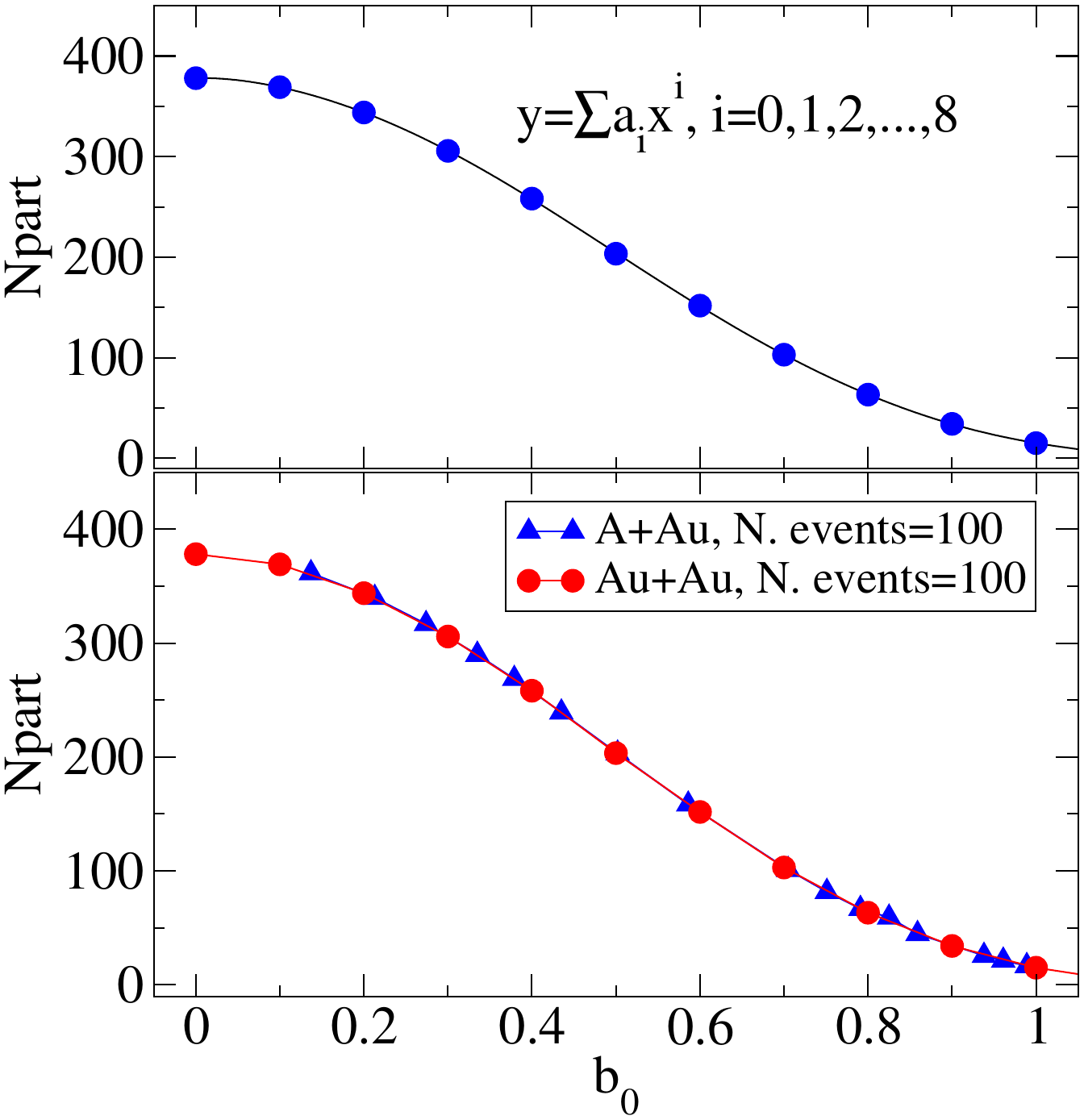}}
		\caption{ (color online)
			The top plot shows a polynomial fit, based on the data from Tab. \ref{Npart-b0_relation}, which relates an impact parameter of the Au+Au collision with an average number of participants. The fit has been used to obtain an equivalence between A, the number of nucleons in a given nucleus in asymmetric A+Au head on collisions, and $b_0$ for the same values of $N_{part}$ in symmetric Au+Au collision (see Tab. \ref{Npart-A-b0_relation}). In the bottom plot, we show $N_{part}$ as function of the impact parameter obtained in symmetric Au+Au collisions (dashed line) and in asymmetric A+Au head on collisions (solid line).} 
		\label{NpartVis}
	\end{center}
\end{figure} 

\begin{table*}
	\centering
	\begin{tabular}{|c|c|c|c|c|c|c|c|c|c|c|c|c|}
		\hline
		$b_0$ & 0.0 & 0.1 & 0.2 & 0.3 & 0.4 & 0.5 & 0.6 & 0.7 & 0.8 & 0.9 & 1.0  \\
		\hline
		$N_{part}$ & 378.0 & 369.7 & 343.8 & 305.7 & 258.0 & 203.5 & 151.8 & 103.0 & 63.3 & 34.2 & 15.1\\
		\hline
	\end{tabular}
	\caption{Number of participant nucleons ($N_{part}$) obtained averaging over $N=100$ events in symmetric Au+Au collisions at different impact parameters.}
	\label{Npart-b0_relation}
\end{table*}

In order to be able to compare the results for peripheral Au+Au with asymmetric A+Au head  on collisions, we, first of all, perform a polynomial fit to the average number of participants for symmetric Au+Au collisions, presented in Tab. \ref{Npart-b0_relation}, as a function of the corresponding impact parameter. The resulting  fit is shown in top plot of Fig. \ref{NpartVis}.
Having the analytical function $N_{part}(b_0)=\sum_{i=0}^{i=8} a_i b_0^i $ (parameters $a_i$ are given in Tab. \ref{fit_par}) one can invert this relation and find $b_0$ as a function of $N_{part}$. This allow us to calculate for each mass number A of the asymmetric A+Au head on collision an equivalent parameter $b_0^{equiv}$ from the average number of participant nucleons (see Tab. \ref{Npart-A-b0_relation}). For example, for A$=$2 we obtained $N_{part}\approx$16.72 (averaging over N$=100$ events), which is very similar to the number of participant nucleons obtained to $b_0=1.0$ in symmetric Au+Au collisions. The bottom plot of Fig. \ref{NpartVis} shows $N_{part}$ as a function of $b_0$ for both sets of data,  Tab. \ref{Npart-b0_relation} and Tab. \ref{Npart-A-b0_relation}.

\begin{table*}
	\centering
	\begin{tabular}{|c|c|c|c|c|c|c|c|c|}
		\hline
		$a_0$ & $a_1$ & $a_2$ & $a_3$ & $a_4$ & $a_5$ & $a_6$ & $a_7$ & $a_8$ \\
		\hline
		378.026 & 18.570 & -1357.451 & 3720.078 & -12366.451 & 
		24885.382 & -26442.638 & 14346.405 & -3166.771 \\
		\hline
	\end{tabular}
	\caption{Parameters of the polynomial fit to the average number of participants for Au+Au collisions $N_{part}(b_0)=\sum_{i=0}^{i=8} a_i b_0^i $. }
	\label{fit_par}
\end{table*}

\begin{table*}
\centering
\begin{tabular}{|c|c|c|c|c|c|c|c|c|c|c|c|c|c|c|c|c|}
\hline
A & 2 & 3 & 4 & 7 & 10 & 13 
& 16 & 20 & 40 & 60 & 80 & 100 
& 120 & 140 & 160 & 180\\
\hline
$N_{part}$ & 16.7 & 21.4 & 25.7 & 44.7 & 55.2 & 66.8 
& 81.7 & 101.1 & 158.5 & 203.2 & 239.2 & 268.6 
& 289.8 & 316.8 & 339.8 & 361.4\\
\hline
$b_0^{equiv}$ & 0.989 & 0.961 & 0.938 & 0.859 & 0.825 & 0.791  
& 0.751 & 0.705 & 0.586 & 0.502 & 0.435 & 0.379 
& 0.335 & 0.274 & 0.213 & 0.137\\
\hline
\end{tabular}
\caption{Number of participant nucleons ($N_{part}$) obtained averaging over $N_{\mathrm{events}}=100$ events in asymmetric A+Au head on collisions for different nuclei A. The equivalent impact parameters ($b_0^{equiv}$) are obtained for the same number of participant nucleons in symmetric Au+Au peripheral reactions. }
\label{Npart-A-b0_relation}
\end{table*}
\begin{figure}[ht!]  
	\begin{center}
		\resizebox{1.01\columnwidth}{!}
		{\includegraphics[width=\linewidth]{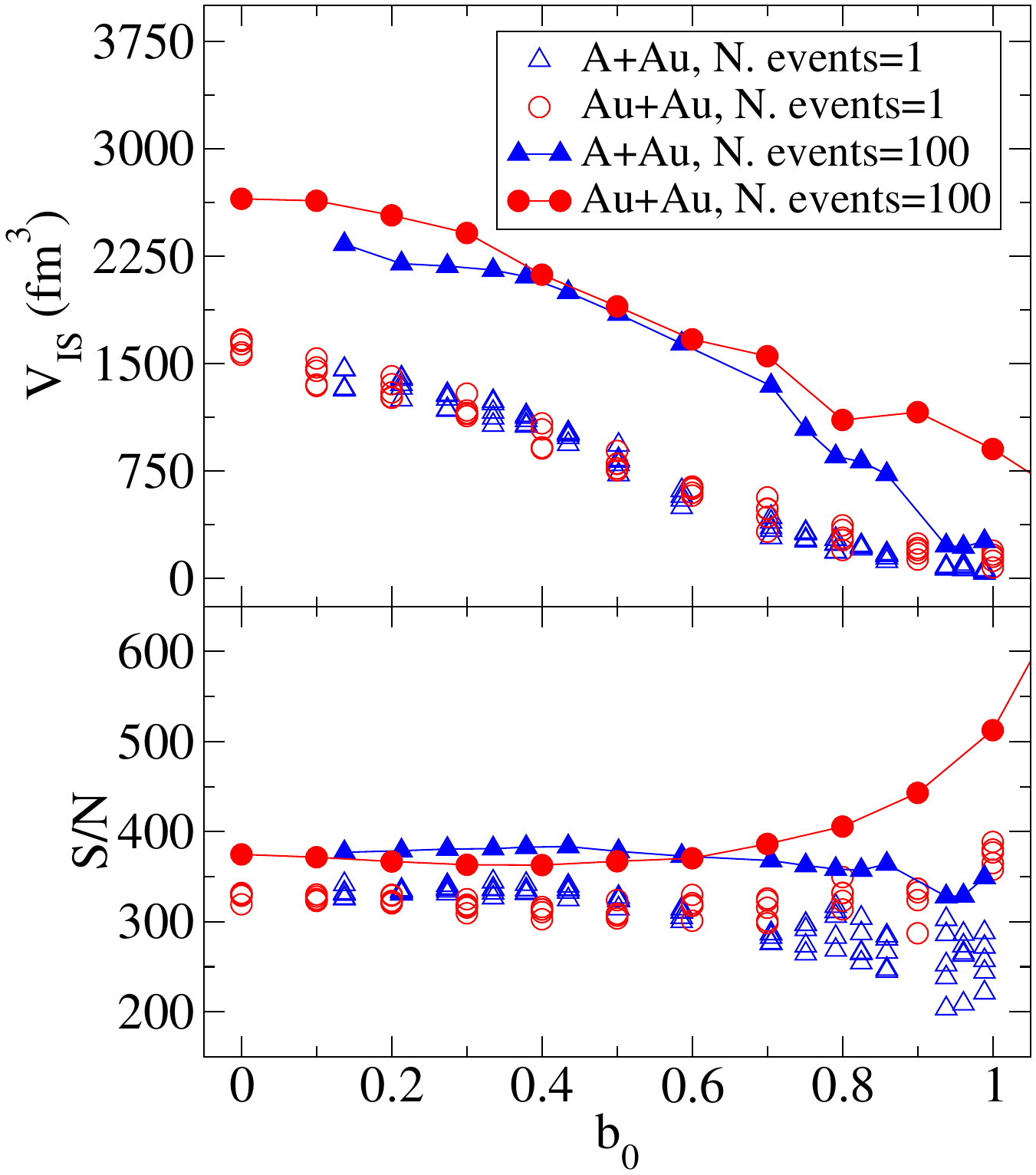}}
		\caption{ (color online)
			Reaction volume, top plot, and entropy per nucleon, $S/N$, bottom plot, obtained in symmetric Au+Au collisions at different impact parameters (dashed line) and in asymmetric A+Au head on collisions (solid line), at $\sqrt{S_{NN}}=200$ GeV and $t_{fin}=$5 fm.
		}
		\label{SpNA}
	\end{center}
\end{figure} 
Now, when we know that the average number of participant nucleons in the corresponding peripheral Au+Au and A+Au head on collisions is the same, we 
can look at the reaction volume and the entropy per nucleon of these initial states. The corresponding quantities are shown in Fig. \ref{SpNA}. 
As we can see,  for both these observables the increase of the impact parameter in symmetric reaction is equivalent to the reduction of the projectile size in asymmetric reaction for central and semicentral collisions, i.e. for $b_0 \leq 0.7$ or in other words for $A \geq 20$. And this equivalence is true not only for the averaged over 100 events initial state, but also for single events. 

For very peripheral (lowest A) collisions we do observe some difference. For the reaction volume this difference in not very significant. The fact that for the three smallest A the reaction volume does not grow when we average over 100 events has to do with the fact that for these nuclei their A nucleons are distributed homogeneously within a  sphere of a rather small radius. Thus, the fluctuations of initial positions of their nucleons can not increase the maximal volume of the fireball, as this happens for the distributions which do not have  well defined limits.  

On the other hand, the entropy per nucleon production in very peripheral symmetric collisions is principally different from that in central  asymmetric collision with the same average number of participants (since $N$ is the same, all the observed difference in $S/N$ has to be attributed to the entropy production). In the peripheral Au+Au collisions $S/N$ shows a very strong rise, which we have seen and discussed in section \ref{mu_T_S} (see Fig. \ref{SPN}), while in asymmetric central collisions of small nuclei with Au the $S/N$ remains more or less constant. 

We can also see in the bottom plot of Fig. \ref{SpNA} again an important effect that the $S/N$  production for the initial state averaged over many events strongly increases with respect to that for a single event, similarly as it appear in the bottom plot of Fig. \ref{SPN}.

\begin{figure}[ht!]  
	\begin{center}
		\resizebox{1.01\columnwidth}{!}
		{\includegraphics[width=\linewidth]{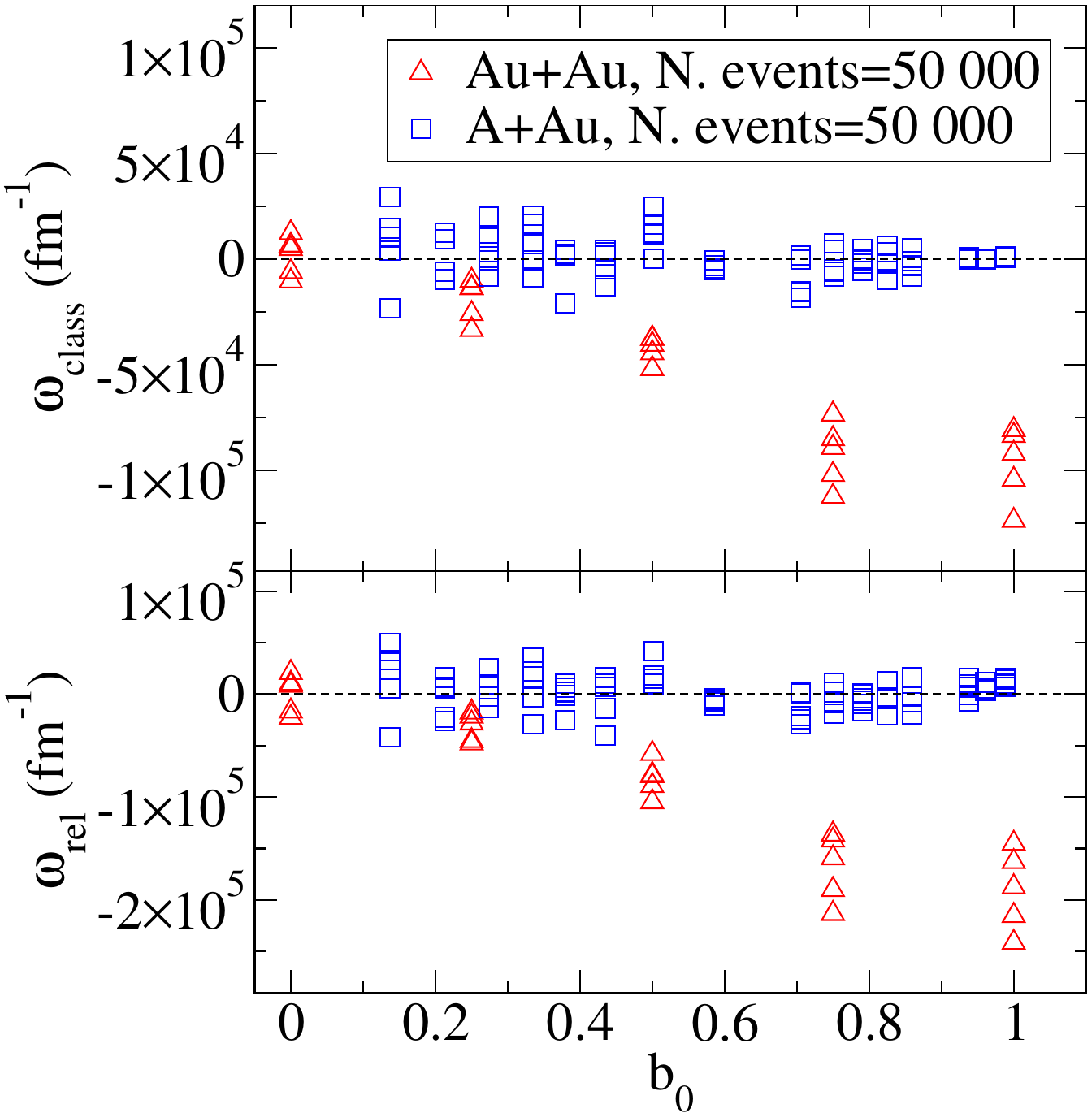}}
		\caption{ (color online)
			The classical, top plot, and relativistic, bottom plot, vorticities summed over all the non-empty cells obtained in symmetric Au+Au collisions and in asymmetric A+Au head on collisions, at $\sqrt{S_{NN}}=200$ GeV and $t_{fin}=$5 fm.
		}
		\label{CRVA}
	\end{center}
\end{figure}

However, if we compare the classical and relativistic vorticities (summed over all the cells), which are shown in Fig. \ref{CRVA}, we will observe that the results obtained in asymmetric A+Au head on collisions are completely different from those obtained in symmetric Au+Au collisions at different impact parameters. While in the latter the vorticity decreases with impact parameter until it reaches some minimum value, in asymmetric A+Au head on  collisions it oscillates around zero, as it should be due to symmetry reasons. Again, as in the previous section, we observe a qualitatively correct (expected) behavior of the vorticity accompanied by very strong fluctuations. 

It is interesting to see that for the large and medium size A (what is equivalent to $b_0\le 0.5$) the typical deviations are actually stronger for A+Au central collisions. On the other hand, for the small A (large $b_0$) the vorticity fluctuations around $0$ for asymmetric central collisions become small, while for peripheral Au+Au they grow.

\section{SUMMARY AND DISCUSSIONS}\label{CD}

We have presented a generalized ESRM, which takes into account fluctuations in the initial state of relativistic heavy ion collisions following the Glauber Monte Carlo approach. The random distribution of nucleons leads to fluctuations in the geometry of the collision, what may lead to an enhancement of the different components of the azimuthal distribution of emitted particles; the odd components,  such as direct flow ($v_1$) and triangular flow ($v_3$), will be particularly affected. 

By averaging our initial state over a large number of events we observed that the average values of the conserved quantities, like $N_{part}$, rapidly converge to some value, determined by the initial energy and impact parameter of the simulated reaction; and these values can be compared with the corresponding values in the initial state model without randomization. Qualitatively, such a comparison is  quite fair, although they never will be absolutely identical. For example, even for the $N_{part}$, which converges very rapidly, there is some discrepancy -   $<N_{part}(b=0)>$ will never be equal to $A_1+A_2$, as it was shown in section \ref{reacVol}.  The energy, baryon charge and flow velocity distributions generated in the GESRM, averaged over many events, also look qualitatively similar to those of the ESRM,  although the overall volume of the fireball is much bigger - such an averaging adds many cells with a very little amount of matter.  

On the other hand we have shown in section \ref{mu_T_S} that there is a principle difference in using single event initial state and the one averaged over many events with randomization. This difference mainly has to do with an entropy per nucleon, which strongly grows in an averaged initial state. Similar trend has also been observed for asymmetric A+Au head on collisions. We would like to stress that the $S/N$ in a single event in the GESRM with randomization and in the original ESRM, considering colliding nuclei as homogeneous spheres, are fairy comparable. Thus, the GESRM initial state, averaged over many events, only looks like smoothed over corresponding initial state obtained in the ESRM (Fig. \ref{ENERDENSOLDNEW}) - due to much higher initial entropy it will lead to differences in hadron production.

The GESRM, as well as the ESRM, generates a strong sheer and vorticity in the initial state. We have seen in section \ref{IV.vort}, the initial state fluctuations, introduced via classical Glauber Monte Carlo approach,  lead to an extremely strong fluctuations of the initial state vorticity, what tells us that we should be careful interpreting any results, based on vorticity, and/or comparing vorticities in different models. 

The new GESRM, as well the original ESRM, should be used to simulate initial state for further  fluid dynamical evolution for relativistic heavy ion collisions at high energies,  
$\sqrt{S_{NN}}\gtrsim 50$ GeV, i.e. at RHIC and LHC.

\section{Acknowledgments}\label{CD}
A.R.R. and V.K.M. acknowledge support from the Ministerio de Ciencia e Innovaci\'on of Spain through the “Unit of Excellence María de Maeztu 2020-2023” award to the Institute of Cosmos Sciences (CEX2019-000918-M) and under~the project PID2020-118758GB-I00.
V.K.M. also acknowledge support from the EU STRONG-2020 project under the program H2020-INFRAIA-2018-1, grant agreement no.\ 824093. 
The work of L.P.C. is supported in part by the Frankfurt Institute for Advanced Studies, Germany, the E\"otv\"os Lóránd Research Network of Hungary, the Research Council of Norway, grant no. 255253, and the National Research, Development and Innovation Office of Hungary, via the project NKFIH-468-3/2021.  


\end{document}